\documentclass[11pt]{article}
\usepackage[a4paper,margin=1in]{geometry}
\usepackage{amsmath,amssymb,amsfonts,amsthm,mathtools,bm}
\usepackage{booktabs,tabularx,array,multirow}
\usepackage[table]{xcolor}
\usepackage{graphicx}
\usepackage{adjustbox}
\usepackage{subcaption}
\usepackage{float}
\usepackage{pgfplots}
\pgfplotsset{compat=1.18}
\usepgfplotslibrary{groupplots}
\usepgfplotslibrary{statistics}
\usepackage{tikz}
\usepackage{hyphenat}
\usetikzlibrary{
    arrows.meta,
    positioning,
    fit,
    backgrounds,
    calc,
    decorations.pathreplacing
}
\usepackage[normalem]{ulem}

\definecolor{gapblue}{HTML}{0B5FD9}
\definecolor{revcolor}{HTML}{0A7D32}   


\usepackage[section]{placeins}
\usepackage{tcolorbox}
\tcbuselibrary{skins,breakable}
\newcounter{algctr}

\AtBeginDocument{}

\usepackage{enumitem}
\usepackage{microtype}
\usepackage[numbers,sort&compress]{natbib}
\usepackage{hyperref}
\hypersetup{
  pdftitle={Pareto-Optimal Hypersonic Design with Canonical-Coordinate
            Neural Operator Surrogates Under Ablation-Driven Shape Change
            and Incidence Variation},
  pdfauthor={Khemraj Shukla and Jeremy Redding and Luis Bravo and
             George Em Karniadakis},
  pdfsubject={Operator learning for hypersonic aerothermodynamics},
  pdfkeywords={DeepONet, operator learning, hypersonics, waverider,
               multi-objective optimization, Pareto front}
}
\usepackage[nameinlink,noabbrev]{cleveref}

\hypersetup{
  colorlinks=true,
  linkcolor=blue!55!black,
  citecolor=blue!55!black,
  urlcolor=blue!55!black
}

\AtBeginDocument{}
\newenvironment{keyword}{\par\medskip\noindent\textbf{Keywords: }}{\par\medskip}
\providecommand{\sep}{,\ }
\definecolor{fixedcolor}{RGB}{31,119,180}
\definecolor{adaptivecolor}{RGB}{255,127,14}
\definecolor{sensorcolor}{RGB}{44,160,44}
\definecolor{twostepcolor}{RGB}{214,39,40}
\definecolor{vanillacolor}{RGB}{148,103,189}
\definecolor{fnocolor}{RGB}{23,23,23}

\title{Pareto-Optimal Hypersonic Vehicle Design with Canonical-Coordinate Neural Operator Surrogates Under Ablation-Driven Shape Change and Incidence Variation}
 \author{
Khemraj Shukla\thanks{%
Corresponding author:
\href{mailto:khemraj_shukla@brown.edu}
{khemraj\_shukla@brown.edu}%
}
\\
\small Division of Applied Mathematics, Brown University\\
\small Providence, Rhode Island 02912, USA
\and
Jeremy Redding
\\
\small Oak Ridge Associated Universities\\
\small Oak Ridge, Tennessee 37830, USA
\and
Luis Bravo
\\
\small DEVCOM Army Research Laboratory\\
\small Adelphi Laboratory Center, Maryland 20783, USA
\and
George Em Karniadakis
\\
\small Division of Applied Mathematics, Brown University\\
\small Providence, Rhode Island 02912, USA
}
\date{}
\begin{document}
\maketitle
\vspace*{-1cm}
\begin{abstract}
High-fidelity computational fluid dynamics (CFD) is essential for resolving
the coupled aerodynamic and aerothermal phenomena of hypersonic flight, but
its cost precludes direct use in multi-objective design optimization. We
develop three DeepONet-based operator surrogates across two hypersonic
configurations that pose distinct learning problems, unified by one device:
the trunk is given a coordinate system in which the feature that varies is
held fixed. The first configuration is a family of $37$ spherically blunted
tangent ogives whose profile is varied through the ogive radius of
curvature. The family is a prescribed geometric parameterisation chosen
to span the forebody change that nose-tip ablation produces; no ablation
model, material response or surface recession is computed. Because each geometry
occupies a different computational domain, a body-fitted canonical
transformation maps every solution onto a common $(s,\eta)$ reference space
while retaining the physical coordinates needed to evaluate forces and heat
transfer. A parameter-dependent branch network predicts reduced coefficients
and a Transolver trunk, whose Physics-Attention captures long-range flow
interactions, learns the spatial basis; a two-step strategy optimizes
case-wise coefficients and the trunk jointly, orthonormalizes the basis by QR
factorization, and trains a second-stage branch to predict the transformed
coefficients at unseen geometries. Surrogates are trained for density,
pressure, velocity, the two-temperature-model temperatures and thermal
conductivities, and chemical-species mass fractions. Reconstructed fields
recover the drag coefficient, the mean wall heat flux and the peak wall heat
flux to $1.16\%$, $1.62\%$ and $5.29\%$ in leave-one-out testing, and a sweep
of $1001$ candidate geometries yields a Pareto front of $200$ designs
on which $6.68\%$ in drag is exchanged for$4.42\%$ in peak
heating, the last reported as a diagnostic rather than adopted as a design
objective. The second configuration is the three-dimensional AEDC waverider
at $29$ angles of attack, for which the mesh is static but the leading-edge
shock moves with incidence, and for which two surrogates are built. For the
surface quantities, the same two-step construction applied to the wall heat
flux and the total traction predicts lift, drag and thermal loading to within
$0.22\%$ in the integrated wall heat load and to absolute errors below
$2.6\times10^{-5}$ in the force coefficients across the operating envelope,
from which Pareto fronts identify the angles of attack that best trade
minimum drag, maximum lift and minimum heat load. For the volume field a
Cartesian trunk averages neighbouring shock positions into a smeared front,
and two additions are required: a frozen ShockNet supplying a shock-aligned
coordinate that holds the discontinuity at approximately the same value for
every incidence, and a hotspot-aware mixture of experts correcting the
localized residual that remains. Together these recover the density field to
a relative $L_2$ error of $12.28\%$ at a held-out incidence on
$5.08\times10^{7}$ nodes. The $1001$-design sweep completes in approximately
three minutes on a single CPU, placing field-resolved multi-objective
hypersonic design at a cost that allows the design space to be explored
rather than sampled.
\end{abstract}

 
 
 \section{Introduction}
\label{sec:intro}

Hypersonic flight combines strong shock waves, thin viscous boundary
layers, steep pressure and temperature gradients, and severe surface
heating \cite{anderson2006hypersonic}. Shock compression and viscous
dissipation convert freestream kinetic energy into internal energy, and
at high enthalpy the post-shock temperature becomes large enough to
excite vibrational modes and drive dissociation and finite-rate
chemistry \cite{scoggins2020mutation++}. The flow then departs from
thermal and chemical equilibrium, and a calorically perfect-gas model
can no longer represent how energy is shared among translational,
rotational, vibrational and chemical modes
\cite{gnoffo1989nonequilibrium,passiatore2022thermochemical}. This
distinction between a high Mach number and high-enthalpy
thermochemistry matters for aerothermal prediction: composition and
energy partitioning set the temperature and transport properties, and
through them the heat delivered to the wall.

Surface heating is a primary design constraint because it fixes the
thermal-protection requirements, allowable material temperatures and
feasible flight envelope. It is also strongly coupled to geometry: the
forebody curvature and nose radius govern the bow-shock stand-off,
post-shock state and near-wall gradients. The Sutton--Graves relation
\cite{sutton1971},
\begin{equation}
    \dot{q}_{s} \propto V_{\infty}^{3}\sqrt{\rho_{\infty}/R_n},
    \label{eq:intro_sutton_graves}
\end{equation}
shows that a blunter nose lowers stagnation heating, but the added
bluntness also raises pressure drag and alters the shock structure.
Hypersonic shape design is therefore an aerodynamic--aerothermal
trade-off rather than a purely geometric problem, as earlier CFD-based
optimization of blunt bodies has shown \cite{seager2017shape}.

High-fidelity computational fluid dynamics (CFD) resolves these coupled
effects, but at a cost that limits how many configurations can be
explored. Shocks and wall layers demand highly clustered meshes, and
for explicit time integration the step size is bounded by both the
convective and the diffusive stability limits. For a high-order
discretization of polynomial degree $N$, these scale as \cite{chan2018discretely, ranocha2021adaptive}
\begin{equation}
    \Delta t_{\mathrm{con}} \sim C_{\mathrm{con}}\,
    \frac{\Delta x}{N^{2}\,\big(|u|+c\big)},
    \qquad
    \Delta t_{\mathrm{vis}} \sim C_{\mathrm{vis}}\,
    \frac{\Delta x^{2}}{N^{4}\,\nu},
    \label{eq:intro_timestep}
\end{equation}
where $\Delta x$ is the local element size, $c$ the speed of sound, $\nu$
the kinematic viscosity, and $C_{\mathrm{con}}$, $C_{\mathrm{vis}}$
method-dependent constants \cite{karniadakis2013spectral}. The additional
factors $N^{2}$ and $N^{4}$, combined with the small $\Delta x$ of
near-wall elements, severely restrict the admissible time step.
Finite-rate chemistry further adds stiff source terms that require
implicit, semi-implicit or operator-split integration
\cite{wu2019efficient}. Shape optimization, uncertainty quantification
and sensitivity analysis may each need tens to thousands of such
solutions, each with its own geometry generation, meshing, convergence
and post-processing. Embedding the solver directly in a design loop is
therefore often prohibitive, which motivates surrogates that reproduce
high-fidelity results at low online cost
\cite{forrester2009recent}.

Most surrogates map a few design parameters to a few scalar outputs
such as drag, lift or peak heating. For hypersonic aerothermodynamics
this is insufficient: wall heat flux and viscous stress depend on local
gradients, species composition, internal-energy modes and spatially
varying transport properties. A \emph{field-level} surrogate is more
useful, since any number of engineering quantities can then be
recovered from the predicted state without training a new regression
model for each objective. Neural operators learn exactly such
parameter-to-field maps \cite{kovachki2023neural, shukla2026fixed, de2025deep,jiang2026complex,bora2026inception,shukla2024comprehensive}. The Deep Operator
Network (DeepONet) \cite{lu2021deeponet} represents a field $q$ as
\begin{equation}
    \widehat{q}(\boldsymbol{x};\xi)
    = \sum_{j=1}^{r} b_j(\xi)\,\phi_j(\boldsymbol{x}),
    \label{eq:intro_deeponet}
\end{equation}
where a branch network gives the coefficients $b_j$ of the parameter
$\xi$ and a trunk network gives the spatial basis $\phi_j$. DeepONets
have already been used to predict coupled flow and finite-rate
chemistry behind hypersonic shocks \cite{mao2021deepmmnet}. Training
branch and trunk simultaneously, however, poses a difficult nonlinear
optimization problem. Lee and Shin \cite{lee2024training} proposed a
two-step procedure that first learns the trunk together with free
per-case coefficients, orthogonalizes the resulting basis, and only
then trains the branch to reproduce the coefficients. This separation
of spatial representation from parameter dependence suits the present
problems, where complex flow structures vary with a single geometric
or flow parameter.

Three difficulties then stand in the way. The first is geometry:
changing the vehicle shape changes the physical domain and the points at
which the solution is known, which has motivated architectures that learn
a deformation to a latent regular grid \cite{li2023geofno,li2021fno} or
that fuse geometric information into the trunk \cite{peyvan2026fusion}.
The second is localization: shocks, boundary layers and regions of strong
chemical activity occupy a small fraction of the domain, global Fourier
layers tend to over-smooth such features
\cite{liu_schiaffini2024localized}, and discontinuities whose position
moves with the input are a known hard case for operator learning
\cite{lanthaler2023discontinuities}, reflecting the spectral bias of
neural networks toward smooth functions \cite{rahaman2019spectral}. The
third is data: thermochemical CFD is so expensive that only a handful of
solutions are available \cite{zhang2026turbulence,cobelli2023physics, shukla2025neurosem,shukla2026uncertainty}, so the surrogate must extract as much shared
spatial structure as possible from very few samples. How the existing
work on the first two difficulties relates to the hypersonic, reacting
setting considered here is taken up in detail at the end of this section.

This work addresses these difficulties across two hypersonic
configurations, for which three operator surrogates are constructed. The
first configuration is high-enthalpy flow over a family of spherically
blunted tangent ogives parameterized by $\xi_g$, and carries one
surrogate; the second is the three-dimensional AEDC waverider at varying
incidence, and carries two --- one for the surface quantities and one for
the volume field.

\paragraph{The ogive surrogate.}
Because the correspondence between ogive geometries is known explicitly,
the operator does not need to learn a domain deformation. Instead, every
CFD solution is mapped from its physical coordinates $(x,y)$ to a
common body-fitted domain $\widehat{\Omega}$ with coordinates
$(s,\eta)$, and the learning problem becomes
\begin{equation}
    \mathcal{G}:\ \xi_g \longmapsto q(s,\eta;\xi_g),
    \qquad (s,\eta)\in\widehat{\Omega},
    \label{eq:intro_canonical_operator}
\end{equation}
while the physical coordinates are kept for post-processing. The
geometric transformation thus precedes operator learning, and the same
spatial coordinates are supplied for every geometry. The spatial basis is
learned by a Transolver \cite{wu2024transolver}, whose Physics-Attention
groups mesh points into learned physical states and attends among them,
within the two-step procedure of \cite{lee2024training}. The formulation
thereby separates geometric transformation, spatial representation and
geometry-to-solution dependence. Within this surrogate, independent
branch--trunk models are trained for fourteen fields: the flow variables
$\{\rho,u,v,p,T\}$, the five air species
$\{Y_{\mathrm{N_2}},Y_{\mathrm{O_2}},Y_{\mathrm{NO}},
Y_{\mathrm{N}},Y_{\mathrm{O}}\}$, and the internal-mode and transport
variables $\{T_m,k,k_m,\mu\}$. Since the turbulent conductivity is zero
in these cases, the wall heat flux is reconstructed as
\begin{equation}
    q_{w,\mathrm{cond}}
    = k_w \left.\frac{\partial T}{\partial n}\right|_w
    + k_{m,w}\left.\frac{\partial T_m}{\partial n}\right|_w ,
    \label{eq:intro_wall_heat_flux}
\end{equation}
from the predicted temperatures, their wall-normal gradients and the
predicted conductivities, rather than from a temperature field with an
externally prescribed conductivity; $\mu$ is used likewise for the
viscous wall stress. Generalization is assessed by leave-one-out (LOO)
validation within each of the integer and half-step geometry families
separately, so each model sees only 19 or 16 training geometries, and
the held-out case is excluded from every data-dependent step,
including normalization, basis learning and orthogonalization. The
validated surrogate is then used to construct Pareto fronts with $\xi_g$
as the design variable and the reconstructed surface heat transfer as the
aerothermal objective.

\paragraph{The two waverider surrogates.}
The second configuration is the three-dimensional waverider tested in
the Arnold Engineering Development Center (AEDC) Hypervelocity Wind
Tunnel No.~9, a vehicle designed to ride an attached leading-edge
shock \cite{nonweiler1959aerodynamic}. Building on earlier DeepONet
predictions for this configuration over a range of angles of attack
\cite{shukla2024waverider}, the geometry and the mesh are now fixed and
the angle of attack $\alpha$ varies, so the difficulty is no longer
deformation but a three-dimensional shock whose position and strength
change with $\alpha$. Two separate surrogates are built. For the surface
quantities, two-step DeepONet surrogates of the wall heat flux and the
total traction give lift, drag and thermal loading across the operating
range, from which surrogate-based Pareto fronts \cite{deb2002nsga2}
identify the angles of attack that best balance minimum drag, maximum
lift and minimum heating; the fixed mesh and the smooth variation of the
surface fields with incidence mean that no special coordinate treatment
is needed here. For the volume field it is, because a trunk expressed in
Cartesian coordinates alone must place the same discontinuity differently
at every incidence and in practice averages neighbouring shock positions
into a smeared front. A shock sensor built from the training solutions
alone therefore supervises a ShockNet that predicts the distance to the
shock surface; this shock-aligned coordinate is supplied to the trunk,
and a hotspot-aware mixture of experts \cite{jacobs1991adaptive} corrects
the residual that remains concentrated near the shock. That the surface
surrogate needs none of this machinery and the volume surrogate does not
converge without it is itself a result, and we return to it in
\Cref{sec:conclusions}.

The main contributions of this work are:
\begin{enumerate}
    \item a body-fitted canonical representation that gives every
    geometry in a shape family the same spatial coordinates, combined
    with a two-step Transolver--DeepONet for data-scarce,
    geometry-dependent operator learning;

    \item a fourteen-field thermochemical surrogate for the ogive family,
    including species and transport properties, from which wall heat flux
    and viscous stress are reconstructed consistently with the CFD model;

    \item a two-step surrogate for the waverider surface quantities ---
    the wall heat flux and the components of the total traction --- which
    recovers the integrated loads to within $0.22\%$ in the wall heat load
    across the operating envelope and supplies the objectives for the
    incidence design study;

    \item a shock-aligned DeepONet with hotspot-aware mixture-of-experts
    correction for the three-dimensional waverider volume field, where the
    moving shock defeats a Cartesian trunk; and

    \item surrogate-driven multi-objective optimization over shape for
    the ogive family and over angle of attack for the waverider.
\end{enumerate}

\paragraph{Relation to prior work.}
Operator learning for flows on varying domains is an active area, and it is
worth being precise about the regimes in which it has been demonstrated and
about the objectives against which it has been trained, because the
present problem differs on both counts. Geometry-aware Fourier neural
operators~\cite{li2023geofno} handle varying domains through a learned
deformation to a latent regular grid; their reported cases are elasticity,
plasticity, advection on the sphere, incompressible pipe flow, and an airfoil
in the inviscid Euler system at $M_\infty=0.8$, which the authors describe as
transonic. Localised integral and differential
kernels~\cite{liu_schiaffini2024localized} improve the representation of
sharp local features and are demonstrated on Darcy flow, incompressible
Navier--Stokes at $\mathrm{Re}=5000$, a diffusion--reaction system, the
spherical shallow-water equations and incompressible flow past a cylinder; no
Mach number appears in that work. The theory of operator approximation for
solutions with discontinuities~\cite{lanthaler2023discontinuities} is
developed for inviscid problems --- advection, inviscid Burgers and the
compressible Euler equations --- in which a shock is a surface of measure
zero. None of these studies contains a supersonic or hypersonic case, and
none addresses a detached bow shock whose standoff and inclination are set by
the body and move as the body changes.

The closest prior work is Fusion-DeepONet~\cite{peyvan2026fusion}, which does
treat a family of blunt geometries at $M_\infty=10$ and reports relative
$L_2$ errors of $3.70\%$ on a uniform grid and $6.39\%$ on an irregular one.
Those calculations solve the hyperbolic part of the compressible
Navier--Stokes equations --- the inviscid Euler system --- for a calorically
perfect gas with $\gamma=1.4$ and $p=\gamma\rho T$. The absence of
thermochemistry is not an implementation detail. For a normal shock in a
calorically perfect gas the static-temperature ratio is
\begin{equation}
  \frac{T_2}{T_1}
  = \frac{\big[2\gamma M_\infty^2-(\gamma-1)\big]
           \big[(\gamma-1)M_\infty^2+2\big]}
          {(\gamma+1)^2 M_\infty^2},
  \label{eq:normal_shock_temperature_ratio}
\end{equation}
which gives $T_2/T_1=20.4$ at $M_\infty=10$, so a cold freestream implies a
post-shock temperature near $6000$~K, and $T_2/T_1=24.5$ at the $M_\infty=11$
of the present study. Molecular oxygen is essentially fully dissociated by
$4000$~K and nitrogen dissociates above it; in a perfect-gas calculation the
dissociation energy has nowhere to go and is carried as translational
temperature instead, so the post-shock state, the shock standoff and the wall
heat flux all depart from those of the real gas. A perfect-gas computation at
these Mach numbers is a high-speed compressible-flow problem, and a demanding
one, but the aerothermodynamics that governs surface heating --- and
therefore the thermal objective of any hypersonic design study --- is absent
from it by construction. The present work solves the viscous Navier--Stokes
equations with a five-species finite-rate air-chemistry model and a
two-temperature thermal-nonequilibrium formulation, and learns the resulting
fourteen-field state rather than a four-field perfect-gas one.

\paragraph{What is being represented.}
A second difference concerns the object to be learned. In the inviscid limit
a shock is a surface across which the state jumps, and the natural remedy for
a moving discontinuity is to transport a coordinate onto that surface; this is
the setting of the theoretical and localised-kernel work cited above. In the
compressible Navier--Stokes equations the shock is instead a layer of finite
thickness, and under reacting hypersonic conditions it is followed by a
substantially thicker relaxation region in which vibrational energy exchange
and finite-rate dissociation proceed towards equilibrium. The feature to be
learned is therefore a zone with internal structure, across which the
thermodynamic state, the two temperatures and the species composition vary on
different length scales, rather than a line at which they jump. The
representation adopted here is built for that object: the shock coordinate
$\widehat d = d/\delta$ is normalised by a shock-layer thickness rather than
measured as a signed distance to a surface, the bounded coordinate
$\chi=\tanh\widehat d$ varies smoothly across the layer instead of switching
at it, and the training objective of \Cref{eq:waverider_r1_loss} carries
separate shock-core and shock-band terms so that the interior of the zone is
resolved and not merely bracketed. A shock-aligned frame is not new. Roohi and
Mahdavi~\cite{roohi2025shockaware,roohi2026shockcentered} normalise a
shock-centred coordinate by a layer thickness for rarefied micro-nozzle
flows, raising the leading proper orthogonal decomposition mode from $83.3\%$
to $98.3\%$ of the fluctuation energy. Their shock is a station on a
centreline, placed by an affine fit in the control variable that presumes
translation, in a monatomic non-reacting gas; $\Gamma_s(\alpha)$ here is a
curved three-dimensional surface that reorients with the parameter, placed by
a learned map, in a viscous reacting flow with a relaxation region behind it.
More broadly, ShockNet is a learned registration map, the device used against
slow Kolmogorov $n$-width decay by shifted proper orthogonal
decomposition~\cite{reiss2018spod,krah2024robustspod}, registration-based
reduction~\cite{taddei2020registration,gowrachari2025registration} and
implicit shock tracking~\cite{zahr2018shocktracking}. It differs in being
learned rather than prescribed, in acting on an operator trunk rather than a
snapshot matrix, and in being available at an unseen parameter with no
reference solution to register against.

\paragraph{What the training objective controls.}
The remaining difference is methodological, and it is the one that decides
whether a surrogate can support a design decision at all. Validation in the
studies cited above is reported as a field-wise relative error, and the
objectives minimised to obtain it are correspondingly field-wise averages
over the computational domain. Such an objective does not control the
quantities on which a hypersonic design decision rests, for three distinct
reasons. The first is one of measure: every engineering quantity is defined
on the wall, and the wall is a set of measure zero in the volume. Discretely,
the present waverider mesh carries $N_s=266{,}848$ surface nodes against
$5.08\times10^{7}$ in the volume, so a uniformly weighted field objective
assigns the entire region that determines drag and heating a weight of
roughly one node in $190$. A surrogate may therefore report an excellent
volume-averaged error while being substantially wrong about every quantity a
designer uses. The second is one of differential order: the wall heat flux
and the viscous traction are functionals of wall-normal derivatives,
$q_w=-k\,\partial T/\partial n$ and $\tau_w=\mu\,\partial u/\partial n$, so a
loss imposed on field values is a $C^0$ objective that places no bound on the
$C^1$ error. Differentiation multiplies each error mode by its wavenumber and
therefore amplifies precisely the high-wavenumber residual that a smooth
global basis is least able to remove, which is why gradient fidelity is
reported here as a validation quantity in its own right
(\Cref{tab:loo_gradient_correlation}) rather than inferred from field
accuracy. The third concerns the direction of the error rather than its size:
integration over the wall suppresses error only to the extent that the error
is zero-mean over the surface, whereas a near-wall error produced by an
unconstrained surrogate is systematic --- it carries a sign, set by the
basis's inability to turn the velocity and temperature profiles over within
the boundary layer --- and systematic error integrates coherently into the
force and the integrated heat load. A field-wise metric cannot distinguish
the two cases, which is the precise sense in which it fails to certify a
design quantity.

The boundary condition makes the same point in its sharpest form. In an
inviscid formulation the no-slip condition does not exist: the wall condition
is tangency, $\mathbf{u}\cdot\mathbf{n}=0$, and with it there is no boundary
layer, no skin friction and no conductive wall heat flux. Both the
geometry-family study at $M_\infty=10$ and the airfoil case of the
geometry-aware operator cited above are inviscid, so neither has been
required to confront the one boundary condition on which the thermal
objective of a hypersonic design study depends. In the present framework that
condition is imposed rather than hoped for. Wall-normal derivatives are
reconstructed from the near-wall grid layers by a local quadratic fit in which
the no-slip condition $U=V=0$ is enforced exactly at the surface
(\Cref{app:traction}), and the known isothermal wall temperature is enforced
in the same way in the reconstruction of the heat flux
(\Cref{app:heat_flux}), so the surface traction and the wall heat flux are
consistent with the physical boundary conditions irrespective of what the
network returns at the wall node itself. The reduction from predicted fields
to engineering quantities is the solver's own post-processing, applied
unchanged, so the surrogate is held to the same reduction as the
high-fidelity data; and accuracy is then reported on the wall functionals
themselves (\Cref{subsec:res_wall_qoi}), not only on the fields from which
they are derived.

Measured on the four fields the two studies share, and by the same relative
$L_2$ metric, the present surrogate is also the more accurate on the
comparable configuration --- the two-dimensional tangent-ogive family:
$2.84\%$ in $\rho$, $0.56\%$ in $u$, $3.48\%$ in $v$ and $3.27\%$ in $p$ over
the interpolation folds, against $3.41$, $0.68$, $5.91$ and $4.80\%$
respectively for the uniform-grid case cited above, and lower still relative
to the irregular-grid case. Even this comparison is not controlled, since the
two problems differ in geometry, in data and in sampling; what it establishes
is that representing the reacting, thermally nonequilibrium state, and
constraining the surrogate at the wall, have not been paid for in field
accuracy. The three-dimensional waverider volume reported in
\Cref{sec:results_waverider_volume} is not of this class at all and no
published result is available to compare it with: it is a reacting,
thermally nonequilibrium field on $5.08\times10^{7}$ nodes about a complete
three-dimensional configuration, with a shock surface that reorients in
three dimensions as the incidence changes. It is reported here on its own
terms, together with the architectural devices required to make it tractable.

\paragraph{What is new here.}
Taken together these observations locate the gap this paper addresses. The
geometry-aware operator literature has not been exercised in the supersonic
or hypersonic regime; the work that has reached high Mach number has done so
without chemistry, without viscosity and therefore without surface heating;
the methods developed for moving discontinuities treat them as surfaces
rather than as zones; and validation throughout is reported as a field-wise
average, which neither weights nor constrains the wall on which every design
quantity is defined. To our knowledge no operator-learning study has
predicted the geometry-dependent, chemically reacting and thermally
nonequilibrium hypersonic field, subject to the wall conditions that make its
surface quantities meaningful, and carried that prediction through to a
design decision.

Three features distinguish the present framework from the work above. It
treats the two ways such a design problem can be parameterised --- a family
of geometries on distinct domains, and a fixed domain whose dominant flow
feature moves --- within one formulation, and shows both to be instances of a
single principle: give the trunk a coordinate in which the varying feature is
stationary. The three surrogates built here also delimit when that principle
is needed. It is required for the ogive family, where the domain itself
changes, and for the waverider volume, where the shock moves through a fixed
domain; it is not required for the waverider surface, where the mesh is fixed
and the fields vary smoothly with incidence, and where a direct two-step
construction is the most accurate surrogate in this work. The contrast
between the second and third cases, which differ in nothing but the quantity
predicted, is the clearest evidence available that the additional machinery
earns its place rather than merely being available. Second, the framework
carries the surrogates through to a decision rather than stopping at field
accuracy, reducing the predicted fields to engineering quantities by the same
post-processing applied to the solver output and driving a multi-objective
optimisation whose front is reported, tested for robustness and used to
select an operating point. Third, it sets the accuracy of the surrogate
against the size of the design differences it is asked to resolve --- a
comparison that is favourable for the drag and the integrated heat load and
unfavourable for the peak heat flux, which we therefore report as a
diagnostic rather than adopt as an objective. A surrogate study that omits
this last comparison cannot establish whether its Pareto front is a result or
an artefact of its own error.

\paragraph{Outline.}
\Cref{sec:prob_setup} describes the two configurations and the generation of
the high-fidelity data. \Cref{sec:computational_methodology} sets out the
three surrogates: the canonical-coordinate Transolver--DeepONet for the ogive
family, the two-step construction for the waverider surface quantities, and
the shock-aligned DeepONet with hotspot-aware correction for the waverider
volume field. The results follow in the same order ---
\Cref{sec:loo_assessment} for the ogive family and \Cref{sec:moo} for its
design study, \Cref{sec:results_waverider_surface} for the waverider surface
quantities and the selection of incidence, and
\Cref{sec:results_waverider_volume} for the volume field.
\Cref{sec:conclusions} summarises the findings and states the limitations.

\section{Problem Setup and High-Fidelity Data Generation}
\label{sec:prob_setup}
 
This section sets out the physical configurations, freestream conditions,
thermochemical models and numerical procedures from which the high-fidelity
CFD databases of this study are generated. Two hypersonic configurations are
treated, and between them they supply the training data for the three
surrogates of \Cref{sec:computational_methodology}. The first is a
parameterised two-dimensional spherically blunted tangent ogive, used for
geometry-dependent thermochemical flow prediction and aerothermal shape
optimisation, and supporting one surrogate. The second is the
three-dimensional AEDC waverider at varying incidence, which supports two: one
for the surface quantities from which the aerodynamic and thermal loads are
formed, and one for the volume field, where the leading-edge shock moves with
the operating condition.

\subsection{Parameterised Two-Dimensional Spherically Blunted Tangent Ogive}
\label{sec:ogive_data}
 
The first configuration is a two-dimensional spherically blunted tangent ogive
in a Mach-$11$ freestream at
\begin{equation}
    p_{\infty}=15{,}027.5~\mathrm{Pa},
    \qquad
    T_{\infty}=293~\mathrm{K},
    \label{eq:ogive_freestream}
\end{equation}
with the incoming gas modelled as a five-species air mixture of initial
composition
\begin{equation}
    Y_{\mathrm{N_2},\infty}=0.767,
    \qquad
    Y_{\mathrm{O_2},\infty}=0.233 .
    \label{eq:ogive_composition}
\end{equation}
 
Tangent-ogive forebodies have been studied extensively in connection with
sounding-rocket configurations \cite{keener1977side,wing1971tangent}. At
hypersonic speeds, however, a sharp forebody concentrates severe aerodynamic
heating near the nose. A spherical blunt section alters the bow-shock
structure and spreads the thermal load over a larger area, which makes the
spherically blunted tangent ogive a useful canonical configuration for
examining how forebody geometry, shock structure, thermochemical behaviour
and surface heating are coupled.
 
The geometry is parameterised so as to generate a family of related
forebodies. The variation is motivated by the nose-tip recession that
accompanies high-temperature flight, but it is important to be clear
about what is and is not modelled. The geometries are \emph{prescribed}: the
shape parameter $\xi_g$ is swept over a chosen range to generate a family of
forebodies spanning the profile change that ablation produces. No ablation
physics enters the computation --- there is no surface energy balance, no
pyrolysis or char model, no material thermal response, no surface blowing and
no recession rate. The wall is treated as a fixed, non-receding isothermal
boundary in every case, and each geometry is an independent steady solution
rather than a stage of a time-evolving shape. What the family provides is a
controlled setting in which the flow depends on forebody geometry alone,
which is what the operator surrogate and the subsequent aerothermal shape
optimisation require.

\subsubsection{Geometry Parameterisation}
\label{sec:ogive_geometry}
 
The geometry is built from a fixed reference length, base radius and
spherical nose radius,
\begin{equation}
    L=1.0~\mathrm{m},
    \qquad
    R=0.25~\mathrm{m},
    \qquad
    r_n=0.055~\mathrm{m},
    \label{eq:ogive_fixed}
\end{equation}
together with a dimensionless shape factor $\xi_g$ varied over
\begin{equation}
    0.50 \leq \xi_g \leq 0.95 .
    \label{eq:ogive_range}
\end{equation}
 
Within the geometry-generation procedure, $\xi_g$ scales the radius of the
circular arc that defines the tangent-ogive section. Denoting that radius by
$R_o$,
\begin{equation}
    R_o
    =
    \xi_g\,\frac{R^2+L^2}{2R},
    \label{eq:ogive_radius}
\end{equation}
so that a single parameter sets the curvature of the ogive while the
reference dimensions and the spherical nose radius stay fixed.
 
The axial position of the centre of the spherical blunt section follows as
\begin{equation}
    x_0
    =
    \sqrt{
        \left(R_o-r_n\right)^2
        -
        \left(R_o-R\right)^2
    },
    \label{eq:ogive_x0}
\end{equation}
the tangency point between the spherical nose and the ogive arc lies at
\begin{equation}
    y_t
    =
    r_n\,
    \frac{R_o-R}{R_o-r_n},
    \label{eq:ogive_yt}
\end{equation}
with axial coordinate
\begin{equation}
    x_t
    =
    x_0
    +
    \sqrt{
        r_n^2-y_t^2
    },
    \label{eq:ogive_xt}
\end{equation}
and the nose reaches forward to
\begin{equation}
    x_a=x_0+r_n .
    \label{eq:ogive_xa}
\end{equation}
 
The upper surface is therefore
\begin{equation}
y_{\mathrm{upper}}(x)
=
\begin{cases}
\sqrt{R_o^2-x^2}+R-R_o,
& x \leq x_t, \\[2mm]
\sqrt{r_n^2-(x-x_0)^2},
& x_t < x \leq x_a ,
\end{cases}
\label{eq:ogive_surface_upper}
\end{equation}
and the lower surface its reflection about the centreline,
\begin{equation}
    y_{\mathrm{lower}}(x)
    =
    -y_{\mathrm{upper}}(x).
    \label{eq:ogive_surface_lower}
\end{equation}
 
\begin{figure}
    \centering
    \begin{subfigure}{\linewidth}
        \centering
        \includegraphics[width=\linewidth]{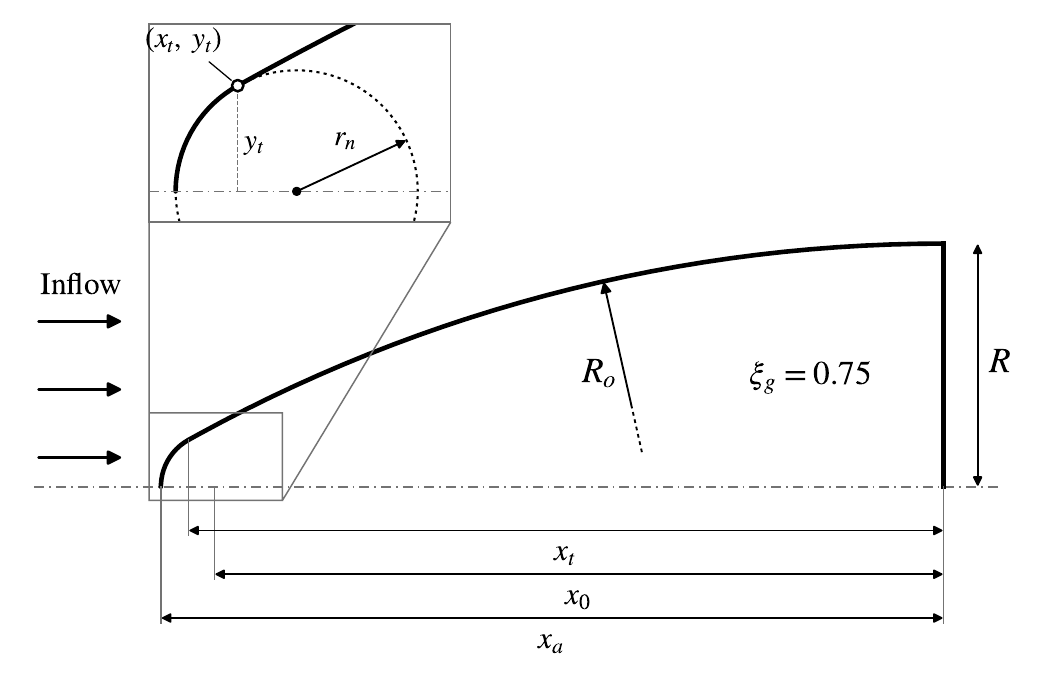}
        \caption{Reference geometry ($\xi_g=0.75$). The ogive arc of radius
        $R_o$ meets the spherical nose of radius $r_n$ tangentially at
        $(x_t,y_t)$; axial positions are measured from the base. The inset
        enlarges the nose region.}
        \label{fig:ogive_ref}
    \end{subfigure}
 
    \vspace{0.6em}
 
    \begin{subfigure}{\linewidth}
        \centering
        \includegraphics[width=\linewidth]{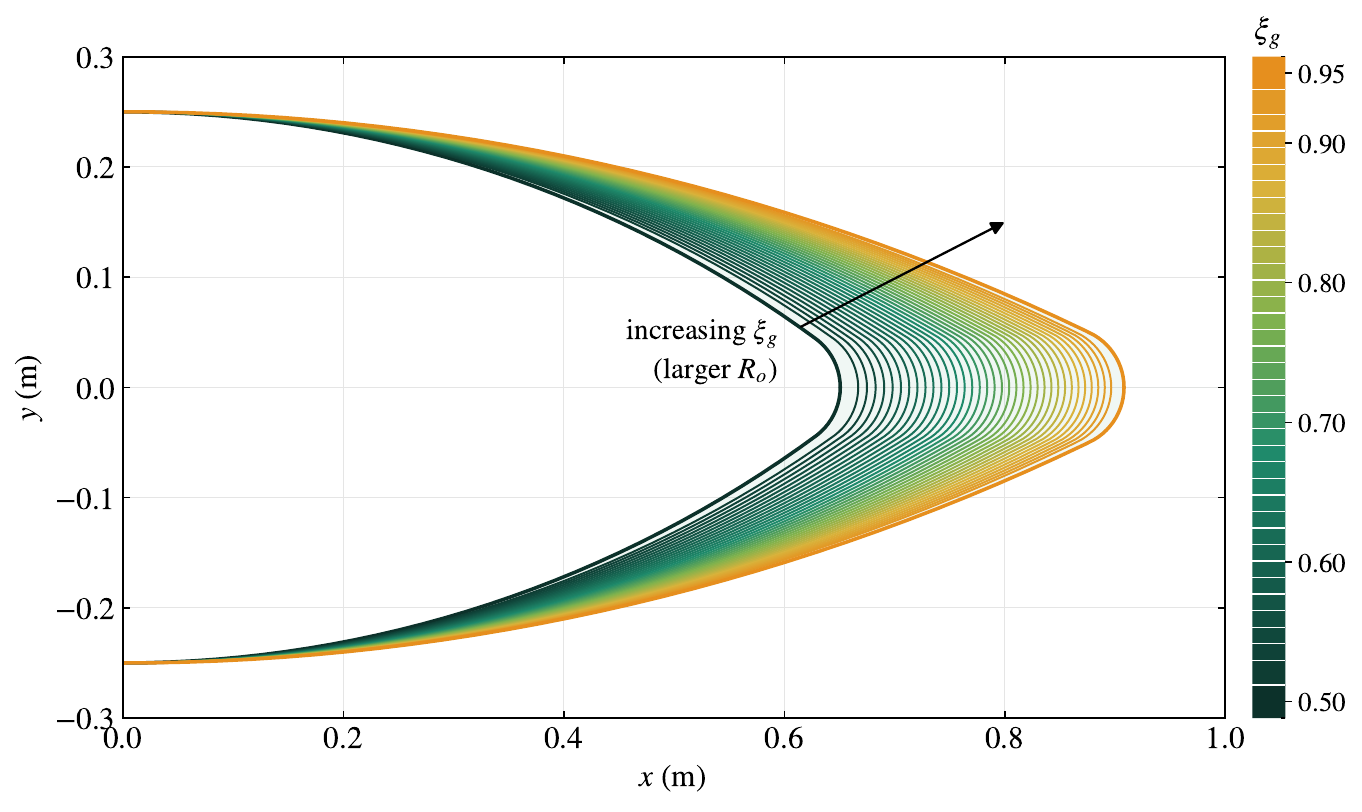}
        \caption{The $37$ geometries, coloured by $\xi_g\in[0.50,\,0.95]$:
        $20$ uniformly spaced cases and $17$ midpoint cases.}
        \label{fig:ogive_family}
    \end{subfigure}
    \caption{Parameterisation of the spherically blunted tangent ogive. The
    base radius $R=0.25$~m, nose radius $r_n=0.055$~m and reference length
    $L=1$~m are fixed. The single shape parameter $\xi_g$ sets the ogive
    radius $R_o=\xi_g\,(R^2+L^2)/(2R)$, so increasing $\xi_g$ lengthens the
    body.}
    \label{fig:ogive_param}
\end{figure}

The result is a one-parameter family of smoothly related spherically blunted
tangent ogives. The reference geometry is shown in
\Cref{fig:ogive_ref} and the $37$ geometries used for data generation in
\Cref{fig:ogive_family}.
 
\subsubsection{Flow Solver and Numerical Configuration}
\label{sec:ogive_numerics}
 
The high-fidelity tangent-ogive simulations use the Eilmer finite-volume CFD
framework \cite{gibbons2023eilmer}, which is built for compressible
high-speed and hypersonic flow and supplies the thermochemical modelling that
nonequilibrium reacting flows require. For every member of the geometry
family, the two-dimensional unsteady compressible Navier--Stokes equations
are solved together with the species and internal-energy equations.

The formulation is planar rather than axisymmetric, and this bears on
how the configuration should be read. The profile of
\Cref{eq:ogive_surface_upper} is that of a spherically blunted tangent ogive,
and the family is named for the body of revolution from which the profile is
taken; the computation, however, resolves the flow about the planar section,
so the nose is a circular arc of radius $r_n$ rather than a sphere, and the
afterbody is an ogival wedge rather than a body of revolution. The
quantitative consequences are known and systematic. For the same nose radius
and freestream state, a two-dimensional blunt body carries a larger shock
stand-off distance than the corresponding axisymmetric one, because the flow
relieves in one transverse direction rather than two; by the same mechanism
the stagnation-point velocity gradient is smaller, and the stagnation heating
is correspondingly lower. Absolute values of
stand-off, stagnation heat flux and drag computed here should therefore not
be transferred directly to a three-dimensional vehicle.

What does transfer is the comparison across the family. Every geometry is
computed with the same formulation, the same mesh topology and the same
thermochemical model, so the differences between designs --- which is what
the surrogate learns and what the Pareto front ranks --- are consistently
resolved even though the absolute levels are those of the planar problem. The
surrogate construction itself is dimension-agnostic: the canonical mapping,
the two-step training and the quantity-of-interest reduction carry over to an
axisymmetric or three-dimensional family without modification, as the
waverider cases of \Cref{sec:waverider_data} demonstrate in three dimensions.
Reproducing the present optimisation for a true body of revolution is
therefore a matter of regenerating the database, not of changing the method.

The flow is treated as laminar and no turbulence model is used. Each mesh is
strongly clustered towards the solid surface to resolve the thin hypersonic
boundary layer and the steep thermal and velocity gradients that develop
near the wall. The initial target spacing is set at roughly ten times the
Kolmogorov length scale $\ell_K$, and further near-wall clustering brings the
wall-normal spacing to approximately
\begin{equation}
    \Delta n_w
    \approx
    \frac{\pi}{1.5}\,\ell_K .
    \label{eq:ogive_wallspacing}
\end{equation}
The governing equations and the associated numerical models are documented in
earlier thermochemical-flow and grid-motion studies
\cite{redding2023thermochemical,redding2025vortex,redding2025investigation}.

\subsubsection{Chemical Nonequilibrium}
\label{sec:ogive_chemistry}
 
Chemical nonequilibrium is represented with the five-species, six-reaction
air-chemistry model of Gupta \emph{et al.} \cite{gupta1990review,peyvan2023high,peyvan2026h3pc}, for which
the reacting mixture is
\begin{equation}
    \mathcal{S}
    =
    \left\{
    \mathrm{N_2},\;
    \mathrm{O_2},\;
    \mathrm{NO},\;
    \mathrm{N},\;
    \mathrm{O}
    \right\}.
    \label{eq:species_set}
\end{equation}
For species $i$ the conservation equation reads
\begin{equation}
    \frac{\partial \rho_i}{\partial t}
    +
    \nabla\cdot
    \left(
        \rho_i\mathbf{u}
    \right)
    =
    -
    \nabla\cdot\mathbf{J}_i
    +
    \dot{\omega}_i,
    \label{eq:massspeciesconservation}
\end{equation}
in which $\rho_i$ is the partial density of species $i$, $\mathbf{J}_i$ its
diffusive mass flux and $\dot{\omega}_i$ its net rate of chemical production.
The reactions retained are listed in \Cref{tab:chemicalmechanisms}.
 
\begin{table}[htbp]
    \centering
    \caption{Five-species, six-reaction chemical-kinetics mechanism used in
    the tangent-ogive simulations \cite{gupta1990review}.}
    \label{tab:chemicalmechanisms}
    \begin{tabular}{cl}
        \toprule
        \textbf{Reaction} & \textbf{Chemical mechanism} \\
        \midrule
        R1 & $\mathrm{O_2 + M \rightleftharpoons O + O + M}$ \\
        R2 & $\mathrm{N_2 + M \rightleftharpoons N + N + M}$ \\
        R3 & $\mathrm{N_2 + N \rightleftharpoons N + N + N}$ \\
        R4 & $\mathrm{NO + M \rightleftharpoons N + O + M}$ \\
        R5 & $\mathrm{NO + O \rightleftharpoons O_2 + N}$ \\
        R6 & $\mathrm{N_2 + O \rightleftharpoons NO + N}$ \\
        \bottomrule
    \end{tabular}
\end{table}

\subsubsection{Thermal Nonequilibrium and Transport}
\label{sec:ogive_thermal}
 
Thermal nonequilibrium is represented with a two-temperature formulation. The
first temperature, $T$, describes the translational--rotational mode; the
second, $T_m$, the additional molecular mode energy. Ionisation is omitted,
being insignificant at the conditions considered here.
 
The total-energy equation takes the general form
\begin{equation}
\begin{aligned}
\frac{\partial (\rho e)}{\partial t}
+
\nabla\cdot
\left[
    (\rho e+p)\mathbf{u}
\right]
={}&
\nabla\cdot
\left[
    k\nabla T
    +
    \sum_{s=1}^{N_v}
    k_{v,s}\nabla T_{v,s}
\right]
\\
&+
\nabla\cdot
\left[
    \sum_i h_i\mathbf{J}_i
\right]
-
\nabla\cdot
\left(
    \boldsymbol{\tau}\cdot\mathbf{u}
\right)
-
Q_{\mathrm{rad}},
\end{aligned}
\label{eq:energyconservation}
\end{equation}
where $e$ is the total specific energy, $T$ the translational--rotational
temperature, $T_{v,s}$ the temperature of vibrational mode $s$, $k$ and
$k_{v,s}$ the corresponding thermal conductivities, $h_i$ the species
enthalpy, $\mathbf{J}_i$ the species diffusive flux, $\boldsymbol{\tau}$ the
viscous stress tensor and $Q_{\mathrm{rad}}$ the radiative contribution.
 
The vibrational energy of mode $i$ obeys
\begin{equation}
\begin{aligned}
\frac{\partial (\rho_i e_{v,i})}{\partial t}
+
\nabla\cdot
\left(
    \rho_i e_{v,i}\mathbf{u}
\right)
={}&
\nabla\cdot
\left(
    k_{v,i}\nabla T_{v,i}
\right)
-
\nabla\cdot
\left(
    e_{v,i}\mathbf{J}_i
\right)
\\
&+
Q_{T-V_i}
+
Q_{V-V_i}
+
Q_{\mathrm{Chem}-V_i}
+
Q_{\mathrm{rad},i},
\end{aligned}
\label{eq:vibrationalenergyconservation}
\end{equation}
in which $Q_{T-V_i}$ is the translational--vibrational exchange, $Q_{V-V_i}$
the vibrational--vibrational exchange, $Q_{\mathrm{Chem}-V_i}$ the coupling
between chemical reaction and vibrational energy, and $Q_{\mathrm{rad},i}$
the radiative contribution. Fuller detail on these thermochemical models is
given in Ref.~\cite{redding2023thermochemical}.
 
Temperature-dependent transport properties are retained throughout: the
thermal conductivities and the dynamic viscosity vary with the local
thermodynamic state through the transport treatment implemented in the
solver. These quantities matter here beyond their role in the CFD, because
the conductivities and the viscosity are themselves among the spatial fields
represented by the surrogate, and the wall heat transfer and viscous surface
quantities are reconstructed from them.
 
The resulting set of $37$ geometry-dependent thermochemical CFD solutions
constitutes the high-fidelity database for the first of the three surrogates,
the canonical-coordinate operator of \Cref{subsec:ogive_method}.

\subsection{Three-Dimensional AEDC Waverider}
\label{sec:waverider_data}
 
The second configuration is the three-dimensional AEDC waverider. The
geometry originates in the design-validation campaign of Kammeyer and
Gillum~\cite{kammeyer1994waverider}, who tested a realistic hypersonic
waverider at Mach $10$, $14$ and $16.5$ in what was then the Naval Surface
Warfare Center Hypervelocity Wind Tunnel No.~9; the facility subsequently
transferred to the Arnold Engineering Development Complex and is now AEDC
Tunnel~9, so the two designations refer to the same installation at White
Oak. Later measurements on this configuration in that facility are reported
by Norris~\cite{norris2006mach}. The present study takes the vehicle
geometry from that lineage and the freestream state from a Tunnel~9
operating point, but it is a computational database rather than a
reproduction of any particular run: the solutions are generated at the single
condition of \Cref{eq:waverider_freestream} and swept over angle of attack,
whereas the design-validation campaign covered a higher-Mach part of the
facility envelope. The freestream and surface boundary conditions of
\Cref{eq:waverider_freestream} were instead selected to be consistent with a
reported Tunnel~9 operating condition at $Ma_\infty=7.36$, and the isothermal
wall temperature likewise follows the experimental value; no experimental
measurement is used as a validation target here, the inheritance being the
vehicle geometry and the operating state. The geometry and its
parameterisation as used in the present work are described in
Refs.~\cite{shukla2024waverider,shukla2024deepII}. Where the tangent-ogive
problem changes the physical geometry between CFD realisations, the waverider
geometry stays fixed and the aerodynamic operating condition is varied
through the angle of attack. This gives a complementary operator-learning
configuration, in which the principal difficulty is that the spatial
structure of a strongly three-dimensional, shock-dominated flow moves with
the input parameter. A single database serves two surrogates: the surface
quantities support the aerodynamic and thermal loads used in the design study
of \Cref{sec:results_waverider_surface}, and the volume field supports the
shock-aware operator of \Cref{sec:results_waverider_volume}.
 
The high-fidelity aerothermodynamic solutions are generated with the US3D
solver, a massively parallel finite-volume framework for compressible
high-speed flow on unstructured meshes that has been applied extensively to
hypersonic configurations involving strong shocks, shock--boundary-layer
interaction, viscous effects and thermochemical processes \cite{US3D}. The
freestream conditions follow the corresponding operating point of the AEDC
Hypervelocity Wind Tunnel No.~9,
\begin{equation}
    \rho_{\infty}= 0.5644~\mathrm{kg\,m^{-3}},
    \qquad
    T_{\infty}=72.77~\mathrm{K},
    \qquad
    V_{\infty}=1279.25~\mathrm{m\,s^{-1}},
    \label{eq:waverider_freestream}
\end{equation}
corresponding to a freestream Mach number $Ma_{\infty}=7.36$. The test gas
is nitrogen, as in Tunnel~9: the stated conditions imply a speed of sound
$a_\infty=V_\infty/Ma_\infty=173.8~\mathrm{m\,s^{-1}}$ and hence a specific
gas constant of $296.5~\mathrm{J\,kg^{-1}K^{-1}}$, which is that of
$\mathrm{N_2}$ rather than of air. The vehicle surface is isothermal at
$T_w=300~\mathrm{K}$, consistent with the experimental condition.
 
Unlike the laminar tangent-ogive calculations, turbulence is modelled with
the Menter shear-stress-transport (SST) Reynolds-averaged Navier--Stokes
formulation with a vorticity source term, and a five-species finite-rate
chemistry model accounts for non-equilibrium chemical effects. The mesh is generated with LINK3D and contains approximately
$50.4\times10^{6}$ unstructured volume cells, of which $N_s=266{,}848$
quadrilateral cells lie on the vehicle surface, with near-wall refinement
giving a wall spacing corresponding to $y^{+}<1$ .

The surface therefore carries approximately one cell in $190$ of the mesh, a
ratio that bears directly on how the two waverider surrogates are trained and
is returned to in \Cref{subsec:waverider_surface}. The wake is excluded from
the computational domain, which terminates at the rear of the vehicle. Each
case is advanced for more than $20$ flow-through times, ensuring that the
dominant shock and boundary-layer structures are fully established and the
solution stable before the surrogate-training fields are extracted.
 
The parametric input is the angle of attack $\alpha$. Twenty-nine
high-fidelity realisations are generated by varying $\alpha$ from
$-14^{\circ}$ to $+14^{\circ}$ in increments of $1^{\circ}$,
\begin{equation}
    \alpha
    \in
    \left\{
    -14^{\circ},\,
    -13^{\circ},\,
    \ldots,\,
    -1^{\circ},\,
    0^{\circ},\,
    1^{\circ},\,
    \ldots,\,
    13^{\circ},\,
    14^{\circ}
    \right\}.
    \label{eq:waverider_aoa}
\end{equation}
Every realisation shares the same vehicle geometry and computational mesh and
differs only in operating condition. Varying $\alpha$ changes the location,
orientation and strength of the shock structure, producing a broad range of
three-dimensional aerothermodynamic states.
 
The resulting $29$-case database is used to train and evaluate both waverider
surrogates. Where the tangent-ogive dataset tests whether a surrogate can
generalise across changing physical geometries, the waverider dataset
isolates the dependence of a three-dimensional solution on a continuously
varying aerodynamic parameter over a fixed spatial domain --- and, because
the surface and volume surrogates are built from the same solutions and
differ only in the quantity predicted, it also allows the two to be compared
directly.

\section{Neural-Operator Surrogate Methodology}
\label{sec:computational_methodology}
 
Three surrogates are constructed, each learning a map from a scalar parameter
$\lambda$ to a spatial field,
\begin{equation}
    \mathcal{G}_q:\ \lambda \longmapsto q(\,\cdot\,;\lambda),
    \label{eq:generic_operator}
\end{equation}
with $\lambda=\xi_g$ for the tangent ogive and $\lambda=\alpha$ for the two
waverider surrogates, and with a separate operator trained for each physical
field. All three share the two-step DeepONet procedure of
\Cref{subsec:two_step_operator}; they differ in what, if anything, makes the
spatial representation difficult.
 
For the tangent ogive the physical domain itself changes with $\xi_g$, which
is resolved by a body-fitted canonical mapping
(\Cref{subsec:ogive_method}). For the waverider surface quantities the mesh
is fixed and the fields vary smoothly with incidence, and the two-step
construction is applied directly to the physical surface coordinates with no
additional device (\Cref{subsec:waverider_surface}). For the waverider volume
field the domain is likewise fixed, but the shock moves through it with
$\alpha$, and a shock-aligned coordinate together with a residual correction
concentrated near the shock is required (\Cref{subsec:waverider_method}).
 
Where a device is needed, it is the same idea in different clothing: the
trunk is given a coordinate in which the feature that varies between cases
does not move, leaving the branch to represent only the parametric dependence
that remains. The surface surrogate is the case in which no feature moves
through the domain and no such coordinate is needed, and because it is built
from the same solutions as the volume surrogate and differs only in the
quantity predicted, the pair delimits when the additional construction is
necessary.
 
Throughout, every data-dependent quantity --- normalisation statistics, the
learned basis and its orthogonalisation, shock masks and hotspot thresholds
--- is computed from the training cases alone. A held-out case enters only at
evaluation, so the reported accuracy measures generalisation to an unseen
parameter value rather than interpolation among points of a solution the
model has already seen.
 
\begin{table}[htbp]
  \centering
  \caption{Principal symbols of \Cref{sec:computational_methodology}.}
  \label{tab:method_notation}
  \begin{tabular}{llll}
    \toprule
    Symbol & Meaning & Space & Introduced \\
    \midrule
    $\lambda$ & parametric input ($\xi_g$ or $\alpha$) & $\mathbb{R}$
      & \eqref{eq:generic_operator} \\
    $q$ & a physical field & --- & \eqref{eq:generic_operator} \\
    $\mathbf{\Psi}_{\theta}$ & trunk output, the learned basis
      & $\mathbb{R}^{N_{\mathrm{pts}}\times r}$ & \eqref{eq:trunk_basis} \\
    $\mathbf{c}_n$ & free coefficients of case $n$ (Step 1)
      & $\mathbb{R}^{r}$ & \eqref{eq:stage1_optimization} \\
    $\mathbf{Q},\mathbf{R}$ & thin QR factors of $\mathbf{\Psi}_{\theta}$
      & --- & \eqref{eq:trunk_qr} \\
    $\mathbf{b}_n$ & orthogonalised coefficients, $\mathbf{R}\mathbf{c}_n$
      & $\mathbb{R}^{r}$ & \eqref{eq:trunk_qr} \\
    $\mathcal{B}_{\phi}$ & branch network (Step 2) & --- &
      \eqref{eq:branch_training} \\
    $(s,\eta)$ & canonical coordinates, ogive & $[0,1]^2$ &
      \eqref{eq:canonical_physical_map} \\
    $\mathbf{\Phi}_{\xi_g}$ & canonical-to-physical map & --- &
      \eqref{eq:canonical_physical_map} \\
    $J_{\xi_g}$ & Jacobian of $\mathbf{\Phi}_{\xi_g}$ & --- &
      \eqref{eq:mapping_jacobian} \\
    $N_s$ & waverider surface cells, $266{,}848$ & --- &
      \Cref{subsec:waverider_surface} \\
    $\widehat{d}$ & ShockNet normalised distance to the shock & $\mathbb{R}$ &
      \eqref{eq:shocknet} \\
    $\chi$ & bounded shock coordinate, $\tanh\widehat{d}$ & $(-1,1)$ &
      \eqref{eq:shocknet} \\
    $m_\rho,\varsigma_\rho$ & mean and s.d.\ of $\log\rho$ over training &
      --- & \eqref{eq:waverider_log_density} \\
    $\mathbf{f}$ & feature vector of the hotspot gate & --- &
      \eqref{eq:hotspot_prob} \\
    \bottomrule
  \end{tabular}
\end{table}

\subsection{Two-Step Operator Learning}
\label{subsec:two_step_operator}
 
Following Lee and Shin \cite{lee2024training}, learning the spatial structure
of a field is separated from learning its dependence on $\lambda$. The
separation matters here because the two are learned from different amounts of
information: the spatial structure is common to every training case and can be
fitted from all of them together, whereas the parametric dependence is a map
from one scalar and is supported by as many samples as there are cases. This
subsection states the construction once; the three surrogates that follow
differ only in the spatial representation on which it acts.
 
Let a spatial trunk with parameters $\theta$ evaluate $r$ basis functions at
the $N_{\mathrm{pts}}$ points of the spatial representation,
\begin{equation}
    \mathbf{\Psi}_{\theta}
    = \begin{bmatrix}\boldsymbol{\psi}_1 & \cdots & \boldsymbol{\psi}_r\end{bmatrix}
    \in \mathbb{R}^{N_{\mathrm{pts}}\times r}.
    \label{eq:trunk_basis}
\end{equation}
 
\paragraph{Step 1: spatial basis and free coefficients.}
Each training case $\lambda^{(n)}$, $n\in\mathcal{I}_{\mathrm{train}}$, is
given an independent trainable coefficient vector
$\mathbf{c}_n\in\mathbb{R}^{r}$, so that
$\widehat{\mathbf{q}}_n=\mathbf{\Psi}_{\theta}\mathbf{c}_n$, and trunk and
coefficients are optimised jointly,
\begin{equation}
    \min_{\theta,\{\mathbf{c}_n\}}
    \sum_{n\in\mathcal{I}_{\mathrm{train}}}
    \mathcal{L}_{q}\big(\mathbf{\Psi}_{\theta}\mathbf{c}_n,\ \mathbf{q}_n\big),
    \label{eq:stage1_optimization}
\end{equation}
where $\mathcal{L}_q$ is the field-specific reconstruction loss. No network
maps $\lambda$ to $\mathbf{c}_n$ at this stage; the coefficients are free
parameters, one vector per case. Each update draws points from several cases,
so the basis is pushed towards structure shared across the family rather than
towards any single realisation.
 
\paragraph{Orthogonalisation.}
The learned basis is factorised by a thin QR decomposition and the
coefficients transformed consistently,
\begin{equation}
    \mathbf{\Psi}_{\theta}=\mathbf{Q}\mathbf{R},\quad
    \mathbf{Q}^{\mathsf T}\mathbf{Q}=\mathbf{I}_r,\qquad
    \mathbf{b}_n=\mathbf{R}\mathbf{c}_n,
    \label{eq:trunk_qr}
\end{equation}
so that $\mathbf{\Psi}_{\theta}\mathbf{c}_n=\mathbf{Q}\mathbf{b}_n$ exactly:
the reconstruction is unchanged and only the coordinates in which it is
expressed have moved. \Cref{eq:stage1_optimization} determines
$\mathbf{\Psi}_{\theta}$ and $\{\mathbf{c}_n\}$ only up to a common invertible
factor, and the arbitrary scaling and mixing among basis functions that this
permits would otherwise be inherited by the regression of Step~2. Fixing the
basis to be orthonormal removes it.
 
\begin{figure}[t]
  \centering
  \includegraphics[width=\textwidth]{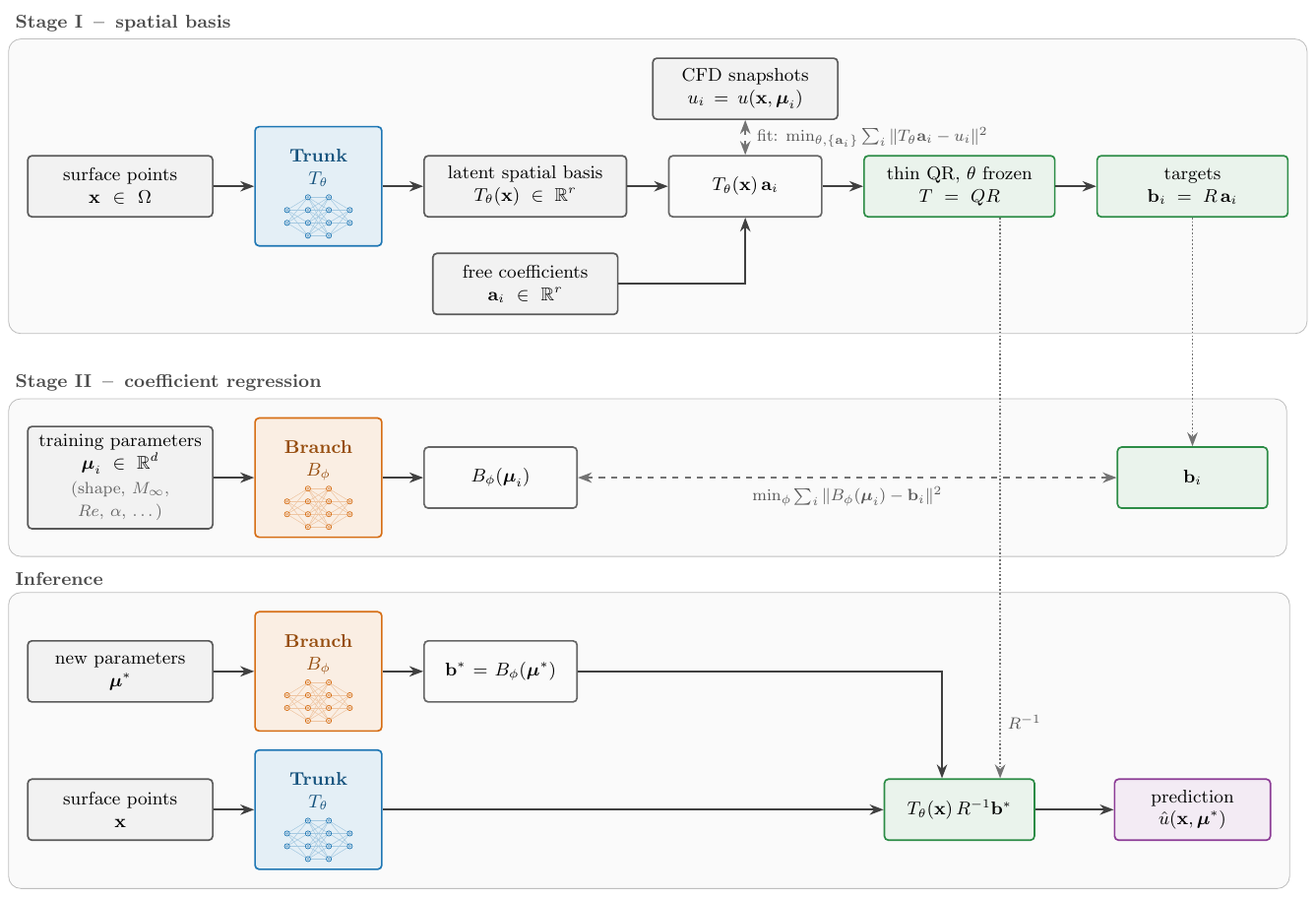}
  \caption{Two-step DeepONet, the construction shared by all three
  surrogates. The trunk $\mathbf{\Psi}_{\theta}$ maps a point
  $\mathbf{x}$ of the spatial representation to $r$ latent basis functions;
  the branch $\mathcal{B}_{\phi}$ maps the parametric input $\lambda$ to the
  corresponding coefficients.
  \emph{Step~1 (spatial basis):} the trunk is trained jointly with free
  coefficients $\mathbf{c}_n$, one vector per training case, by minimising
  $\sum_n\|\mathbf{\Psi}_{\theta}\mathbf{c}_n-\mathbf{q}_n\|^2$ over the
  training solutions $\mathbf{q}_n=q(\cdot\,;\lambda^{(n)})$. With $\theta$
  frozen, a thin QR factorisation $\mathbf{\Psi}_{\theta}=\mathbf{Q}\mathbf{R}$
  orthonormalises the basis and defines the regression targets
  $\mathbf{b}_n=\mathbf{R}\mathbf{c}_n$.
  \emph{Step~2 (coefficient regression):} the branch is trained to map the
  training parameters to those targets,
  $\min_\phi\sum_n\|\mathcal{B}_{\phi}(\lambda^{(n)})-\mathbf{b}_n\|^2$.
  \emph{Inference:} for an unseen $\lambda^\star$ the field is reconstructed as
  $\widehat{\mathbf{q}}=\mathbf{\Psi}_{\theta}\mathbf{R}^{-1}
  \mathcal{B}_{\phi}(\lambda^\star)=\mathbf{Q}\,\mathcal{B}_{\phi}(\lambda^\star)$.
  Solid arrows: data flow; dashed double arrows: training objectives; dotted
  arrows: quantities frozen and carried between steps. The network sketches
  are illustrative and do not show the actual layer widths.}
  \label{fig:twostep}
\end{figure}

\paragraph{Step 2: parameter-to-coefficient map.}
With $\mathbf{Q}$ frozen, a branch network $\mathcal{B}_{\phi}$ is trained to
reproduce the orthogonalised coefficients,
\begin{equation}
    \min_{\phi}\sum_{n\in\mathcal{I}_{\mathrm{train}}}
    \big\|\mathcal{B}_{\phi}(\lambda^{(n)})-\mathbf{b}_n\big\|_2^2,
    \label{eq:branch_training}
\end{equation}
and the prediction at an unseen parameter $\lambda^\star$ is
\begin{equation}
    \widehat{\mathbf{q}}(\lambda^\star)
    =\mathbf{Q}\,\mathcal{B}_{\phi}(\lambda^\star).
    \label{eq:two_step_prediction}
\end{equation}
Written in terms of the unfactorised trunk this is
$\widehat{\mathbf{q}}=\mathbf{\Psi}_{\theta}\mathbf{R}^{-1}
\mathcal{B}_{\phi}(\lambda^\star)$, which is the form shown in
\Cref{fig:twostep}; the two are identical, and $\mathbf{Q}$ is used in what
follows because it is what is stored.
 
This decomposition suits the present data-scarce setting. The
high-dimensional spatial structure is learned collectively from every
training solution in Step~1, and Step~2 is left with a map from one scalar
into $\mathbb{R}^{r}$ --- a problem whose sample requirement is governed by
$r$ rather than by $N_{\mathrm{pts}}$.
 
\paragraph{Coefficient-interpolation baseline.}
To separate the quality of the learned basis from that of the branch network,
piecewise cubic Hermite interpolation (PCHIP) \cite{fritsch1980monotone} is
applied independently to each component of $\mathbf{b}_n$,
\begin{equation}
    \widehat{\mathbf{q}}_{\mathrm{PCHIP}}(\lambda^\star)
    =\mathbf{Q}\,\mathcal{P}_{\mathrm{PCHIP}}
    \big(\lambda^\star;\{(\lambda^{(n)},\mathbf{b}_n)\}\big).
    \label{eq:pchip_prediction}
\end{equation}
Both predictors use the same frozen basis $\mathbf{Q}$, so the difference
between them isolates the parameter-to-coefficient map. Where the held-out
parameter lies outside the range of the training parameters, PCHIP
extrapolates by first-order Hermite continuation from the end interval;
the comparison in those folds therefore tests the branch network against a
deliberately simple alternative rather than against a competitive one.
 
\subsection{Surrogate I --- Tangent Ogive: Canonical-Coordinate Operator}
\label{subsec:ogive_method}
 
Changing $\xi_g$ changes the physical domain $\Omega(\xi_g)$. A trunk trained
directly on physical coordinates would have to learn the changing grid and the
changing flow at once, and would have no fixed set of spatial inputs on which
to define a basis at all. Each solution is therefore first mapped onto a
common logical domain, the operator is learned there, and predictions are
returned to the physical geometry for evaluation.
 
\subsubsection{Body-Fitted Canonical Mapping}
\label{subsec:canonical_mapping}
 
\begin{figure}[t]
    \centering
    \includegraphics[width=\textwidth]{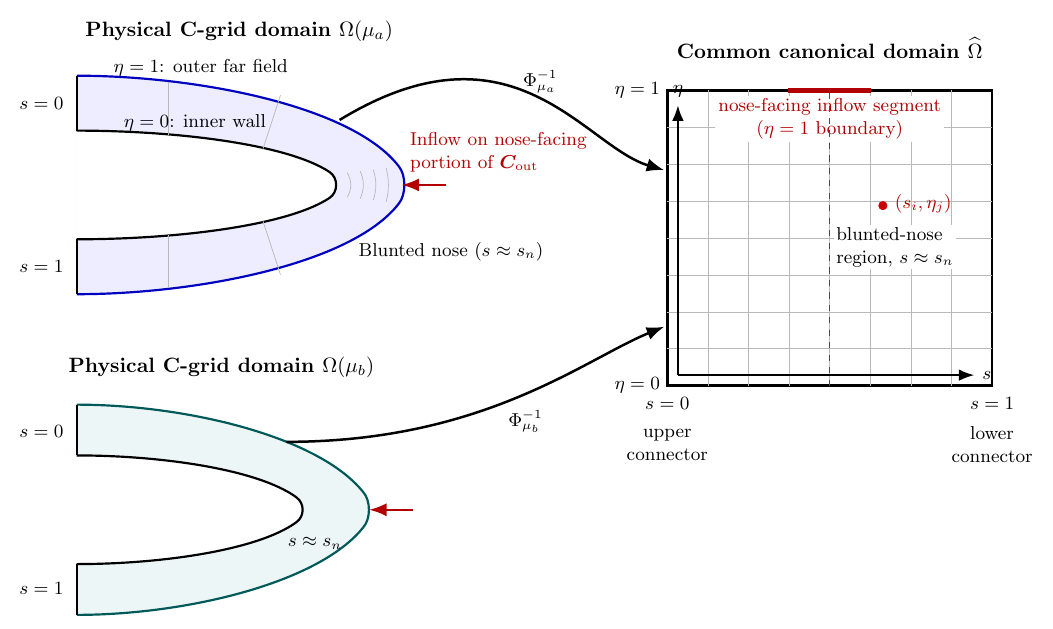}
    \caption{Canonical mapping of the tangent-ogive C-grids. The coordinate
    $s$ follows the inner and outer boundaries from the upper connector to the
    lower, and $\eta$ runs from the body ($\eta=0$) to the far field
    ($\eta=1$). Every geometry is mapped to the same logical $(s,\eta)$
    domain, so the trunk sees one set of spatial inputs for the whole family.}
    \label{fig:canonical_mapping}
\end{figure}

Each domain is a two-dimensional C-grid bounded by an inner body-fitted curve
and an outer far-field curve joined by two short connectors
(\Cref{fig:canonical_mapping}). The two curves,
\begin{equation}
    \mathbf{C}_{\mathrm{in}}(s;\xi_g),\qquad
    \mathbf{C}_{\mathrm{out}}(s;\xi_g),\qquad 0\le s\le1,
    \label{eq:cgrid_boundaries}
\end{equation}
are both parameterised by normalised arc length and oriented alike, from the
upper connector around the nose to the lower connector. That common
orientation is what puts equivalent locations on different geometries into
correspondence, and it is the only place where the family is tied together
geometrically.
 
Every domain is mapped onto $\widehat{\Omega}=[0,1]\times[0,1]$ with
coordinates $(s,\eta)$, where $s$ runs along the boundary from the upper
connector ($s=0$) to the lower ($s=1$) and $\eta$ from the body ($\eta=0$) to
the far field ($\eta=1$). Both curves are sampled at the same uniform logical
points $s_i=i/(N_s^{\mathrm{og}}-1)$, $\eta_j=j/(N_\eta-1)$, and the physical
location of each logical point follows by linear blending between them,
\begin{equation}
     \mathbf{\Phi}_{\xi_g}(s_i,\eta_j)
    =(1-\eta_j)\,\mathbf{C}_{\mathrm{in}}(s_i;\xi_g)
    +\eta_j\,\mathbf{C}_{\mathrm{out}}(s_i;\xi_g)
    =\begin{bmatrix}x_{ij}^{(\xi_g)}\\[2pt] y_{ij}^{(\xi_g)}\end{bmatrix}.
    \label{eq:canonical_physical_map}
\end{equation}
With $N_s^{\mathrm{og}}=600$ and $N_\eta=300$, every geometry is represented
by the same $N_{\mathrm{pts}}=180{,}000$ logical points. The physical
coordinates $(x_{ij},y_{ij})$ change with $\xi_g$; the logical coordinates
$(s_i,\eta_j)$ do not. A single set of spatial inputs therefore serves the
whole family, which is precisely what \Cref{eq:trunk_basis} requires.
 
The mapping is invertible only where its Jacobian
\begin{equation}
    J_{\xi_g}=\det\frac{\partial(x,y)}{\partial(s,\eta)}
    =x_s y_\eta-x_\eta y_s
    \label{eq:mapping_jacobian}
\end{equation}
keeps one sign throughout the domain, since a change of sign would mean a
folded logical cell. This was verified for every member of the family.
 
\subsubsection{Canonical Fields and Physical Gradients}
\label{subsec:field_pullback}
 
A CFD field $q_{\xi_g}(x,y)$ is sampled at the mapped points,
\begin{equation}
    \widetilde{q}(s_i,\eta_j;\xi_g)
    =q_{\xi_g}\big(\mathbf{\Phi}_{\xi_g}(s_i,\eta_j)\big),
    \label{eq:field_pullback}
\end{equation}
which yields the same $N_\eta\times N_s^{\mathrm{og}}$ array for every
geometry, and the learning problem \eqref{eq:generic_operator} becomes
\begin{equation}
    \mathcal{G}_q:\ \xi_g\longmapsto\widetilde{q}(s,\eta;\xi_g),
    \qquad (s,\eta)\in\widehat{\Omega}.
    \label{eq:canonical_operator}
\end{equation}
 
Canonicalisation changes the coordinate representation and nothing else: each
logical point retains its physical location through $\mathbf{\Phi}_{\xi_g}$.
The gradients required for wall heat flux and viscous stress are accordingly
computed in $(s,\eta)$ and converted by the inverse metric,
\begin{equation}
    \begin{bmatrix} q_x\\ q_y\end{bmatrix}
    =\frac{1}{J_{\xi_g}}
    \begin{bmatrix} y_\eta & -y_s\\ -x_\eta & x_s\end{bmatrix}
    \begin{bmatrix} q_s\\ q_\eta\end{bmatrix}.
    \label{eq:physical_gradient_transform}
\end{equation}
Wall quantities are thus evaluated on the physical geometry, from fields
predicted on the logical one, with no interpolation between the two.
 
\subsubsection{Transolver Spatial Trunk}
\label{subsec:transolver}
 
The trunk of \Cref{eq:trunk_basis} is a Transolver \cite{wu2024transolver}
whose only inputs are the logical coordinates
$\mathbf{p}_k=(s_k,\eta_k)$, $k=1,\ldots,N_{\mathrm{pts}}$. The geometry
enters nowhere in the trunk; it reaches the prediction only through the
branch. This is the division of labour the canonical mapping was built to
permit, and it is what allows one basis to serve all $37$ geometries.
 
\begin{figure}[t]
    \centering
    \includegraphics[width=\linewidth]{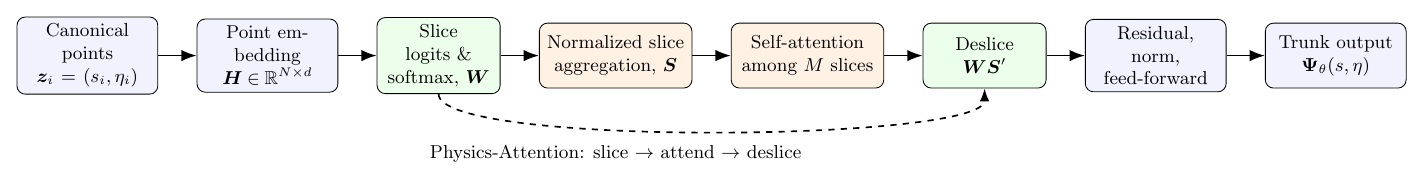}
    \caption{Transolver spatial trunk. The canonical coordinates $(s,\eta)$
    are lifted to point features, grouped into learned physical-state tokens,
    updated by attention among the tokens, and projected back to the points to
    form the spatial basis $\mathbf{\Psi}_{\theta}$ used in the two-step
    procedure.}
    \label{fig:transolver_architecture}
\end{figure}

Physics-Attention compresses the $N_{\mathrm{pts}}$ point features
$\mathbf{H}\in\mathbb{R}^{N_{\mathrm{pts}}\times d}$ into $M\ll
N_{\mathrm{pts}}$ learned physical-state tokens, attends among the tokens, and
projects the result back to the points. The soft point-to-slice assignment and
the slice tokens are
\begin{equation}
    \mathbf{W}=\operatorname{softmax}\!\Big(\frac{\mathbf{H}\mathbf{A}}{\tau}\Big)
    \in\mathbb{R}^{N_{\mathrm{pts}}\times M},
    \qquad
    \mathbf{S}_m=\frac{\sum_{k}W_{km}\mathbf{H}_k}{\sum_{k}W_{km}+\varepsilon},
    \label{eq:transolver_slice_weights}
\end{equation}
the softmax being taken over the $M$ slices, $\mathbf{A}$ the learned
projection to slice logits, $\tau$ a temperature controlling how sharply
points are assigned, and $\varepsilon$ a small constant guarding the
normalisation of an empty slice. After self-attention among the tokens
produces $\mathbf{S}'$, the point features are updated as
$\mathbf{H}'=\mathbf{W}\mathbf{S}'$; residual connections, normalisation and
feed-forward layers complete each block (\Cref{fig:transolver_architecture}).
Attending among $M$ tokens rather than $N_{\mathrm{pts}}$ points reduces the
cost of attention from quadratic in the number of points to quadratic in $M$,
while still representing the non-local correlations that matter here --- above
all that between the shock and the wall, which are far apart in $(s,\eta)$ but
coupled in the physics.
 
\subsubsection{Prediction on an Unseen Geometry}
\label{subsec:physical_output}
 
For a held-out $\xi_g^\star$, \Cref{eq:two_step_prediction} --- or its PCHIP
counterpart \eqref{eq:pchip_prediction} --- gives the field on the canonical
grid, which is then attached to the stored physical coordinates of that
geometry, $\{\widehat{q}_{ij},\,x_{ij}^{(\xi_g^\star)},\,
y_{ij}^{(\xi_g^\star)}\}$. The result lives on the actual tangent-ogive domain
and is used there for visualisation, error analysis, differentiation and
aerothermal post-processing. The complete chain is
\begin{equation}
    q_{\xi_g}
    \xrightarrow{\ \mathbf{\Phi}_{\xi_g}^{-1}\ }
    \widetilde{q}(s,\eta;\xi_g)
    \xrightarrow{\ \text{Transolver}+\text{two-step}\ }
    \widehat{\widetilde{q}}(s,\eta;\xi_g^\star)
    \xrightarrow{\ \mathbf{\Phi}_{\xi_g^\star}\ }
    \widehat{q}(x,y;\xi_g^\star),
    \label{eq:complete_workflow}
\end{equation}
in which the first arrow is applied to the training data and the last to the
prediction, so that the canonical domain is an internal representation and
never appears in a reported result.

\subsection{Surrogate II --- Waverider Surface Quantities}
\label{subsec:waverider_surface}
\label{sec:twostep_deeponet}
 
The surface quantities supply the objectives of the incidence design study,
and they are the case in which the two-step construction of
\Cref{subsec:two_step_operator} applies without modification. The surface mesh
is common to all $29$ angles, so the spatial representation is fixed; and the
surface fields vary smoothly with $\alpha$, so no feature migrates through the
domain and no auxiliary coordinate is required. The trunk therefore acts on
the physical cell centroids directly,
\begin{equation}
    \mathbf{x}=(x,y,z)\in\mathbb{R}^{3},
    \qquad
    \mathcal{G}_u:\ \alpha\longmapsto u(\mathbf{x};\alpha),
    \label{eq:surface_operator}
\end{equation}
with $N_{\mathrm{pts}}=N_s=266{,}848$ surface cells. Independent surrogates
are built for the wall heat flux $q_w$ and the three components
$(t_x,t_y,t_z)$ of the total surface traction supplied by the solver, which
includes both the pressure and the viscous contributions
(\Cref{eq:total_traction}), from which
the integrated aerodynamic loads and thermal objectives are subsequently
formed.
 
\paragraph{Architecture and training.}
Step~1 and Step~2 follow
\eqref{eq:stage1_optimization}--\eqref{eq:two_step_prediction} exactly, with
$\mathbf{\Psi}_\theta$ a fully connected trunk of layer widths
$3\rightarrow128\rightarrow128\rightarrow128\rightarrow128\rightarrow r$ and
$\mathcal{B}_\phi$ a branch of widths $1\rightarrow32\rightarrow32\rightarrow
r$, both with hyperbolic-tangent activations and a latent dimension $r=16$.
The latent coefficients $\mathbf{b}_n$ are standardised mode by mode before
Step~2, so that directions of differing magnitude contribute comparably to
\eqref{eq:branch_training}. The prediction at an unseen incidence
$\alpha^\star$ is \eqref{eq:two_step_prediction}, applied independently to
each of the four fields.
 
\paragraph{Sampling and weighting.}
Because the surface carries more than $2.6\times10^{5}$ cells, the Step-1
objective \eqref{eq:stage1_optimization} is evaluated on randomly sampled
cells at each iteration: $65{,}536$ cells, drawn together with a subset of the
training angles. Uniform spatial weighting is used for the traction
components. For $q_w$ a mild hotspot-aware weighting is applied, so that the
high-heat-flux region --- which is small in cell count but is the region the
thermal objective is formed from --- is not overwhelmed by the far more
numerous low-magnitude cells. This is the one place where the measure
argument of \Cref{sec:intro} enters the surface surrogate directly: the
quantity of interest is concentrated where the cells are few, and a uniformly
weighted objective would under-represent it.
 
\paragraph{Validation protocol.}
Entire angles are held out rather than individual cells. Since all $29$
solutions share one surface mesh, a cell-wise split would place neighbouring
cells of the same solution on both sides of the partition and would measure
interpolation within a known field rather than generalisation to an unseen
incidence.

\subsection{Surrogate III --- Waverider Volume Field: Shock-Aware Operator}
\label{subsec:waverider_method}
 
The volume field uses the same geometry, the same mesh and the same $29$
solutions as the surface surrogate, and differs only in the quantity
predicted. That difference is nonetheless decisive. No canonical mapping is
required, and the trunk could in principle act directly on the physical
coordinates $\mathbf{x}=(x,y,z)$; for the density field,
\begin{equation}
    \mathcal{G}_{\rho}:\ \alpha\longmapsto\rho(\mathbf{x};\alpha).
    \label{eq:waverider_operator}
\end{equation}
 
In practice it cannot. The shock's position, orientation and strength all
change with $\alpha$, yet it occupies a small fraction of the mesh, so uniform
sampling rarely sees it and a global loss under-weights it; a trunk given only
Cartesian coordinates must represent the same discontinuity at a different
place for every incidence, and averages neighbouring shock positions into a
smeared front. The volume surrogate therefore proceeds in three stages
(\Cref{fig:pipeline}). A training-only shock sensor and a ShockNet locate the
shock (\Cref{subsec:waverider_shock_sensor,subsec:shocknet}); a shock-aligned
two-step DeepONet is trained with shock-focused sampling and loss weighting
(\Cref{subsec:waverider_two_step}); and a hotspot-aware mixture of experts
corrects what remains (\Cref{subsec:hotspot_moe}). As for the surface
surrogate, entire $\alpha$ realisations are held out and points from a
held-out angle are never mixed into training.
 
\subsubsection{Density Transformation}
\label{subsec:waverider_density_transform}
 
Density is positive and varies over a large range across the shock, so it is
learned as a standardised log-density,
\begin{equation}
    z_{\rho}=\frac{\log\rho-m_{\rho}}{\varsigma_{\rho}},
    \qquad
    \widehat{\rho}=\exp\!\big(\varsigma_{\rho}\,\widehat{z}_{\rho}+m_{\rho}\big),
    \label{eq:waverider_log_density}
\end{equation}
with $m_{\rho}$ and $\varsigma_{\rho}$ the mean and standard deviation of
$\log\rho$ over the training angles. Working in the logarithm makes the
relative error the quantity being minimised, which is the appropriate measure
when the same field spans the freestream and the post-shock region.
 
\subsubsection{Connectivity-Based Shock Sensor}
\label{subsec:waverider_shock_sensor}
 
For each hexahedral element $e$ of the mesh, a local density-jump indicator is
\begin{equation}
    S_e=\frac{\rho_{\max,e}-\rho_{\min,e}}{\rho_{\max,e}+\rho_{\min,e}+\varepsilon},
    \label{eq:waverider_shock_sensor}
\end{equation}
where $\rho_{\max,e}$ and $\rho_{\min,e}$ are the extreme nodal densities of
the element. Each node takes the largest value among its incident elements,
and the nodal sensor is then maximised over the training angles
$\mathcal{A}_{\mathrm{train}}$,
\begin{equation}
    S_i^{\mathrm{train}}=\max_{\alpha\in\mathcal{A}_{\mathrm{train}}}S_i(\alpha).
    \label{eq:waverider_training_sensor}
\end{equation}
A prescribed upper percentile of $S_i^{\mathrm{train}}$ defines the
shock-region mask. Taking the maximum over training angles before thresholding
is what makes the mask cover every shock position seen in training, including
those at angles adjacent to a held-out one, while using no held-out solution.
 
\subsubsection{ShockNet and the Shock-Aligned Coordinate}
\label{subsec:shocknet}
 
The sensor identifies the shock surface $\Gamma_s(\alpha)$ at each training
angle. ShockNet is trained on those angles to predict the normalised distance
to that surface,
\begin{equation}
    \widehat{d}(\mathbf{x};\alpha)\approx
    \frac{\mathrm{dist}\big(\mathbf{x},\Gamma_s(\alpha)\big)}{\delta},
    \label{eq:shocknet}
\end{equation}
in which $\delta$ sets the width of the shock layer, and is then frozen. The
bounded shock coordinate
\begin{equation}
    \chi=\tanh\widehat{d}
    \label{eq:shock_coordinate}
\end{equation}
saturates towards $\pm1$ away from the shock and varies rapidly across it.
Normalising by a layer width rather than taking a signed distance to a
surface is deliberate: in the viscous, reacting flow considered here the shock
is a zone of finite thickness followed by a relaxation region, and $\chi$ is
accordingly a coordinate that varies smoothly across that zone rather than
one that switches at a surface. Because $\widehat{d}$ is predicted from
$\alpha$ and $\mathbf{x}$ alone, $\chi$ is available at an unseen angle
without any CFD information --- which is the property that makes it usable as
a trunk input rather than merely as an analysis tool.
 
\subsubsection{Shock-Aligned Two-Step DeepONet}
\label{subsec:waverider_two_step}
 
The stage-one model applies the procedure of
\Cref{subsec:two_step_operator} to $z_\rho$, with the trunk input augmented by
the shock coordinate,
\begin{equation}
    \boldsymbol{\zeta}=(x,y,z,\chi).
    \label{eq:waverider_trunk_input}
\end{equation}
In this input the discontinuity sits near $\chi=0$ for every $\alpha$, so the
basis is not required to spend functions on representing the same front at
many locations --- the role played for the ogive by the canonical mapping, and
the role that is unnecessary for the surface quantities of
\Cref{subsec:waverider_surface}.
 
Because the mesh is too large for full-field updates, each Step-1 update uses
a spatial mini-batch
$\mathcal{S}=\mathcal{S}_{\mathrm{global}}\cup\mathcal{S}_{\mathrm{shock}}$,
with a fraction $f_{\mathrm{shock}}$ of its $N_b$ points drawn from the shock
mask and the remainder from the whole domain. The reconstruction loss combines
global, shock-core and shock-band contributions,
\begin{equation}
    \mathcal{L}_{R1}
    =\lambda_g\mathcal{L}_{\mathrm{global}}
    +\lambda_c\mathcal{L}_{\mathrm{core}}
    +\lambda_b\mathcal{L}_{\mathrm{band}},
    \label{eq:waverider_shock_loss}
\end{equation}
so that the shock receives both a higher sampling probability and a larger
loss weight, and the interior of the shock zone is resolved rather than merely
bracketed by the core and band terms. This is a data-driven bias and not a
physics-informed constraint: neither the Rankine--Hugoniot relations nor the
governing equations are imposed as residuals, and the model is therefore not
guaranteed to produce a physically admissible jump. After QR orthogonalisation
and Step~2 \eqref{eq:trunk_qr}--\eqref{eq:two_step_prediction}, the stage-one
prediction at an unseen angle is
\begin{equation}
    \widehat{z}_{R1}(\alpha^\star)=\mathbf{Q}\,\mathcal{B}_{\phi}(\alpha^\star).
    \label{eq:waverider_prediction}
\end{equation}
 
\subsubsection{Hotspot-Aware Mixture-of-Experts Correction}
\label{subsec:hotspot_moe}
 
The stage-one residual remains concentrated in a few regions near the shock,
which is a case for a correction that can vary its behaviour by region rather
than one applied uniformly. Training points whose residual
\begin{equation}
    r=z_{\rho,\mathrm{CFD}}-\widehat{z}_{R1}
    \label{eq:stage1_residual}
\end{equation}
exceeds in magnitude the $95$th percentile $\tau_{95}$ of $|r|$ over the
training set are labelled hotspots, $y_h=\mathbf{1}(|r|\ge\tau_{95})$. A
classifier $f_h$, trained with binary cross-entropy, predicts the hotspot
probability
\begin{equation}
    p_h=\operatorname{sig}\big(f_h(\mathbf{f}_h)\big),\qquad
    \mathbf{f}_h=(\alpha,\mathbf{x},\chi,\widehat{z}_{R1}),
    \label{eq:hotspot_prob}
\end{equation}
where $\operatorname{sig}$ is the logistic function. Every entry of
$\mathbf{f}_h$ is available at inference, so $p_h$ stands in for the residual
itself, which is unknown at an unseen angle. With
$\mathbf{f}=(\mathbf{f}_h,p_h)$, a softmax gate blends four residual experts
\cite{jacobs1991adaptive},
\begin{equation}
    \Delta z_{\mathrm{MoE}}
    =\sum_{m=1}^{4}\pi_m(\mathbf{f})\,\Delta z_m(\mathbf{f}),
    \qquad
    \boldsymbol{\pi}=\operatorname{softmax}\big(g(\mathbf{f})\big),
    \label{eq:moe}
\end{equation}
allowing different experts to specialise in shock-dominated and in smooth
regions (\Cref{fig:hotspot_moe}). Both the residual and the correction are
formed in the standardised log variable, so the two are commensurate, and the
final prediction is
\begin{equation}
    \widehat{\rho}(\mathbf{x};\alpha^\star)
    =\exp\!\Big[\varsigma_{\rho}\big(\widehat{z}_{R1}
      +\Delta z_{\mathrm{MoE}}\big)+m_{\rho}\Big].
    \label{eq:waverider_complete_operator}
\end{equation}

\medskip
 
The three surrogates thus span the difficulties that arise in hypersonic
operator learning and, between them, delimit when each remedy is needed. The
tangent ogive addresses \emph{geometry-dependent domains}, mapping every
member of the family onto one canonical representation on which a Transolver
basis is learned. The waverider surface quantities are the control case: the
domain is fixed, no feature moves through it, and the two-step construction
applies unaltered on the physical coordinates. The waverider volume field
addresses a \emph{parameter-dependent moving discontinuity} on that same fixed
domain, through a learned shock-aligned coordinate, shock-focused training and
a hotspot-aware residual correction. Where a feature does move, the trunk is
supplied with a coordinate in which it is stationary; where none does, no such
coordinate is introduced. In all three the parametric dependence is then
carried by a branch network acting on a single scalar.

\section{Computational Experiments}
\label{sec:computational_experiments}
 
This section describes the datasets, training protocols, validation
strategies and error measures used to evaluate the three surrogates. The
tangent-ogive experiments assess geometry-dependent prediction of a
fourteen-field thermochemical state on changing physical domains. The
waverider surface experiments assess prediction of the wall heat flux and
traction components, and of the integrated loads formed from them, on a fixed
surface mesh. The waverider volume experiments assess density-field
reconstruction in the presence of an angle-of-attack-dependent moving shock.
In all three, complete CFD realizations are held out during evaluation, and
spatial samples from a held-out realization are never used during surrogate
construction. The error measures are common to all three and are stated
first.

\subsection{Error Measures}
\label{subsec:exp_field_metrics}
 
For a reference CFD field $q$ and its surrogate prediction $\widehat{q}$
evaluated at $N$ spatial points, the primary accuracy measure throughout this
work is the percentage relative $L_2$ error,
\begin{equation}
E_{L_2}(q)
=
100\,
\frac{
\left\|\widehat{q}-q\right\|_2
}{
\left\|q\right\|_2
}.
\label{eq:exp_relative_l2_error}
\end{equation}
Every relative error quoted in this paper is this quantity, evaluated on the
point set appropriate to the surrogate concerned: the canonical grid for the
tangent ogive, the surface cells for the waverider surface fields, and the
volume mesh for the waverider density field. The normalized root-mean-square
error is
\begin{equation}
E_{\mathrm{NRMSE}}(q)
=
100\,
\frac{
\sqrt{
\frac{1}{N}
\sum_{i=1}^{N}
\left(\widehat{q}_i-q_i\right)^2
}
}{
q_{\max}-q_{\min}
},
\label{eq:exp_nrmse}
\end{equation}
and the normalized maximum error is
\begin{equation}
E_{\infty}(q)
=
100\,
\frac{
\max_i
\left|
\widehat{q}_i-q_i
\right|
}{
q_{\max}-q_{\min}
}.
\label{eq:exp_linf_error}
\end{equation}
The coefficient of determination is evaluated as
\begin{equation}
R^2(q)
=
1-
\frac{
\displaystyle\sum_{i=1}^{N}
\left(\widehat{q}_i-q_i\right)^2
}{
\displaystyle\sum_{i=1}^{N}
\left(q_i-\overline{q}\right)^2
},
\qquad
\overline{q}
=
\frac{1}{N}\sum_{i=1}^{N}q_i.
\label{eq:exp_r2}
\end{equation}
This is $R^2$ about the \emph{line of equality} $\widehat{q}=q$, not about a
regression line fitted to the scatter of $\widehat{q}$ against $q$. The
distinction matters for the diagnostic use made of it later: a surrogate
whose predictions are perfectly correlated with the reference but
systematically offset or rescaled scores well on the fitted-line measure and
poorly on \Cref{eq:exp_r2}, and it is the latter that reflects whether the
prediction can be used in place of the reference.
 
For the two-dimensional tangent-ogive velocity field, the two components are
also assessed jointly using
\begin{equation}
E_{\mathbf{u}}
=
100\,
\frac{
\sqrt{
\left\|\widehat{u}-u\right\|_2^2
+
\left\|\widehat{v}-v\right\|_2^2
}
}{
\sqrt{
\left\|u\right\|_2^2
+
\left\|v\right\|_2^2
}
}.
\label{eq:exp_velocity_vector_error}
\end{equation}
 
Gradient accuracy is evaluated in physical coordinates, because wall heating
depends on physical rather than canonical derivatives. Canonical derivatives
are transformed using \Cref{eq:physical_gradient_transform}, and the gradient
error is
\begin{equation}
E_{\nabla q}
=
100\,
\frac{
\left\|
\nabla\widehat{q}
-
\nabla q
\right\|_2
}{
\left\|
\nabla q
\right\|_2
}.
\label{eq:exp_gradient_error}
\end{equation}
This is reported as a validation quantity in its own right rather than
inferred from field accuracy, for the reason given in \Cref{sec:intro}: a
loss imposed on field values bounds the error in $q$ but not in $\nabla q$,
and every wall functional is a functional of the latter.
 
Finally, a global error can be dominated by the comparatively large
smooth-flow portion of a domain and can under-represent localized
discrepancies near a shock or a wall. Errors are therefore also reported over
restricted regions $\Omega_r$,
\begin{equation}
E_{L_2}^{(r)}(q)
=
100\,
\frac{
\left\|
\widehat{q}-q
\right\|_{2,\Omega_r}
}{
\left\|
q
\right\|_{2,\Omega_r}
},
\label{eq:exp_regional_error}
\end{equation}
with the regions defined per configuration in the subsections that follow.

\subsection{Surrogate I --- Tangent-Ogive Dataset and Predicted Quantities}
\label{subsec:exp_tangent_dataset}
 
The tangent-ogive database contains $37$ CFD realizations organized into two
independent geometry families. The integer-indexed family contains the $20$
configurations \texttt{run.1}--\texttt{run.20}, whereas the fractional family
contains the $17$ intermediate configurations
\texttt{run.2\_5}--\texttt{run.18\_5}. The intermediate cases were generated
at half the spatial resolution of the integer-indexed cases to reduce the cost
of database construction. Following canonicalization, the fields from both
families are represented on the common body-fitted coordinate system
$(s,\eta)$. Because the two families correspond to distinct sampling
trajectories and spatial resolutions, they are retained as separate data
clusters during validation rather than pooled into a single random split.
 
Independent operator surrogates are constructed for the primary flow
variables, reacting-species mass fractions, internal-mode temperature, and
transport quantities required for aerodynamic and aerothermal
post-processing. The complete set of modeled scalar fields is
\begin{equation}
\mathcal{Q}
=
\left\{
\rho,\,
u,\,
v,\,
p,\,
T,\,
Y_{\mathrm{N_2}},\,
Y_{\mathrm{O_2}},\,
Y_{\mathrm{NO}},\,
Y_{\mathrm{N}},\,
Y_{\mathrm{O}},\,
T_m,\,
k,\,
k_m,\,
\mu
\right\}.
\label{eq:exp_surrogate_variables}
\end{equation}
Here $u$ and $v$ are the axial and transverse velocity components; $T$ is the
translational--rotational temperature; $T_m$ is the additional
molecular-mode temperature; $k$ and $k_m$ are the corresponding thermal
conductivities; and $\mu$ is the dynamic viscosity. A separate branch--trunk
model is trained for each scalar field. This permits field-specific
preprocessing and prevents variables with large numerical magnitudes from
dominating quantities with smaller scales or more localized spatial support.
 
\begin{table}[htbp]
    \centering
    \caption{Primary flow and reacting-species fields represented by
    independent branch--trunk models within the tangent-ogive surrogate.}
    \label{tab:exp_cfd_species_variables}
    \begin{tabular}{llll}
        \toprule
        \textbf{Category}
        & \textbf{Variable}
        & \textbf{Symbol}
        & \textbf{Physical quantity} \\
        \midrule
        Flow & Density & $\rho$ & Mixture density \\
        Flow & Axial velocity & $u$ & Streamwise velocity component \\
        Flow & Transverse velocity & $v$ & Transverse velocity component \\
        Flow & Pressure & $p$ & Static pressure \\
        Thermal & Temperature & $T$ & Translational--rotational temperature \\
        \midrule
        Chemistry & Molecular nitrogen & $Y_{\mathrm{N_2}}$ & $\mathrm{N_2}$ mass fraction \\
        Chemistry & Molecular oxygen & $Y_{\mathrm{O_2}}$ & $\mathrm{O_2}$ mass fraction \\
        Chemistry & Nitric oxide & $Y_{\mathrm{NO}}$ & $\mathrm{NO}$ mass fraction \\
        Chemistry & Atomic nitrogen & $Y_{\mathrm{N}}$ & $\mathrm{N}$ mass fraction \\
        Chemistry & Atomic oxygen & $Y_{\mathrm{O}}$ & $\mathrm{O}$ mass fraction \\
        \bottomrule
    \end{tabular}
\end{table}
 
The additional mode-temperature and transport fields used for heat-transfer
and viscous-force reconstruction are summarized in
\Cref{tab:exp_transport_variables}. The mode temperature $T_m$ is
standardized directly. Because $k$, $k_m$ and $\mu$ are strictly positive,
these quantities are logarithmically transformed before standardization. For
$q\in\{k,k_m,\mu\}$,
\begin{equation}
q^{*}
=
\frac{
\ln(q)-\mu_{\ln q}
}{
\sigma_{\ln q}
},
\label{eq:exp_transport_log_normalization}
\end{equation}
where $\mu_{\ln q}$ and $\sigma_{\ln q}$ are computed using only the training
realizations in the current validation fold. The physical prediction is
recovered as
\begin{equation}
\widehat{q}
=
\exp\left(
\widehat{q}^{*}\sigma_{\ln q}
+
\mu_{\ln q}
\right),
\label{eq:exp_transport_inverse_transform}
\end{equation}
which preserves positivity of the reconstructed transport coefficients.
 
\begin{table}[htbp]
    \centering
    \caption{Mode-temperature and transport fields represented by independent
    models and their use in physical post-processing.}
    \label{tab:exp_transport_variables}
    \resizebox{\textwidth}{!}{%
    \begin{tabular}{lllll}
        \toprule
        \textbf{Variable}
        & \textbf{Symbol}
        & \textbf{CFD field}
        & \textbf{Transformation}
        & \textbf{Primary use} \\
        \midrule
        Mode temperature
        & $T_m$
        & \texttt{T\_modes[0]}
        & Standardization
        & Wall heat flux \\
        Thermal conductivity
        & $k$
        & \texttt{k}
        & Log-standardization
        & Wall heat flux \\
        Mode thermal conductivity
        & $k_m$
        & \texttt{k\_modes[0]}
        & Log-standardization
        & Wall heat flux \\
        Dynamic viscosity
        & $\mu$
        & \texttt{mu}
        & Log-standardization
        & Viscous surface force \\
        \bottomrule
    \end{tabular}%
    }
\end{table}

\subsection{Tangent-Ogive Training Configuration}
\label{subsec:exp_tangent_training}
 
All tangent-ogive fields are represented on the same canonical grid. The
canonical coordinates $(s,\eta)$ are supplied to the Transolver trunk, while
the scalar geometry parameter $\xi_g$ is supplied to the branch model. The
geometry-dependent physical coordinates $(x,y)$ are retained for
visualization, transformation of spatial derivatives, surface-force
integration and wall heat-flux evaluation.
 
The two-step procedure of \Cref{subsec:two_step_operator} is applied
independently to every field in \Cref{eq:exp_surrogate_variables}. In Step~1
the Transolver trunk and a free coefficient vector for each training
realization are optimized jointly; the learned spatial representation is then
orthogonalized by QR factorization. In Step~2 the branch network is trained to
map $\xi_g$ to the transformed coefficient vector. Piecewise cubic Hermite
interpolation is applied componentwise in the same orthogonal coefficient
space as a non-neural baseline. The branch and PCHIP predictions therefore
share an identical frozen spatial basis, so differences between them are
attributable to the parameter-to-coefficient map rather than to the spatial
representation. The complete model and optimization configuration is given in
\Cref{tab:exp_tangent_hyperparameters}.
 
\begin{table}[htbp]
    \centering
    \caption{Tangent-ogive surrogate architecture and training settings.}
    \label{tab:exp_tangent_hyperparameters}
    \begin{tabular}{ll}
        \toprule
        \textbf{Setting} & \textbf{Value} \\
        \midrule
 
        Canonical grid size $(N_s \times N_\eta)$
        & $600\times300$ \\
 
        Latent basis dimension $r$
        & $32$ \\
 
        Transolver feature width $d$
        & $128$ \\
 
        Number of Transolver blocks
        & $6$ \\
 
        Number of physics-state slices $M$
        & $64$ \\
 
        Step-1 spatial batch size
        & full $600\times300$ grid ($180{,}000$ points/geometry) \\
 
        Step-1 geometry batch size
        & $4$ geometries \\
 
        Step-1 optimizer and learning rate
        & AdamW, cosine decay; $5\times10^{-4}$ for $\rho$,
          $3\times10^{-4}$ for $p,u,v,T$ \\
 
        Step-1 training iterations
        & $150{,}000$ optimization steps \\
 
        Step-2 branch architecture
        & $1\!-\!128\!-\!128\!-\!128\!-\!r$ MLP, $\tanh$ \\
 
        Step-2 optimizer and learning rate
        & AdamW, $10^{-3}$ with cosine decay to $10^{-6}$ \\
 
        Step-2 training iterations
        & $40{,}000$ optimization steps \\
 
        Number of random initializations
        & $1$ per LOO fold (seed $2026$) \\
 
        Hardware and precision
        & 1 NVIDIA Blackwell 40 GB GPU, FP32 \\
 
        \bottomrule
    \end{tabular}
\end{table}

\subsection{Family-Wise Leave-One-Out Validation}
\label{subsec:exp_familywise_loo}
 
Generalization to unseen tangent-ogive shapes is assessed using exhaustive
leave-one-out (LOO) validation independently within the integer and fractional
geometry families. Let $\mathcal{D}_{\mathrm{int}}$ and
$\mathcal{D}_{\mathrm{frac}}$ denote the two datasets, with cardinalities $20$
and $17$. For a held-out case $\mathcal{D}_i$ belonging to family
$\mathcal{F}\in\{\mathrm{int},\mathrm{frac}\}$,
\begin{equation}
\mathcal{D}_{\mathrm{train}}^{(i,\mathcal{F})}
=
\mathcal{D}_{\mathcal{F}}
\setminus
\left\{\mathcal{D}_i\right\},
\qquad
\mathcal{D}_{\mathrm{test}}^{(i,\mathcal{F})}
=
\left\{\mathcal{D}_i\right\}.
\label{eq:exp_familywise_loo}
\end{equation}
An integer-family fold therefore contains $19$ training geometries, a
fractional-family fold $16$. The held-out realization is excluded from all
data-dependent stages of surrogate construction, including normalization,
Step-1 trunk-basis learning, free-coefficient optimization, QR
orthogonalization, Step-2 branch fitting and PCHIP construction.
 
A fold is classified as interpolation when the held-out geometry parameter
lies inside the range spanned by the remaining members of the same family.
The minimum- and maximum-parameter members of each family are classified as
endpoint extrapolation cases and are reported separately from the interior LOO
statistics. PCHIP is used in its shape-preserving interpolation mode for
interior folds; outside the available coefficient range, linear coefficient
extrapolation is used. This separation prevents the greater difficulty of
endpoint extrapolation from being hidden within the principal interpolation
statistics.
 
Regional errors \eqref{eq:exp_regional_error} are reported for the
tangent-ogive fields over the shock region $\Omega_{\mathrm{sh}}$, the
near-wall region $\Omega_{\mathrm{w}}$ and the remaining smooth-flow region
$\Omega_{\mathrm{sm}}$.

\subsection{Reacting-Species Closure and Error Measures}
\label{subsec:exp_species_metrics}
 
Each species mass fraction is predicted by an independent field model, so the
raw predictions do not automatically satisfy positivity or the unit-sum
constraint. With $N_{\mathrm{sp}}=5$ the number of species, a
positivity-preserving normalized closure is applied,
\begin{equation}
\widehat{Y}_k^{\mathrm{closed}}
=
\frac{
\max\left(
\widehat{Y}_k^{\mathrm{raw}},
\varepsilon
\right)
}{
\displaystyle
\sum_{j=1}^{N_{\mathrm{sp}}}
\max\left(
\widehat{Y}_j^{\mathrm{raw}},
\varepsilon
\right)
},
\label{eq:exp_species_closure}
\end{equation}
which guarantees
\begin{equation}
\widehat{Y}_k^{\mathrm{closed}}\geq0,
\qquad
\sum_{k=1}^{N_{\mathrm{sp}}}
\widehat{Y}_k^{\mathrm{closed}}=1
\end{equation}
up to floating-point precision. Field errors are evaluated using the closed
mass fractions. To quantify the unconstrained behavior of the independently
trained species models, the raw closure residual is retained as
\begin{equation}
\epsilon_{\Sigma Y}^{\mathrm{raw}}(\mathbf{x})
=
\left|
\sum_{k=1}^{N_{\mathrm{sp}}}
\widehat{Y}_k^{\mathrm{raw}}(\mathbf{x})
-1
\right|,
\label{eq:exp_species_raw_closure_residual}
\end{equation}
with mean and maximum values
\begin{equation}
\overline{\epsilon}_{\Sigma Y}^{\mathrm{raw}}
=
\frac{1}{N}
\sum_{i=1}^{N}
\epsilon_{\Sigma Y}^{\mathrm{raw}}(\mathbf{x}_i),
\qquad
\epsilon_{\Sigma Y,\max}^{\mathrm{raw}}
=
\max_i
\epsilon_{\Sigma Y}^{\mathrm{raw}}(\mathbf{x}_i).
\label{eq:exp_species_raw_closure_metrics}
\end{equation}
 
For trace species a full-domain relative error is difficult to interpret,
because the reference mass fraction approaches zero over a large portion of
the domain. Species accuracy is therefore also evaluated over the chemically
active region
\begin{equation}
\Omega_k^{\mathrm{act}}
=
\left\{
\mathbf{x}:
Y_k(\mathbf{x})>\tau_k
\right\},
\label{eq:exp_active_species_region}
\end{equation}
with $\tau_k$ a species-dependent threshold determined from the CFD data.

\subsection{Surrogate II --- Waverider Surface Validation Strategy}
\label{subsec:exp_waverider_surface_protocol}
 
Two complementary validation strategies are used for the surface surrogate.
In the first, a leave-one-angle-out procedure excludes one CFD snapshot
entirely from both stages of training; the omitted surface field is then
predicted from its angle of attack and compared with the CFD solution, giving
an out-of-sample assessment of the parametric map. In the second, production
surrogates are trained on all $29$ snapshots and used for dense evaluation of
the design space, for which the comparison at the training angles measures
reconstruction rather than generalization. Both are reported, and the
distinction between them is maintained throughout
\Cref{sec:results_waverider_surface}.
 
To separate errors arising from the learned spatial representation from those
introduced by the branch map, the optimal projection of the reference field
onto the frozen Stage-I basis is also computed,
\begin{equation}
\mathbf{u}_{\mathrm{proj}}
=
\mathbf{Q}\mathbf{Q}^{\mathsf T}\mathbf{u}_{\mathrm{CFD}} .
\end{equation}
Evaluating \Cref{eq:exp_relative_l2_error} on $\mathbf{u}_{\mathrm{proj}}$
gives the projection error $E_{L_2}^{\mathrm{proj}}$, a lower bound on what
any branch network can achieve with this basis, and evaluating it on the
surrogate prediction $\widehat{\mathbf{u}}$ gives the full error
$E_{L_2}^{\mathrm{DON}}$. When $E_{L_2}^{\mathrm{DON}}$ approaches
$E_{L_2}^{\mathrm{proj}}$, the residual surrogate error is dominated by the
finite rank of the spatial representation rather than by the
angle-to-coefficient interpolation, and increasing $r$ rather than improving
the branch is the route to further accuracy.
 
Dataset and training settings are given in \Cref{tab:deeponet_config}.
 
\begin{table}[t]
\centering
\caption{Two-step DeepONet dataset and training configuration for the AEDC
waverider surface quantities.}
\label{tab:deeponet_config}
\begin{tabular}{ll}
\hline
Quantity & Value\\
\hline
AoA range & $[-14^\circ,14^\circ]$\\
Number of CFD snapshots & 29\\
AoA spacing & $1^\circ$\\
Surface cells & 266,848\\
Trunk input & $(x,y,z)$\\
Branch input & $\alpha$\\
Output fields & $q_w,t_x,t_y,t_z$\\
Latent rank $r$ & 16\\
Trunk network & $3$--$128$--$128$--$128$--$128$--$16$\\
Branch network & $1$--$32$--$32$--$16$\\
Activation & $\tanh$\\
Stage-I iterations & 100,000\\
Stage-II iterations & 15,000\\
Surface samples/Stage-I step & 65,536\\
Stage-I initial learning rate & $5\times10^{-4}$\\
Stage-II initial learning rate & $10^{-3}$\\
\hline
\end{tabular}
\end{table}

\subsection{Waverider Surface Quantities of Interest and Their Errors}
\label{subsec:exp_waverider_surface_metrics}
\label{sec:metrics}
 
The surface surrogate is evaluated on the $N_c=266{,}848$ surface cells of the
waverider. For a surface field
$u\in\{q_w,t_x,t_y,t_z\}$ at angle of attack $\alpha$, let $u_e$ and
$\widehat{u}_e$ denote the CFD and predicted values in cell $e$; the field
accuracy is \Cref{eq:exp_relative_l2_error} evaluated over those cells,
\begin{equation}
  E_{L_2}\big(u;\alpha\big) =
  100\,\frac{\big(\sum_{e=1}^{N_c} (\widehat u_e - u_e)^2\big)^{1/2}}
       {\big(\sum_{e=1}^{N_c} u_e^2\big)^{1/2}} .
  \label{eq:field_error}
\end{equation}
Here $\mathbf{t}$ is the \emph{total} surface traction, comprising both the
pressure and the viscous contributions,
\begin{equation}
  \mathbf{t}
  = \underbrace{-(p-p_\infty)\,\mathbf{n}}_{\text{pressure}}
  + \underbrace{\boldsymbol{\tau}\,\mathbf{n}}_{\text{viscous}},
  \label{eq:total_traction}
\end{equation}
with $\mathbf{n}$ the outward wall normal, $p_\infty$ the freestream
pressure and $\boldsymbol{\tau}$ the compressible Newtonian viscous stress
tensor, whose components are given in \Cref{app:traction}. Its body-axis
components are written $(t_x,t_y,t_z)$, and each carries both terms of
\Cref{eq:total_traction}. The pressure term dominates the drag of a blunt
hypersonic body, so the two contributions are retained together throughout:
the surrogate is trained on the total traction supplied by the solver, and
the forces of \Cref{eq:cd_cl} are formed from it without separating them.
 
The engineering quantities of interest are computed identically from the CFD
and the predicted fields. The integrated surface force is
$\mathbf{F}=\sum_e \mathbf{t}_e A_e$, with $A_e$ the cell area. Because the
vehicle is laterally symmetric the pitch plane is the $x$--$y$ plane, and
drag and lift are the projections of $\mathbf{F}$ on the freestream and
normal directions,
\begin{equation}
  C_D = \frac{\mathbf{F}\cdot(\cos\alpha,\ \sin\alpha,\ 0)}{q_\infty S_\mathrm{ref}},
  \qquad
  C_L = \frac{\mathbf{F}\cdot(-\sin\alpha,\ \cos\alpha,\ 0)}{q_\infty S_\mathrm{ref}} ,
  \label{eq:cd_cl}
\end{equation}
where $q_\infty=\tfrac12\rho_\infty V_\infty^2$ is the freestream dynamic
pressure and $S_\mathrm{ref}$ the reference area. The thermal quantities are
the integrated wall heat load and the peak wall heat flux,
\begin{equation}
  Q_\mathrm{wall} = \sum_{e=1}^{N_c} q_{w,e} A_e ,
  \qquad
  q_{w,\max} = \max_{e} q_{w,e} .
  \label{eq:thermal_qoi}
\end{equation}
 \begin{table}
    \centering
    \caption{Waverider volume data split and shock-aware DeepONet training
    settings.}
    \label{tab:exp_waverider_configuration}
    \begin{tabular}{ll}
        \toprule
        \textbf{Setting} & \textbf{Value} \\
        \midrule
        Training angles of attack
        &
        $\{-14^\circ,\ldots,14^\circ\}
        \setminus\{-6^\circ,+7^\circ\}$ (27 cases)
        \\
        Validation angle of attack
        &
        $-6^\circ$
        \\ 
        Test/reference angle of attack
        &
        $+7^\circ$ (held historical reference)
        \\
        Latent basis dimension $r$
        &
        $64$
        \\ 
        Spatial mini-batch size $N_b$
        &
        $16{,}384$ points per selected AoA
        \\
        AoA cases per optimization step
        &
        $4$
        \\
        Spatial sampling fractions
        &
        $50\%$ background, $25\%$ shock band, $25\%$ shock core
        \\
        Shock-focused sample fraction $f_{\mathrm{shock}}$
        &
        $0.50$ (band $+$ core)
        \\
        Shock-sensor percentile threshold
        &
        top $5\%$ shock region; top $1\%$ shock core
        \\
        Shock-region loss weights
        &
        $1:2:4:1$ (background:band:core:edge)
        \\
        Shock-edge samples
        &
        $4096$ edge pairs per selected AoA
        \\
        Step-1 optimizer
        &
        AdamW, weight decay $10^{-6}$
        \\
        Step-1 learning rate
        &
        $2\times10^{-4}
        \rightarrow6\times10^{-5}
        \rightarrow1.8\times10^{-5}
        \rightarrow5.4\times10^{-6}$
        \\
        Step-2 optimizer
        &
        AdamW, weight decay $10^{-6}$
        \\
        Step-2 learning rate
        &
        $5\times10^{-4}
        \rightarrow1.5\times10^{-4}
        \rightarrow4.5\times10^{-5}
        \rightarrow1.35\times10^{-5}$
        \\
        Step-1 / Step-2 training epochs
        &
        $100{,}000$ / $100{,}000$
        \\
        QR sensor points
        &
        $10^{6}$
        \\
 
        Number of random initializations
        &
        $1$ (seed $=42$)
        \\
 
        Hardware and precision
        &
        1 NVIDIA Blackwell 40 GB GPU, FP32
        \\
 
        \bottomrule
    \end{tabular}
\end{table}

Errors in these quantities are reported as signed differences between
surrogate and CFD,
\begin{equation}
  \Delta C_D = \hat C_D - C_D , \qquad
  \Delta C_L = \hat C_L - C_L , \qquad
  \Delta Q = \frac{\hat Q - Q}{Q}\times 100\% ,
  \quad Q \in \{Q_\mathrm{wall},\, q_{w,\max}\} ,
  \label{eq:qoi_error}
\end{equation}
whose sign indicates over- or under-prediction. The aerodynamic errors are
absolute rather than relative, because $C_L$ changes sign near
$\alpha\approx-5^\circ$, where a relative error is ill-defined; they are
quoted in units of $10^{-4}$, that is, in drag and lift counts. Tables report
the magnitudes $|\Delta C_D|$ and $|\Delta C_L|$.
 
Two models are compared. The \emph{all-AoA} model is trained on all $29$
snapshots and evaluated at those angles, which measures reconstruction. To
measure generalization, a \emph{leave-one-out} model is trained for each
$\alpha^\star\in\{-12^\circ,-8^\circ,-4^\circ,0^\circ,+4^\circ,+8^\circ,
+12^\circ\}$ with that snapshot removed from both training stages, and is
evaluated only at the held-out $\alpha^\star$. Comparisons between the two
are made at the same angles, since the all-AoA maximum is taken over $29$
angles and the leave-one-out maximum over $7$.

\subsection{Surrogate III --- Waverider Volume Evaluation Protocol}
\label{subsec:exp_waverider_protocol}
 
The volume surrogate is trained on the same $29$ realizations. Complete
angle-of-attack realizations are assigned to training, validation or testing,
and spatial points from a held-out angle are never mixed into the training
set. The logarithmic-density normalization of
\Cref{eq:waverider_log_density} and the connectivity-based shock mask of
\Cref{eq:waverider_training_sensor} are computed exclusively from the
training-angle fields; validation and test fields are used only for
assessment. Accuracy is reported over the complete domain and, separately,
over the training-derived shock mask, using
\Cref{eq:exp_relative_l2_error,eq:exp_regional_error}. The data split and training settings are given in \Cref{tab:exp_waverider_configuration}.

\subsection{Aerodynamic and Thermal Quantities of Interest}
\label{sec:qoi_methodology}
 
The aerodynamic and thermal quantities of interest are evaluated from the CFD
reference fields and from the surrogate predictions by one and the same
post-processing routine, so that the reported differences reflect differences
in the predicted flow fields alone and nothing else. For the tangent ogive the
routine acts on the common structured grid of
\Cref{subsec:canonical_mapping}, taking the eight fields
$p$, $U$, $V$, $T$, $T_m$, $\mu$, $k$ and $k_m$ and returning the surface
traction, the wall heat flux, the Nusselt number, the integrated loads and the
zero-incidence symmetry diagnostics. For the waverider it acts on the surface
mesh and returns the force coefficients in the axis convention of
\Cref{app:axes}.
 
The definitions are given in full in \Cref{app:qoi}: wall geometry and normals
(\Cref{app:wall_geometry}), pressure and viscous traction
(\Cref{app:traction}), wall heat flux (\Cref{app:heat_flux}), Nusselt number
(\Cref{app:nusselt}), the lift and pressure symmetry diagnostics
(\Cref{app:symmetry,app:pressure_symmetry}), the comparison metrics used
throughout the results (\Cref{app:comparison_metrics}), and the waverider
force convention (\Cref{app:axes}). They are collected there because they are
standard reductions whose details are needed for reproduction but not for
reading the results.

\section{Results I: Tangent-Ogive Surrogate and Aerothermal Shape Optimisation}
\label{sec:results_ogive}
 
\label{sec:loo_assessment}%
 
This section collects every result for the tangent-ogive configuration: the
field-level accuracy of the surrogate over the $37$ geometries, the
wall-resolved and global quantities of interest reduced from those fields, and
the multi-objective shape optimisation that the surrogate makes affordable.
The order is one of increasing reduction. The field errors of
\Cref{sec:loo_cfd,sec:loo_species,sec:loo_transport} measure what the
surrogate reproduces pointwise; the wall and global quantities of
\Cref{subsec:res_wall_qoi,subsec:res_global_qoi} measure what survives the
reduction a design loop actually consumes; and
\Cref{sec:pareto_optimization} asks what can be decided with what survives.
 
\subsection{CFD variables}\label{sec:loo_cfd}
 
The geometry database consists of two families of tangent-ogive shapes. The
first, denoted the integer family, contains $20$ geometries whose shape factor
$\xi_g$ decreases uniformly from $1.000$ to $0.500$. The second, the half
family, contains $17$ geometries interleaved between them, spanning
$\xi_g \in [0.536, 0.938]$. The two families were generated and simulated
separately, and they are treated here as two independent ensembles: for each
family, the field normalization, the Transolver trunk, the free case
coefficients, the discrete orthonormalization and both coefficient models are
fitted using members of that family alone. No quantity is shared between the
families, so a surrogate trained on one is never informed by the other.
 
\begin{figure}
    \centering
    \includegraphics[width=\linewidth]{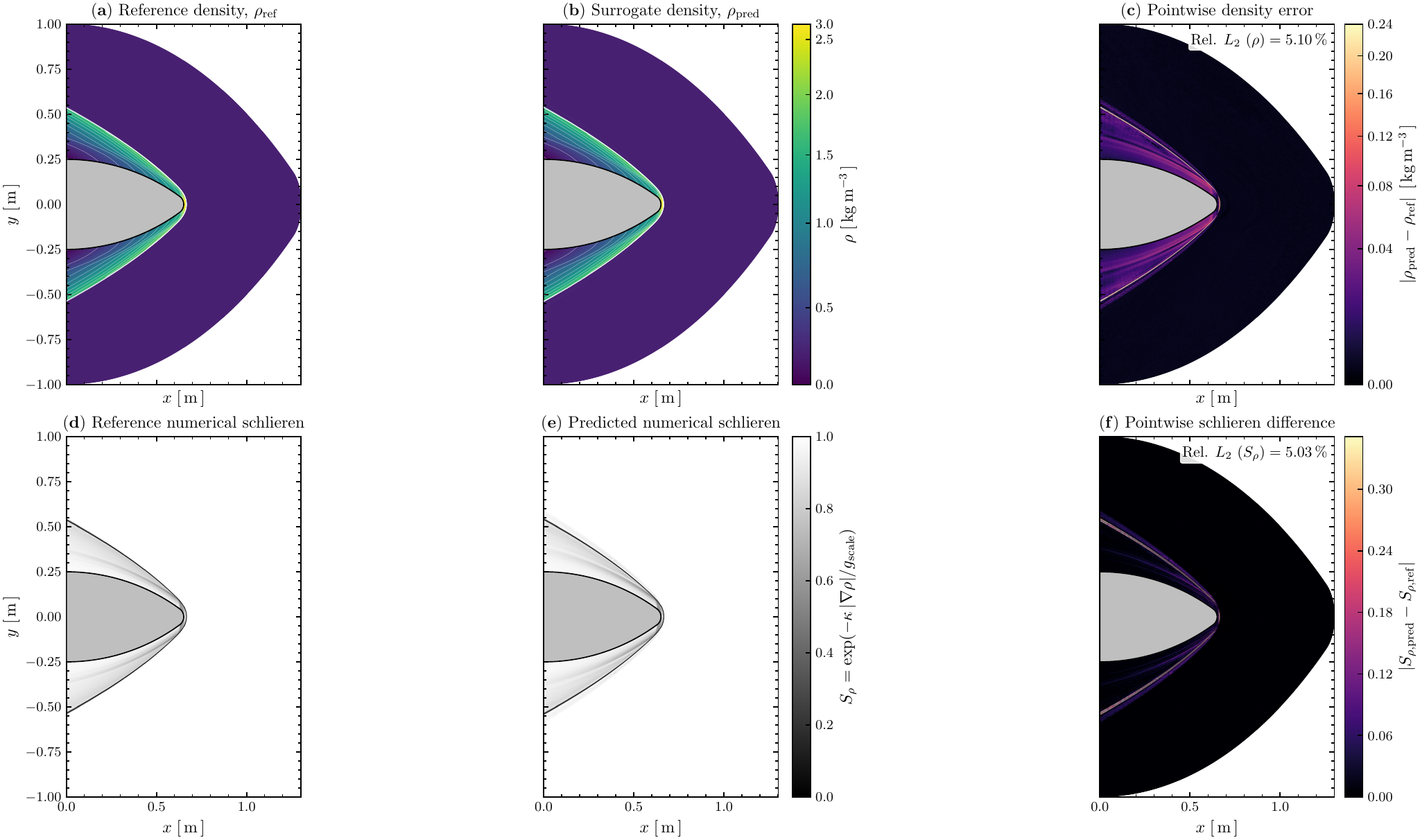}
    \caption{Representative held-out density-field prediction for
    $\xi_g=0.5$.}
    \label{fig:run20_density_holdout}
\end{figure}
 
Within each family the surrogate is assessed by leave-one-out
cross-validation. One geometry is held out, the entire two-step construction
is refitted on the remaining members, and the held-out field is predicted.
Repeating this for every member gives $20$ folds from the integer family and
$17$ from the half family, $37$ in total. The protocol has one consequence
that is central to what follows. When the held-out geometry is the smallest or
largest shape factor of its family, the remaining fit points all lie to one
side of it, and the coefficient model has no bracketing nodes. Those folds are
extrapolation problems rather than interpolation problems. There are four of
them --- $\mathrm{run.1}$ and $\mathrm{run.20}$ at the ends of the integer
family, $\mathrm{run.2\_5}$ and $\mathrm{run.18\_5}$ at the ends of the half
family --- leaving $33$ interpolation folds. The two regimes are reported separately in \Cref{tab:loo_refined_relative_l2} because pooling them would be misleading rather than merely coarse. Which of the two coefficient models is the more accurate depends on the regime: the baseline is better on every field within the sampled range, and worse on every field outside it. A single figure pooled over all 
37
37 folds therefore reflects the 
33
33-to-
4
4 split of the design more than it reflects either model.
 
\begin{figure}[htbp]
    \centering
 
    \begin{subfigure}[b]{0.48\textwidth}
        \centering
        \includegraphics[width=\linewidth]{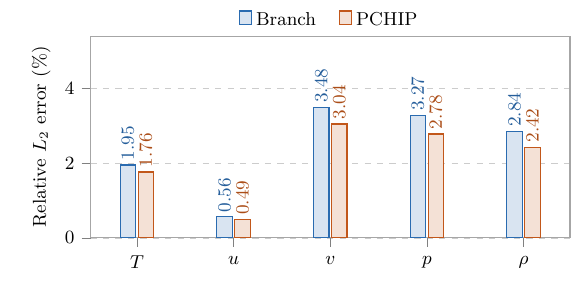}
        \caption{Interpolation folds, mean over the $33$ folds.}
        \label{fig:loo_bar_interp}
    \end{subfigure}
    \hfill
    \begin{subfigure}[b]{0.48\textwidth}
        \centering
        \includegraphics[width=\linewidth]{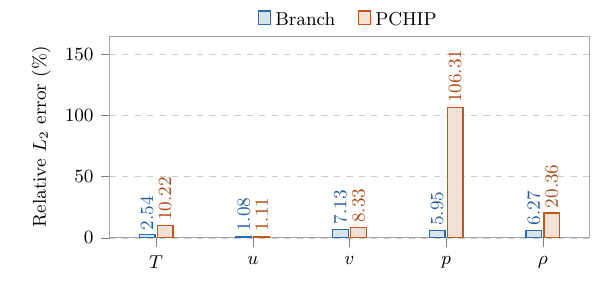}
        \caption{Extrapolation folds, mean over the four folds.}
        \label{fig:loo_bar_extrap}
    \end{subfigure}
 
    \vspace{0.9em}
 
    \begin{subfigure}[b]{\textwidth}
        \centering
        \includegraphics[width=\linewidth]{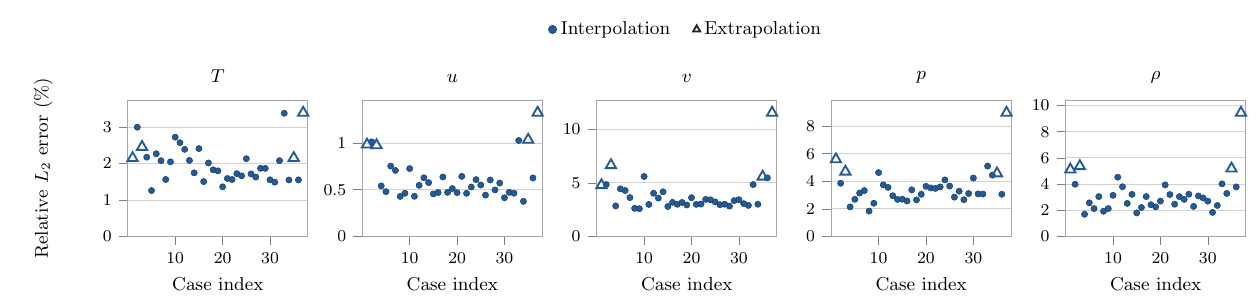}
        \caption{Fold-resolved error of the surrogate.}
        \label{fig:loo_case_branch}
    \end{subfigure}
 
    \vspace{0.6em}
 
    \begin{subfigure}[b]{\textwidth}
        \centering
        \includegraphics[width=\linewidth]{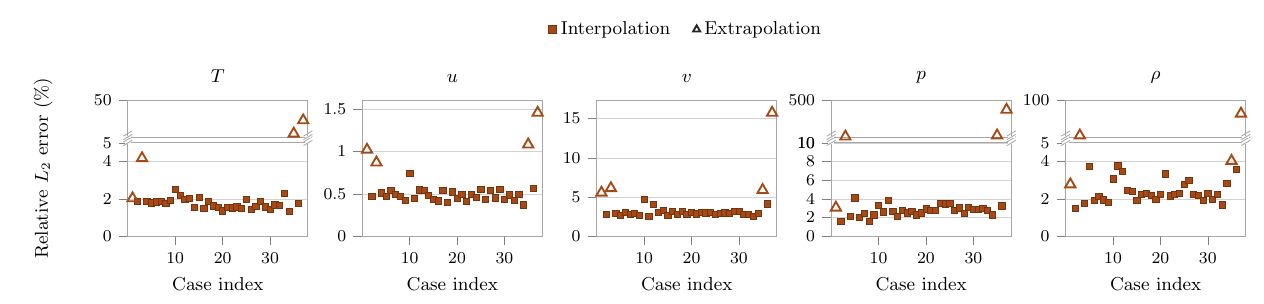}
        \caption{Fold-resolved error of the PCHIP baseline.}
        \label{fig:loo_case_pchip}
    \end{subfigure}
 
    \caption[Leave-one-out accuracy of the canonical-coordinate surrogate]%
{Leave-one-out accuracy for the five primary flow variables over the $37$
folds of the two geometry families: $33$ interpolate within the sampled range
of the shape factor, and four --- the extremes of each family --- extrapolate
beyond it. \subref{fig:loo_bar_interp} and \subref{fig:loo_bar_extrap}~Mean
relative $L_2$ error in each regime, printed above each bar. Within the
sampled range PCHIP is the more accurate on every variable, by $9.6$ to
$15.2\%$; outside it the branch construction is the more accurate on every
variable, by up to a factor of $17.9$ on the pressure. Spread is omitted
because the fold-wise distribution is right-skewed.
\subref{fig:loo_case_branch} and \subref{fig:loo_case_pchip}~The same errors
by fold, ordered by shape factor with the families pooled so that the
extrapolation folds (open triangles) lie at the ends of the axis. The largest
error over the study is $11.55\%$ for the branch construction and $375.23\%$
for PCHIP, both at $\mathrm{run.1}$. \emph{Each panel is scaled to its own
data}, and the three PCHIP panels with outlying folds carry a broken vertical
axis, marked by slashes on the facing spines; \emph{the metric is not
continuous across the break}.}
\label{fig:loo_refined}
\end{figure}

 
Accuracy is measured by the relative $L_2$ error over all valid points of the
held-out field. \Cref{fig:loo_refined} reports the mean of this quantity
across folds within each regime, and
\Cref{tab:loo_refined_relative_l2} adds the median and the maximum, because
for one model the three statistics tell materially different stories. Over the
four extrapolation folds the PCHIP error on the pressure has a mean of
$106.31\%$ but a median of $23.46\%$: a single fold, $\mathrm{run.1}$,
contributes $88\%$ of that sum. A mean of $106\%$ therefore describes no fold
that was actually computed. The same quantity for the branch construction has
a mean of $5.96\%$ and a median of $5.14\%$, a gap of a few per cent rather
than a factor of four. We report all three statistics rather than choosing
between them: the mean is what an optimizer experiences in aggregate, the
median describes a typical fold, and the maximum bounds the worst case. Where
the three disagree, as they do for PCHIP outside the sampled range, that
disagreement is itself the result.
 
 
On the $33$ interpolation folds the two coefficient models are close, and the
PCHIP baseline is consistently the more accurate of the two. It leads on
\emph{every} field, by between $9.6\%$ and $15.2\%$ in relative terms:
$1.76\%$ against $1.95\%$ on $T$, $0.49\%$ against $0.56\%$ on $u$, and
$2.42\%$ against $2.84\%$ on $\rho$. This is the expected outcome. When
bracketing nodes are available, monotone cubic interpolation of a smooth
coefficient trajectory is close to optimal, and the branch network pays a
small price for representing that trajectory as a single global fit rather
than a local one. The ranking of the fields by difficulty is identical for
both models: $u$ is reproduced most accurately, at $0.49$--$0.56\%$, followed
by $T$ and $\rho$, while $p$ and $v$ carry the largest relative errors, $v$
reaching $3.48\%$ for the branch construction. This reflects the
normalization of the metric rather than a weakness of the surrogate, since $v$
and $p$ are small-magnitude quantities over much of the domain and the
reference norm in the denominator is correspondingly small. Neither model
exceeds $5.59\%$ on any interpolation fold of any field.

\begin{figure}
    \centering
    \includegraphics[width=\linewidth]{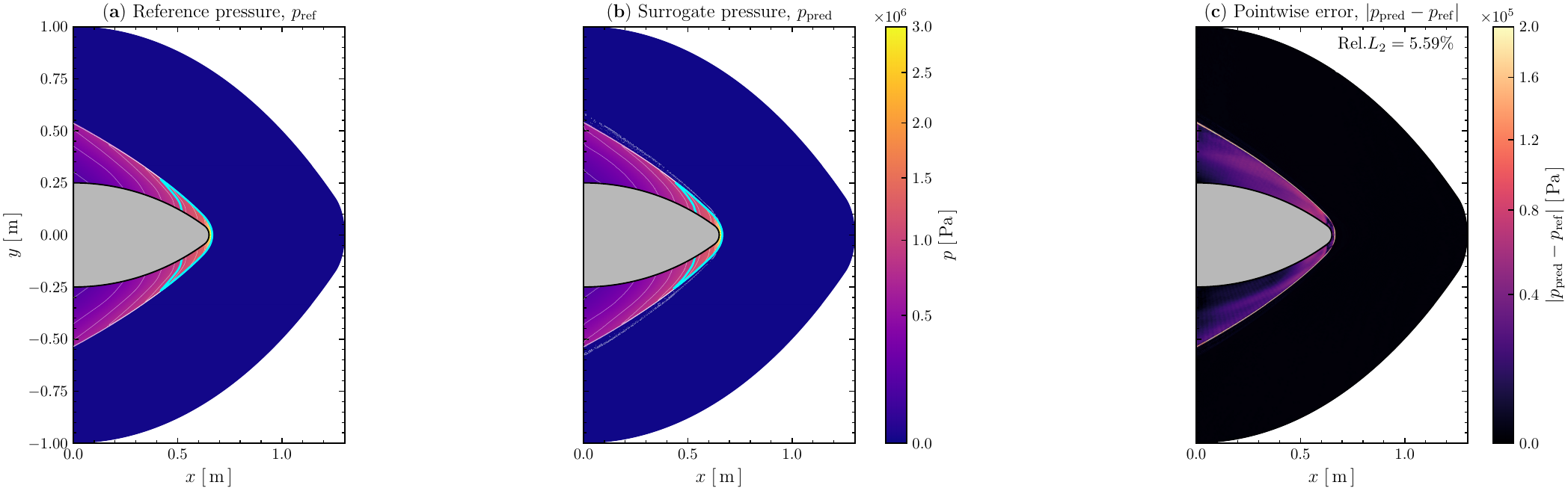}
    \caption{Representative held-out pressure-field prediction for
    $\xi_g=0.5$.}
    \label{fig:run20_pressure_holdout}
\end{figure}
 
On the four extrapolation folds the ordering reverses on every field, and the
size of the reversal, rather than its direction, is the substantive result.
The branch construction degrades gracefully: its mean error rises by a factor
of between $1.3$ and $2.2$ relative to the interpolation regime, and its
largest single error over all $37$ folds and all five fields is $11.55\%$, on
$v$ at $\mathrm{run.1}$. The PCHIP baseline degrades by a factor of between
$2.3$ and $38.3$, and its largest single error is $375.23\%$, on $p$ at the
same fold --- an error almost four times the magnitude of the reference field,
that is, a prediction carrying no usable information.
 
The degradation is not uniform across fields, and the distinction matters. On
the two velocity components the two models remain comparable outside the
sampled range: PCHIP is worse by factors of only $1.02$ on $u$ and $1.17$ on
$v$. The divergence is confined to the thermodynamic variables, where PCHIP is
worse by a factor of $4.0$ on $T$, $3.3$ on $\rho$ and $17.9$ on $p$, with
worst-fold errors of $25.49\%$, $65.60\%$ and $375.23\%$ respectively. The
cause is structural. Outside its node range the PCHIP interpolant reverts to a
first-order Hermite extrapolation, which continues the end-interval slope
indefinitely and places no bound on the coefficient vector; the fields whose
coefficient trajectories have the largest end-interval slope are therefore the
ones that diverge. The branch network instead maps the shape factor through a
smooth global function fitted across the whole family, so its predictions
remain within the range of coefficient behaviour it has learned.
 
 
The two-step construction should therefore be understood as buying bounded
worst-case behaviour rather than uniformly better accuracy. For geometries
inside the sampled range the two coefficient models are equivalent for
engineering purposes, and the baseline is marginally preferable. What the
branch network adds is a guarantee at the edge of the design space: across
every field and every fold its error never exceeds $11.55\%$, whereas the
worst-case error of the baseline ranges from $1.46\%$ on $u$ to $375.23\%$ on
$p$, a factor of $32$ above the branch worst case. This matters for shape
optimization, where candidate geometries are routinely proposed outside the
convex hull of the training set and a single divergent surrogate evaluation
can misdirect an entire search. A bounded worst case is worth more in that
setting than a $10$--$15\%$ improvement in typical interior accuracy. We note
that the extrapolation statistics rest on four folds per field, and report
them as observed behaviour of the two constructions rather than as a
statistically resolved difference.
 
 
%
%
%
 
\begin{table}[htbp]
  \centering
  \caption{Pearson correlation between the predicted and reference field
  gradients, over the $37$ leave-one-out folds of the two geometry families.
  Each entry is the mean over the folds of that regime, with the minimum over
  those folds in parentheses; the minimum is the informative statistic here,
  since a correlation that collapses on a single geometry is what an
  optimisation would encounter. The temperature and the two velocity
  components are reproduced at a correlation above $0.98$ everywhere and are
  insensitive to the regime; the two thermodynamic variables carry the whole
  of the difference between the coefficient models.}
  \label{tab:loo_gradient_correlation}
  \begin{tabular}{llccccc}
    \toprule
    Regime & Model & $\rho$ & $p$ & $T$ & $u$ & $v$ \\
    \midrule
    \multirow{2}{*}{Interpolation}
      & Branch & 0.9781 & 0.9805 & 0.9999 & 0.9999 & 0.9996 \\
      &        & (0.9396) & (0.9434) & (0.9997) & (0.9996) & (0.9990) \\
      & PCHIP  & 0.9837 & 0.9856 & 0.9999 & 0.9999 & 0.9997 \\
      &        & (0.9673) & (0.9709) & (0.9998) & (0.9999) & (0.9993) \\
    \midrule
    \multirow{2}{*}{Extrapolation}
      & Branch & 0.8961 & 0.9209 & 0.9997 & 0.9997 & 0.9987 \\
      &        & (0.8414) & (0.8822) & (0.9995) & (0.9995) & (0.9981) \\
      & PCHIP  & 0.7944 & 0.6023 & 0.9933 & 0.9997 & 0.9981 \\
      &        & (0.4523) & (0.1730) & (0.9812) & (0.9994) & (0.9958) \\
    \bottomrule
  \end{tabular}
\end{table}
 

 
The relative $L_2$ error measures how well the field is reproduced; it says
nothing about whether the shock is in the right place. The gradient
correlation does, and it separates the five variables into two groups
(\Cref{tab:loo_gradient_correlation}). The temperature and both velocity
components are reproduced at a correlation of at least $0.98$ by either
coefficient model, at every fold and in both regimes, so their gradient
structure is fixed by the shared trunk and is essentially unaffected by the
choice of coefficient map. The density and the pressure behave differently,
and they carry the whole of the difference between the two constructions.
 
Within the sampled range both models place the gradients accurately. The
correlation for $\rho$ and $p$ averages $0.978$ and $0.981$ for the branch
construction and $0.984$ and $0.986$ for PCHIP, and no interpolation fold of
any field or either model falls below $0.9396$. As with the field errors, the
baseline is marginally the better of the two here.
 
Outside the sampled range the two separate sharply. The branch construction
retains mean correlations of $0.896$ on $\rho$ and $0.921$ on $p$, with a
worst fold of $0.841$: the gradient field is degraded but still strongly
related to the reference, which is the quantitative statement behind the
visual observation that the predicted shock remains in the right place.
PCHIP falls to mean correlations of $0.794$ and $0.602$, with worst folds of
$0.452$ on $\rho$ and $0.173$ on $p$. A correlation of $0.17$ is not a
degraded shock but an unrelated one: at that fold the predicted gradient
field carries almost no information about where the reference discontinuity
lies. Over every field and every fold the branch construction never drops
below $0.841$, while the baseline reaches $0.173$.
 
\begin{table}
  \centering
  \caption{Relative $L_2$ error of the predicted flow fields over the $37$ leave-one-out folds of both geometry families. The folds are separated into the $33$ that interpolate within the sampled range of the shape factor and the $4$ that extrapolate beyond it, the latter being the extreme shape factors of each geometry family. The median is reported alongside the mean because the fold-wise distribution is right-skewed by the extrapolation folds, so that the mean exceeds the error of a typical fold; the maximum is reported because the worst case, rather than the average, bounds the reliability of the surrogate when it is evaluated inside an optimisation loop. All values are percentages.}
  \label{tab:loo_refined_relative_l2}
  \begin{tabular}{llrrrrrr}
    \toprule
    & & \multicolumn{3}{c}{Branch} & \multicolumn{3}{c}{PCHIP} \\
    \cmidrule(lr){3-5}\cmidrule(lr){6-8}
    Regime & Field & Mean & Median & Max. & Mean & Median & Max. \\
    \midrule
    Interpolation & $T$ & 1.947 & 1.829 & 3.388 & 1.761 & 1.772 & 2.526 \\
     & $u$ & 0.563 & 0.528 & 1.028 & 0.489 & 0.486 & 0.741 \\
     & $v$ & 3.482 & 3.170 & 5.589 & 3.045 & 2.970 & 4.759 \\
     & $p$ & 3.272 & 3.138 & 5.099 & 2.775 & 2.769 & 4.133 \\
     & $\rho$ & 2.841 & 2.820 & 4.519 & 2.421 & 2.238 & 3.777 \\
    \midrule
    Extrapolation & $T$ & 2.543 & 2.308 & 3.401 & 10.220 & 6.680 & 25.487 \\
     & $u$ & 1.081 & 1.011 & 1.326 & 1.107 & 1.051 & 1.456 \\
     & $v$ & 7.134 & 6.109 & 11.548 & 8.333 & 6.014 & 15.745 \\
     & $p$ & 5.955 & 5.141 & 8.974 & 106.306 & 23.461 & 375.234 \\
     & $\rho$ & 6.269 & 5.268 & 9.445 & 20.359 & 6.529 & 65.600 \\
    \midrule
    All folds & $T$ & 2.012 & 1.871 & 3.401 & 2.676 & 1.799 & 25.487 \\
     & $u$ & 0.619 & 0.547 & 1.326 & 0.556 & 0.491 & 1.456 \\
     & $v$ & 3.877 & 3.330 & 11.548 & 3.616 & 2.988 & 15.745 \\
     & $p$ & 3.562 & 3.321 & 8.974 & 13.968 & 2.809 & 375.234 \\
     & $\rho$ & 3.211 & 3.031 & 9.445 & 4.360 & 2.265 & 65.600 \\
    \bottomrule
  \end{tabular}
\end{table}

\begin{table}[htbp]
  \centering
  \caption{Relative $L_2$ error of the predicted flow fields over the $37$ leave-one-out folds of both geometry families. The folds are separated into the $33$ that interpolate within the sampled range of the shape factor and the $4$ that extrapolate beyond it, the latter being the extreme shape factors of each geometry family. The arithmetic mean over the folds in each regime is reported. All values are percentages.}
  \label{tab:loo_species_relative_l2}
  \begin{tabular}{llrr}
    \toprule
    Regime & Field & Branch & PCHIP \\
    \midrule
    Interpolation & $Y_{\mathrm{N}}$ & 3.360 & 3.175 \\
     & $Y_{\mathrm{N_2}}$ & 0.019 & 0.017 \\
     & $Y_{\mathrm{NO}}$ & 10.641 & 10.819 \\
     & $Y_{\mathrm{O}}$ & 5.400 & 5.418 \\
     & $Y_{\mathrm{O_2}}$ & 0.106 & 0.093 \\
    \midrule
    Extrapolation & $Y_{\mathrm{N}}$ & 5.690 & 106.731 \\
     & $Y_{\mathrm{N_2}}$ & 0.036 & 0.039 \\
     & $Y_{\mathrm{NO}}$ & 10.898 & 10.901 \\
     & $Y_{\mathrm{O}}$ & 6.366 & 24.440 \\
     & $Y_{\mathrm{O_2}}$ & 0.219 & 0.203 \\
    \midrule
    All folds & $Y_{\mathrm{N}}$ & 3.612 & 14.370 \\
     & $Y_{\mathrm{N_2}}$ & 0.020 & 0.019 \\
     & $Y_{\mathrm{NO}}$ & 10.669 & 10.828 \\
     & $Y_{\mathrm{O}}$ & 5.505 & 7.474 \\
     & $Y_{\mathrm{O_2}}$ & 0.118 & 0.105 \\
    \bottomrule
  \end{tabular}
\end{table}

\Cref{fig:run7_density_interpolation,fig:run7_pressure_interpolation,%
fig:run20_density_holdout,fig:run20_pressure_holdout} show reconstructed
density and pressure fields for two held-out geometries of the integer family,
both predicted by the branch construction and chosen to illustrate the two
regimes. The first, $\xi_g=0.83156$, lies well inside the sampled
range and is an interpolation fold; its relative $L_2$ errors are $3.21\%$ on
density and $2.84\%$ on pressure. The second, $\xi_g = 0.500$, is the smallest
shape factor of that family and is therefore extrapolated; its errors are
$5.10\%$ and $5.59\%$, the latter at a coefficient of determination of
$0.995$. Neither case is a favourable selection: on the interpolation fold the
density error sits above the mean over all $33$ such folds while the pressure
error sits below it, and on the extrapolation fold the ordering is reversed.
 
The numerical schlieren comparison shows that the reported accuracy is not
obtained by smoothing the discontinuity. In both cases the predicted bow shock
stands off at the correct distance and retains a sharp density jump, and the
gradient correlation confirms it: $0.978$ on $\rho$ and $0.981$ on $p$ over
the interpolation folds, falling only to $0.896$ and $0.921$ over the
extrapolation folds, with a worst fold of $0.841$
(\Cref{tab:loo_gradient_correlation}). The relative $L_2$ error of the density
gradient is nonetheless large --- $18.74\%$ at the interpolation fold and
$38.01\%$ at the extrapolated one --- and the two statements are consistent
rather than contradictory. A front displaced by a fraction of a cell leaves
the correlation almost unchanged, since both gradient fields still peak in
nearly the same place, while producing a large relative error, since the
difference of two narrow offset spikes is itself of the order of the spikes.
A high correlation alongside a large gradient error is therefore the signature
of a correctly placed but slightly displaced shock, not of a smeared one.
 
That the extrapolated case remains a recognisable hypersonic flow --- shock at
the right standoff, stagnation region resolved, gradient correlation still
near $0.9$ --- is the qualitative counterpart of the bounded worst case
quantified in \Cref{tab:loo_refined_relative_l2}. The contrast with the
baseline at the same fold is the point: there the pressure error reaches
$375.23\%$ and the gradient correlation falls to $0.173$, which is not a
degraded shock but an unrelated one.
 
\begin{figure}[htbp]
    \centering
 
    \begin{subfigure}[b]{0.48\textwidth}
        \centering
        \includegraphics[width=\linewidth]{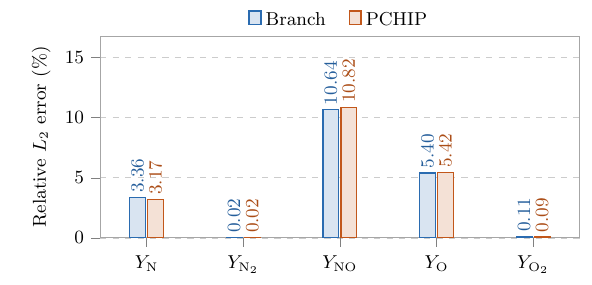}
        \caption{Interpolation folds, mean over the $33$ folds.}
        \label{fig:loo_species_bar_interp}
    \end{subfigure}
    \hfill
    \begin{subfigure}[b]{0.48\textwidth}
        \centering
        \includegraphics[width=\linewidth]{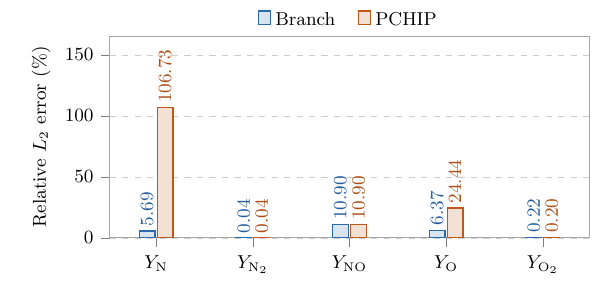}
        \caption{Extrapolation folds, mean over the four folds.}
        \label{fig:loo_species_bar_extrap}
    \end{subfigure}
 
    \vspace{0.9em}
 
    \begin{subfigure}[b]{\textwidth}
        \centering
        \includegraphics[width=\linewidth]{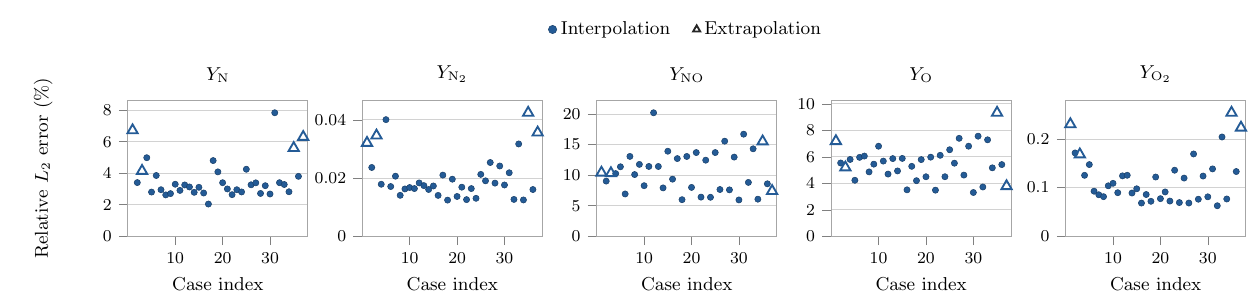}
        \caption{Fold-resolved error of the surrogate.}
        \label{fig:loo_species_case_branch}
    \end{subfigure}
 
    \vspace{0.6em}
 
    \begin{subfigure}[b]{\textwidth}
        \centering
        \includegraphics[width=\linewidth]{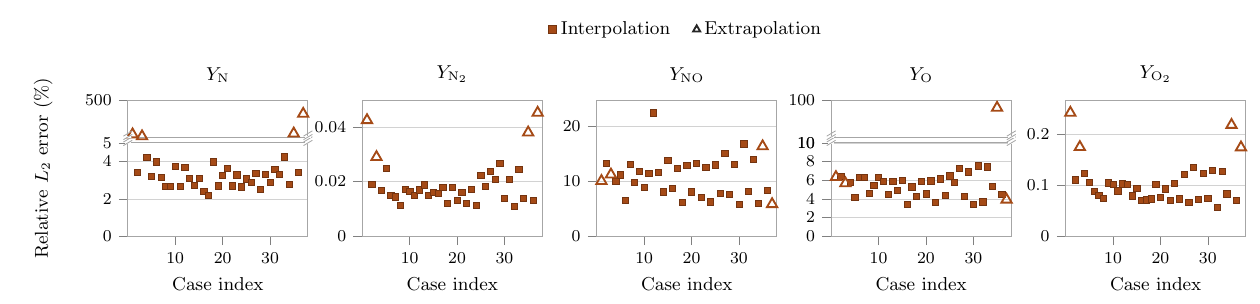}
        \caption{Fold-resolved error of the PCHIP baseline.}
        \label{fig:loo_species_case_pchip}
    \end{subfigure}
 
    \caption[Leave-one-out accuracy for the species mass fractions]%
    {Leave-one-out accuracy for the five species mass fractions over the $37$
    folds of the two geometry families: $33$ interpolate within the sampled
    range of the shape factor, and four --- the extremes of each family ---
    extrapolate beyond it.
    \subref{fig:loo_species_bar_interp} and
    \subref{fig:loo_species_bar_extrap}~Mean relative $L_2$ error in each
    regime, printed above each bar. The five species span three orders of
    magnitude in accuracy: within the sampled range the branch construction
    attains $0.019\%$ on $Y_{\mathrm{N_2}}$ and $0.106\%$ on
    $Y_{\mathrm{O_2}}$, but $10.641\%$ on $Y_{\mathrm{NO}}$, a factor of $560$
    between the best and worst species. Inside the sampled range the two
    coefficient models are close and neither is uniformly better: the baseline
    is the more accurate on $Y_{\mathrm{N}}$, $Y_{\mathrm{N_2}}$ and
    $Y_{\mathrm{O_2}}$, by $5$ to $12\%$ in relative terms, and the branch
    construction on $Y_{\mathrm{NO}}$ and $Y_{\mathrm{O}}$, by $1.6\%$ and
    $0.3\%$. Outside it the branch construction is the more accurate on four
    of the five, decisively on $Y_{\mathrm{N}}$, where the mean error is
    $5.69\%$ against $106.73\%$ for the baseline, and on $Y_{\mathrm{O}}$,
    where it is $6.37\%$ against $24.44\%$; the baseline remains marginally
    ahead on $Y_{\mathrm{O_2}}$. Spread is omitted because the fold-wise
    distribution is right-skewed.
    \subref{fig:loo_species_case_branch} and
    \subref{fig:loo_species_case_pchip}~The same errors by fold, ordered by
    shape factor with the families pooled so that the extrapolation folds
    (open triangles) lie at the ends of the axis. Crossing from the
    interpolation to the extrapolation regime costs the branch construction a
    factor of between $1.02$ and $2.07$, and the baseline a factor of between
    $1.01$ and $33.6$. \emph{Each panel is scaled to its own data, and
    panels whose outlying folds dwarf the rest carry a broken vertical axis,
    marked by slashes on the facing spines; the metric is not continuous
    across such a break}.}
    \label{fig:loo_species}
\end{figure}

\begin{figure}[htbp]
    \centering
 
    \begin{subfigure}[b]{0.48\textwidth}
        \centering
        \includegraphics[width=\linewidth]{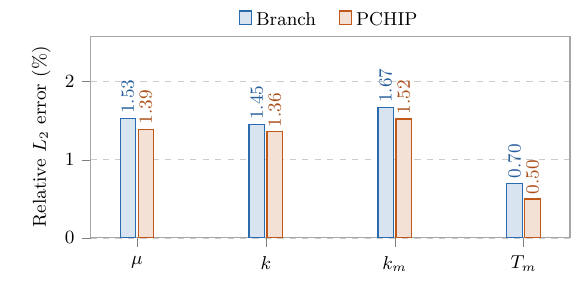}
        \caption{Interpolation folds, mean over the $33$ folds.}
        \label{fig:loo_transport_bar_interp}
    \end{subfigure}
    \hfill
    \begin{subfigure}[b]{0.48\textwidth}
        \centering
        \includegraphics[width=\linewidth]{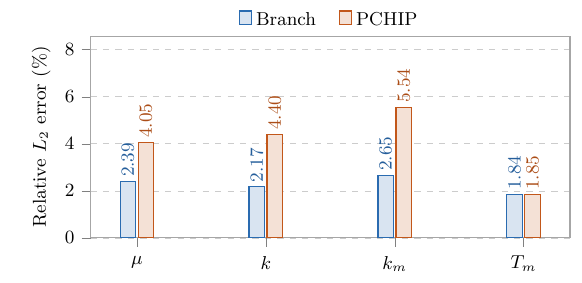}
        \caption{Extrapolation folds, mean over the four folds.}
        \label{fig:loo_transport_bar_extrap}
    \end{subfigure}
 
    \vspace{0.9em}
 
    \begin{subfigure}[b]{\textwidth}
        \centering
        \includegraphics[width=\linewidth]{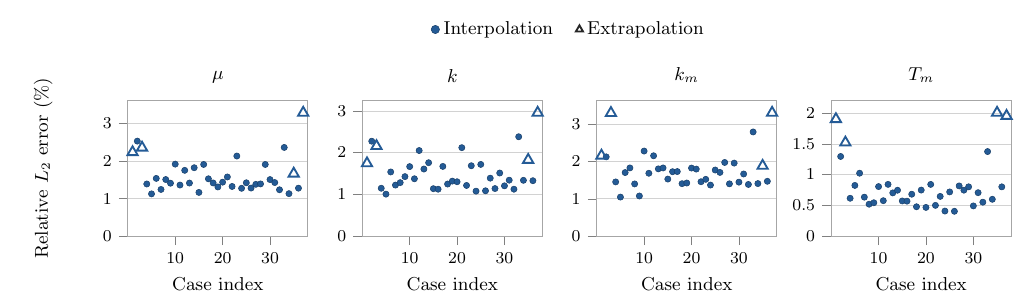}
        \caption{Fold-resolved error of the surrogate.}
        \label{fig:loo_transport_case_branch}
    \end{subfigure}
 
    \vspace{0.6em}
 
    \begin{subfigure}[b]{\textwidth}
        \centering
        \includegraphics[width=\linewidth]{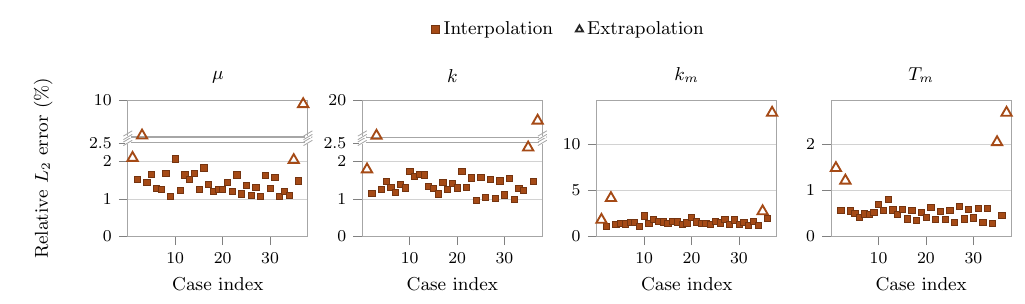}
        \caption{Fold-resolved error of the PCHIP baseline.}
        \label{fig:loo_transport_case_pchip}
    \end{subfigure}
 
    \caption[Leave-one-out accuracy for the transport properties]%
    {Leave-one-out accuracy for the mixture viscosity $\mu$, the
    translational and internal thermal conductivities $k$ and $k_m$, and the
    vibrational--electronic temperature $T_m$. Panels
    (\subref{fig:loo_transport_bar_interp}) and
    (\subref{fig:loo_transport_bar_extrap}) give the mean relative $L_2$ error
    over the $33$ interpolation and the $4$ extrapolation folds; panels
    (\subref{fig:loo_transport_case_branch}) and
    (\subref{fig:loo_transport_case_pchip}) resolve the same errors fold by
    fold, ordered by shape factor, with the extrapolation folds at the ends of
    each family. Within the sampled range the two constructions are within
    $9\%$ of each other on the mixture properties and the baseline is the more
    accurate throughout; outside it the ordering reverses and the baseline
    reaches $4.05\%$, $4.40\%$ and $5.54\%$ against $2.39\%$, $2.17\%$ and
    $2.65\%$. $T_m$ is the exception: the baseline is the more accurate inside
    the range by $29\%$ and the two are indistinguishable outside it
    ($1.845\%$ against $1.847\%$). Note the independent vertical scale on each
    quantity, and the broken vertical axis on the baseline panel, without which
    the extrapolation folds compress every interpolation fold onto the
    axis.}
    \label{fig:loo_transport}
\end{figure}

\begin{table}[htbp]
  \centering
  \caption{Pearson correlation between the predicted and reference gradients
  of the species mass fractions, over the $37$ leave-one-out folds of the two
  geometry families. Each entry is the mean over the folds of that regime,
  with the minimum over those folds in parentheses; the minimum is the
  informative statistic here, since a correlation that collapses on a single
  geometry is what an optimisation would encounter. The two diatomic species
  $Y_{\mathrm{N_2}}$ and $Y_{\mathrm{O_2}}$ stay above $0.92$ for both models
  in both regimes; the difference between the coefficient models is carried
  entirely by $Y_{\mathrm{N}}$, $Y_{\mathrm{NO}}$ and $Y_{\mathrm{O}}$. For
  the branch construction on $Y_{\mathrm{N}}$ the worst fold is an
  interpolation fold rather than an extrapolation one, the only such case in
  the table.}
  \label{tab:loo_species_gradient_correlation}
  \begin{tabular}{llccccc}
    \toprule
    Regime & Model & $Y_{\mathrm{N}}$ & $Y_{\mathrm{N_2}}$
           & $Y_{\mathrm{NO}}$ & $Y_{\mathrm{O}}$ & $Y_{\mathrm{O_2}}$ \\
    \midrule
    \multirow{2}{*}{Interpolation}
      & Branch & 0.9694 & 0.9898 & 0.9599 & 0.9674 & 0.9880 \\
      &        & (0.8588) & (0.9709) & (0.9305) & (0.9127) & (0.9620) \\
      & PCHIP  & 0.9701 & 0.9903 & 0.9608 & 0.9682 & 0.9894 \\
      &        & (0.9228) & (0.9738) & (0.9221) & (0.9338) & (0.9712) \\
    \midrule
    \multirow{2}{*}{Extrapolation}
      & Branch & 0.9240 & 0.9723 & 0.9367 & 0.9276 & 0.9638 \\
      &        & (0.8797) & (0.9504) & (0.9074) & (0.8701) & (0.9361) \\
      & PCHIP  & 0.3823 & 0.9540 & 0.8158 & 0.7904 & 0.9499 \\
      &        & (0.0354) & (0.9255) & (0.5328) & (0.3622) & (0.9228) \\
    \bottomrule
  \end{tabular}
\end{table}

\subsection{Species mass fractions}
\label{sec:loo_species}
 
 
The five-species air model carries the mass fractions $Y_{\mathrm{N_2}}$,
$Y_{\mathrm{O_2}}$, $Y_{\mathrm{NO}}$, $Y_{\mathrm{N}}$ and $Y_{\mathrm{O}}$,
assessed on the same $37$ folds and under the same interpolation--extrapolation
split as the primary flow variables
(\Cref{tab:loo_species_relative_l2}). The first feature of the results is not a difference between the coefficient
models but a difference between the species. Within the sampled range the
branch construction attains a mean relative $L_2$ error of $0.019\%$ on
$Y_{\mathrm{N_2}}$ and $0.106\%$ on $Y_{\mathrm{O_2}}$, but $3.36\%$ on
$Y_{\mathrm{N}}$, $5.40\%$ on $Y_{\mathrm{O}}$ and $10.64\%$ on
$Y_{\mathrm{NO}}$ --- a factor of $560$ between the best and the worst species
for the same model in the same regime. The ordering is the inverse of the
ordering by abundance: the two diatomic species that make up the bulk of the
mixture are reproduced most accurately, the dissociation products less so, and
nitric oxide, present in trace quantities over most of the domain, least of
all. This is a property of the metric rather than of the surrogate. The
relative $L_2$ error normalises by the norm of the reference field, so a
species whose mass fraction is small over most of the domain is divided by a
small number, and the same absolute discrepancy is reported as a much larger
relative one. 
 
Two further observations support that reading. The gradient correlations
(\Cref{tab:loo_species_gradient_correlation}) order the species the same way:
$Y_{\mathrm{N_2}}$ and $Y_{\mathrm{O_2}}$ stay above $0.92$ for both models in
both regimes, while $Y_{\mathrm{N}}$, $Y_{\mathrm{NO}}$ and $Y_{\mathrm{O}}$
carry all of the variation. And $Y_{\mathrm{NO}}$ is, uniquely in this study,
insensitive to the regime: its mean error moves from $10.64\%$ to $10.90\%$
for the branch construction and from $10.82\%$ to $10.90\%$ for the baseline,
factors of $1.02$ and $1.01$, where every other quantity in the paper degrades
measurably at the edges of the range. An error that does not respond to the
extrapolation that governs every other quantity is not being set by the
coefficient model, and is best read as the floor imposed by the normalisation.
 
 
On the $33$ interpolation folds the two coefficient models are
indistinguishable, and unlike the primary flow variables neither is uniformly
better. The baseline is the more accurate on $Y_{\mathrm{N}}$,
$Y_{\mathrm{N_2}}$ and $Y_{\mathrm{O_2}}$, by $5$ to $12\%$ in relative terms,
and the branch construction on $Y_{\mathrm{NO}}$ and $Y_{\mathrm{O}}$, by
$1.6\%$ and $0.3\%$. A $0.3\%$ relative difference between two errors of
$5.4\%$ is not a difference worth attributing, and the gradient correlations
say the same: the two models differ by between $0.0005$ and $0.0014$ on every
species. Within the range of geometries actually sampled, the choice of
coefficient map does not matter for the species fields.
 
 
Outside the sampled range the separation is the same as for the primary
variables, and on one species it is larger than anywhere else in the study.
The branch construction degrades by a factor of between $1.02$ and $2.07$
relative to the interpolation regime, with a largest mean error of $10.90\%$,
on $Y_{\mathrm{NO}}$, which is the same floor it carries inside the range. The
baseline degrades by a factor of between $1.01$ and $33.6$. Its mean error on
$Y_{\mathrm{N}}$ reaches $106.73\%$, against $5.69\%$ for the branch
construction --- a factor of $18.8$ --- and on $Y_{\mathrm{O}}$ it reaches
$24.44\%$ against $6.37\%$, a factor of $3.8$. The gradient correlations make
the character of that failure explicit: on $Y_{\mathrm{N}}$ the baseline falls
to a mean correlation of $0.382$ and a worst fold of $0.035$, which is not a
degraded gradient field but one unrelated to the reference, while the branch
construction holds $0.924$ and never drops below $0.859$ on any species at any
fold. The cause is the one identified for the primary variables: outside its
node range the PCHIP interpolant continues the end-interval slope
indefinitely, placing no bound on the coefficient vector, and the species
whose coefficient trajectories have the steepest end-interval slope are the
ones that diverge.
 
The reversal is not universal, and the exception is worth recording. On
$Y_{\mathrm{O_2}}$ the baseline remains marginally the more accurate outside
the range, at $0.203\%$ against $0.219\%$, and on $Y_{\mathrm{N_2}}$ and
$Y_{\mathrm{NO}}$ the two are within a few per cent of each other. The
advantage of the branch construction is confined to the two species whose
errors are large enough to matter, $Y_{\mathrm{N}}$ and $Y_{\mathrm{O}}$.
 
 
The conclusion matches the one drawn for the primary flow variables. Inside
the sampled range the two coefficient models are equivalent for the species
fields, with no consistent ordering between them. Outside it the branch
construction is bounded where the baseline is not, and it is bounded on
precisely the species that carry the chemistry: atomic nitrogen and atomic
oxygen, whose distributions set the dissociation state and therefore the
wall heat flux. A baseline that returns a prediction with a $106\%$ error and
a gradient correlation of $0.035$ on $Y_{\mathrm{N}}$ has not produced a
degraded estimate of the composition; it has produced no estimate at all, and
an optimisation evaluating a candidate geometry there would be misled rather
than merely imprecise. As with the primary variables, the extrapolation
statistics rest on four folds per species and are reported as observed
behaviour rather than as a statistically resolved difference.
\begin{table}[htbp]
  \centering
  \caption{Relative $L_2$ error of the predicted transport properties over the
  $37$ leave-one-out folds of both geometry families. The folds are separated
  into the $33$ that interpolate within the sampled range of the shape factor
  and the $4$ that extrapolate beyond it, the latter being the extreme shape
  factors of each geometry family; the arithmetic mean over the folds of each
  regime is reported. The mixture properties $\mu$, $k$ and $k_m$ span a factor
  of only $1.15$ within the sampled range, as all three follow from the same
  mixture rules, while the vibrational--electronic temperature $T_m$ is
  reproduced roughly twice as accurately as any of them. The two constructions
  separate only outside the sampled range, and then only on the mixture
  properties: there the baseline is worse by factors of $1.70$, $2.03$ and
  $2.09$, whereas on $T_m$ the two agree to $0.002$ percentage points. The
  pooled means in the final block are dominated by the $33$ interpolation folds
  and should not be read as an overall ranking. All values are percentages.}
  \label{tab:loo_transport_relative_l2}
  \begin{tabular}{llrr}
    \toprule
    Regime & Field & Branch & PCHIP \\
    \midrule
    Interpolation & $\mu$ & 1.529 & 1.392 \\
     & $k$ & 1.449 & 1.358 \\
     & $k_m$ & 1.667 & 1.521 \\
     & $T_m$ & 0.699 & 0.497 \\
    \midrule
    Extrapolation & $\mu$ & 2.387 & 4.047 \\
     & $k$ & 2.170 & 4.395 \\
     & $k_m$ & 2.654 & 5.536 \\
     & $T_m$ & 1.845 & 1.847 \\
    \midrule
    All folds & $\mu$ & 1.622 & 1.679 \\
     & $k$ & 1.527 & 1.686 \\
     & $k_m$ & 1.774 & 1.955 \\
     & $T_m$ & 0.823 & 0.643 \\
    \bottomrule
  \end{tabular}
\end{table}

\begin{table}[htbp]
  \centering
  \caption{Pearson correlation between the predicted and reference gradients
  of the transport properties, over the $37$ leave-one-out folds of the two
  geometry families. Each entry is the mean over the folds of that regime,
  with the minimum over those folds in parentheses; the minimum is the
  informative statistic here, since a correlation that collapses on a single
  geometry is what an optimisation would encounter. The first three
  quantities behave almost identically, as they are computed from the same
  mixture rules: their means differ by at most $0.006$ within the sampled
  range. The vibrational--electronic temperature $T_m$ is the best reproduced
  of the four in both regimes and for both constructions, and is the only
  quantity for which the baseline retains a mean correlation above $0.93$
  outside the sampled range. The worst fold of every series is an
  extrapolation fold. $n = 148$.}
  \label{tab:loo_transport_gradient_correlation}
  \begin{tabular}{llcccc}
    \toprule
    Regime & Model & $\mu$ & $k$ & $k_m$ & $T_m$ \\
    \midrule
    \multirow{4}{*}{Interpolation}
      & Branch & 0.9636 & 0.9654 & 0.9602 & 0.9874 \\
      &        & (0.9037) & (0.9045) & (0.8950) & (0.9642) \\
      & PCHIP  & 0.9697 & 0.9710 & 0.9679 & 0.9907 \\
      &        & (0.9394) & (0.9485) & (0.9312) & (0.9781) \\
    \midrule
    \multirow{4}{*}{Extrapolation}
      & Branch & 0.8902 & 0.9053 & 0.8874 & 0.9363 \\
      &        & (0.8694) & (0.8756) & (0.8758) & (0.8930) \\
      & PCHIP  & 0.8337 & 0.8188 & 0.7961 & 0.9302 \\
      &        & (0.6466) & (0.6272) & (0.5591) & (0.8689) \\
    \bottomrule
  \end{tabular}
\end{table}
 
\subsection{Transport properties}
\label{sec:loo_transport}
 
The mixture viscosity $\mu$, the translational and internal thermal
conductivities $k$ and $k_m$, and the vibrational--electronic temperature
$T_m$ are secondary quantities: none is itself a target of the aerodynamic
objective, but $\mu$ and $k$ enter the wall shear stress and the wall
heat-flux evaluation, and $T_m$ sets the internal energy state on which the
other three depend. They are assessed on the same $37$ folds and under the
same interpolation--extrapolation split as the primary flow variables
(\Cref{tab:loo_transport_relative_l2,tab:loo_transport_gradient_correlation}
and \Cref{fig:loo_transport}).
 
The three mixture properties behave almost as one, and it is worth saying so
before any comparison between the coefficient models. Over the interpolation
folds the branch construction attains mean relative $L_2$ errors of $1.53\%$,
$1.45\%$ and $1.67\%$ on $\mu$, $k$ and $k_m$ --- a spread of a factor of
$1.15$ between the best and the worst --- and their mean gradient correlations
differ by at most $0.006$. This uniformity is expected, since all three follow
from the same mixture rules applied to the same predicted composition and
temperature, and it stands in contrast to the species mass fractions, whose
interpolation errors span a factor of $560$ (\Cref{sec:loo_species}). The
vibrational--electronic temperature stands apart from the three: at $0.70\%$
it is reproduced more than twice as accurately as any of them, and its mean
gradient correlation of $0.987$ is the highest of the four. It is a
temperature rather than a composition-weighted mixture average, and so does
not inherit the sensitivity of the species fractions from which the other
three are assembled.
 
Those errors are small in absolute terms. Every transport property is
reproduced more accurately over the interpolation folds than the temperature,
the two velocity components taken together, the pressure or the density; only
the axial velocity, at $0.56\%$, is competitive, and $T_m$ is more accurate
than any primary flow variable. The derived quantities are therefore not a
weak link in the surrogate.
 
On the $33$ interpolation folds the PCHIP baseline is the more accurate of the
two constructions on all four quantities. On the mixture properties the margin
is small --- $9.0\%$, $6.3\%$ and $8.8\%$ in relative terms, or $1.39\%$
against $1.53\%$ on $\mu$, $1.36\%$ against $1.45\%$ on $k$, and $1.52\%$
against $1.67\%$ on $k_m$ --- and the gradient correlations agree, the
baseline holding a higher mean on all three by between $0.006$ and $0.008$.
On $T_m$ the margin is larger, $0.50\%$ against $0.70\%$, or $29\%$ in
relative terms. This is the same result as for the primary flow variables and
for the same reason: where bracketing nodes exist, monotone cubic interpolation
of a smooth coefficient trajectory is close to optimal, and the branch network
pays a small price for fitting that trajectory globally rather than locally.
The price is largest on the quantity whose trajectory is smoothest.
 
On the four extrapolation folds the ordering reverses on the three mixture
properties. The branch construction degrades by a factor of between $1.50$ and
$1.59$ relative to the interpolation regime, reaching mean errors of $2.39\%$,
$2.17\%$ and $2.65\%$; the baseline degrades by a factor of between $2.91$ and
$3.64$, reaching $4.05\%$, $4.40\%$ and $5.54\%$. The branch construction is
therefore the more accurate by factors of $1.70$, $2.03$ and $2.09$. The
gradient correlations tell the same story with a wider margin: the branch
construction holds means of $0.890$, $0.905$ and $0.887$ with a worst fold of
$0.869$, while the baseline falls to $0.834$, $0.819$ and $0.796$ with worst
folds of $0.647$, $0.627$ and $0.559$. The advantage widens across the three
--- $0.057$ on $\mu$, $0.087$ on $k$, $0.091$ on $k_m$ --- so the internal
conductivity, which depends most strongly on the predicted internal energy
state, is the most exposed of the three.
 
$T_m$ is the exception, and it is the more interesting result. Outside the
sampled range the two constructions are indistinguishable: $1.845\%$ for the
branch construction against $1.847\%$ for the baseline, a difference of two
thousandths of a percentage point, and mean gradient correlations of $0.936$
against $0.930$. Only the worst fold separates them at all, at $0.893$ against
$0.869$. This is the one quantity in any of the three groups on which the
baseline extrapolates as well as the branch construction. It is not that the
baseline degrades less on $T_m$ in relative terms --- its error grows by a
factor of $3.72$, the largest relative degradation it suffers anywhere in this
group --- but that it starts from $0.50\%$, low enough that even a
near-fourfold increase leaves it below $2\%$. The branch construction likewise
degrades most on $T_m$, by a factor of $2.64$. Both models find the same
quantity the hardest to continue and the easiest to get right, and they arrive
at the same place.
 
Read pooled over all $37$ folds, the final block of
\Cref{tab:loo_transport_relative_l2} would report the branch construction ahead
on $\mu$, $k$ and $k_m$ and the baseline ahead on $T_m$. That ranking is an
artefact of the $33$-to-$4$ split and is not evidence about either regime: the
interpolation folds outnumber the extrapolation folds by more than eight to
one, so the pooled mean is very nearly the interpolation mean, and the
quantity on which the baseline's interpolation advantage is largest is the one
where it survives pooling. The two regimes are reported separately for this
reason, and the pooled block is included only for completeness.
 
What distinguishes the transport properties from the other two groups is that
neither construction diverges. The largest mean error anywhere in this group is
$2.65\%$ for the branch construction and $5.54\%$ for the baseline, against
$7.13\%$ and $106.31\%$ for the primary flow variables and $10.90\%$ and
$106.73\%$ for the species. The worst-fold gradient correlation tells the same
story: $0.869$ and $0.559$ on the mixture properties here, against $0.841$ and
$0.173$ for the primary variables and $0.859$ and $0.035$ for the species. The
baseline remains usable outside the sampled range for the transport properties
in a way it does not for the other two groups.
 
That contrast is worth drawing rather than passing over, because it sharpens
the claim made elsewhere in this section. The failure of the baseline outside
the sampled range is not a general property of monotone cubic extrapolation; it
is a property of the particular coefficient trajectories it is asked to
continue. Where those trajectories have a mild end-interval slope, as they do
here --- and most mildly of all for $T_m$ --- the first-order Hermite
continuation stays bounded and the baseline degrades gracefully. Where they are
steep, as for the pressure and for atomic nitrogen, it does not. The branch
construction meanwhile shows the same worst-fold gradient correlation in all
three groups --- $0.841$, $0.859$, $0.869$ --- which is what a bounded
construction should look like, and is stronger evidence for boundedness than
any single group could provide. 
 
\subsection{Wall-Resolved Quantities of Interest}
\label{subsec:res_wall_qoi}
\begin{figure}
    \centering
    \captionsetup{font=small}
    \captionsetup[subfigure]{font=footnotesize}
 
    \begin{subfigure}[b]{0.48\textwidth}
        \centering
        \includegraphics[width=\linewidth]{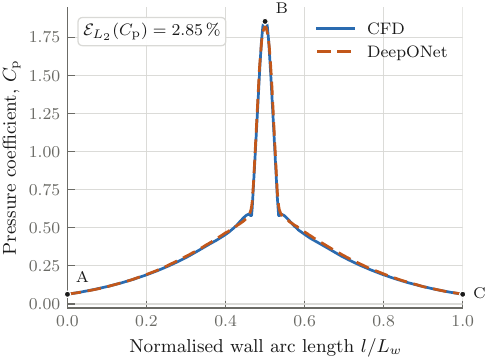}
        \caption{Wall pressure coefficient.}
        \label{fig:wall_qoi_cp}
    \end{subfigure}
    \hfill
    \begin{subfigure}[b]{0.48\textwidth}
        \centering
        \includegraphics[width=\linewidth]{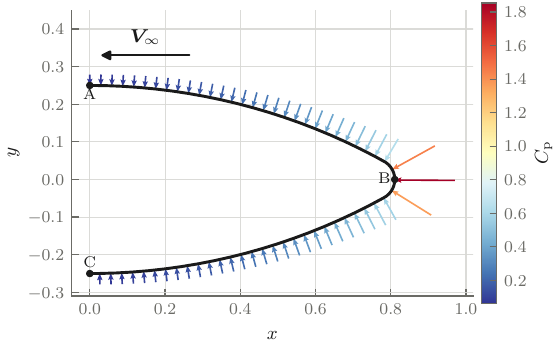}
        \caption{Pressure loading on the body.}
        \label{fig:wall_qoi_cp_surface}
    \end{subfigure}
 
    \vspace{0.9em}
 
    \begin{subfigure}[b]{0.48\textwidth}
        \centering
        \includegraphics[width=\linewidth]{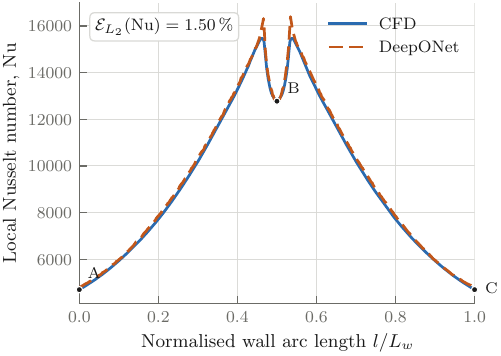}
        \caption{Local Nusselt number.}
        \label{fig:wall_qoi_nusselt}
    \end{subfigure}
    \hfill
    \begin{subfigure}[b]{0.48\textwidth}
        \centering
        \includegraphics[width=\linewidth]{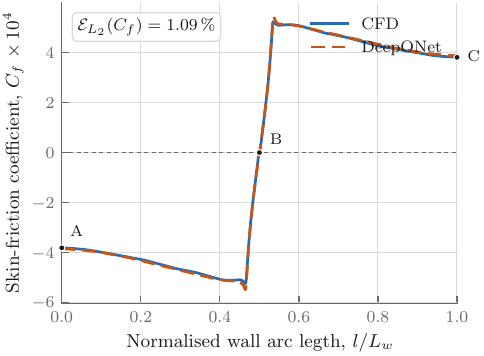}
        \caption{Skin-friction coefficient.}
        \label{fig:wall_qoi_cf}
    \end{subfigure}
 
    \caption[Wall quantities of interest for a representative geometry]%
    {Wall quantities of interest: reference against surrogate. Panels
    (\subref{fig:wall_qoi_cp}), (\subref{fig:wall_qoi_nusselt}) and
    (\subref{fig:wall_qoi_cf}) are plotted against the normalised wall
    coordinate $s$, the stagnation point lying at $s = 0.5$; panel
    (\subref{fig:wall_qoi_cp_surface}) maps $C_{\mathrm{p}}$ onto the body,
    the arrows giving the direction and relative magnitude of the pressure
    loading. Relative $L_2$ errors are quoted on the panels.}
    \label{fig:wall_qoi}
\end{figure}

The field errors of the preceding subsections do not by themselves establish
that the wall quantities are reproduced. None of the quantities reported here
is predicted directly: each is reconstructed from the predicted fields by the
post-processing of \Cref{app:qoi}, which differentiates the temperature and
velocity in the canonical coordinates and converts those gradients through the
inverse metric \eqref{eq:physical_gradient_transform}. Differentiation
amplifies the pointwise error, and it does so most where the field varies most
sharply, so the accuracy of a wall profile is a separate measurement from the
accuracy of the volume field it is derived from.
\Cref{fig:wall_qoi} makes that measurement on a held-out representative geometry.
 
The three profiles are recovered to relative $L_2$ errors of $2.85\%$ in the
pressure coefficient, $1.50\%$ in the local Nusselt number and $1.09\%$ in the
skin-friction coefficient. The ordering is the reverse of what the chain of
derivation would suggest --- $C_{\mathrm{p}}$ requires no gradient at all,
while $\mathrm{Nu}$ and $C_f$ both do --- and the explanation is not the
differentiation but the stagnation point. The pressure profile turns over
within a few per cent of the arc length at $s = 0.5$, and essentially all of
its error is concentrated in that turn; the two gradient-derived quantities
vary more gently there and are reproduced more accurately in spite of the
extra operation. The practical reading is that the surrogate's difficulty on
this body is localised rather than systemic, and that it sits precisely where
the peak heating is assessed.
 
Integrating the surface pressure recovers the drag coefficient to $1.24\%$,
under half the pointwise error in $C_{\mathrm{p}}$ itself. The pointwise
residual oscillates in $s$ and largely cancels when projected onto the
freestream direction, which is the same cancellation that separates the
integrated from the extremal thermal measures in
\Cref{subsec:res_global_qoi}, and it is why a surrogate accurate enough for a
design study need not be accurate enough for a local materials assessment.
 
The skin friction warrants one qualification. Its scale is two orders of
magnitude below that of the pressure coefficient, so its contribution to the
integrated load is negligible: the loading on a blunt hypersonic forebody is
pressure-dominated, and $C_f$ matters here not through the drag but through
the wall heat flux, where the velocity gradient it measures enters alongside
the thermal gradients. Its $1.09\%$ error should therefore be read as
supporting the heat-flux results of \Cref{subsec:res_global_qoi} rather than
the drag.
 
\subsection{Global Quantities of Interest} \label{subsec:res_global_qoi}
\begin{figure}[htbp]
    \centering
    \captionsetup{font=small}
    \captionsetup[subfigure]{font=footnotesize}
 
    \begin{subfigure}[b]{0.48\textwidth}
        \centering
        \includegraphics[width=\linewidth]{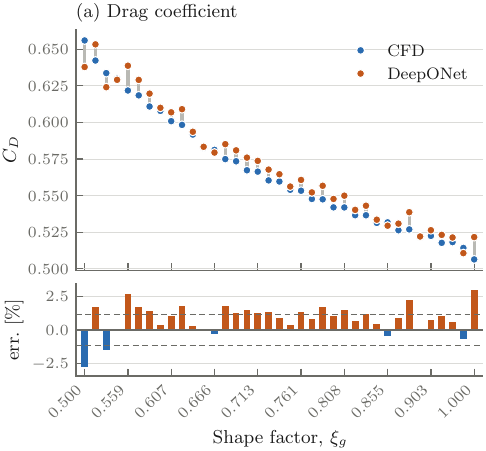}
        \caption{Drag coefficient.}
        \label{fig:qoi_cd}
    \end{subfigure}
    \hfill
    \begin{subfigure}[b]{0.48\textwidth}
        \centering
        \includegraphics[width=\linewidth]{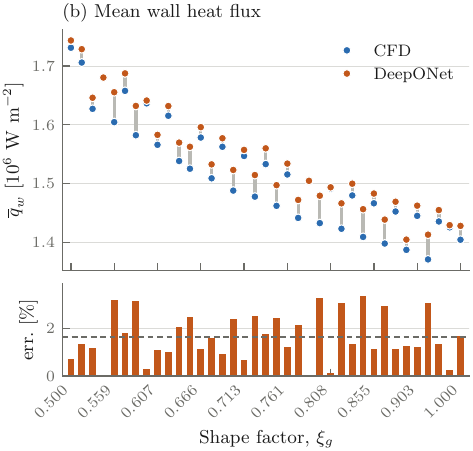}
        \caption{Mean wall heat flux.}
        \label{fig:qoi_qmean}
    \end{subfigure}
 
    \vspace{0.8em}
 
    \begin{subfigure}[b]{0.48\textwidth}
        \centering
        \includegraphics[width=\linewidth]{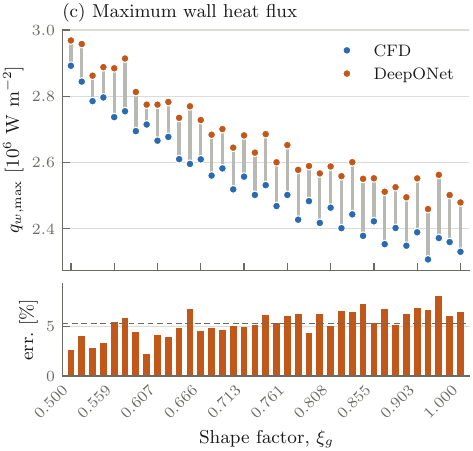}
        \caption{Maximum wall heat flux.}
        \label{fig:qoi_qmax}
    \end{subfigure}
    \hfill
    \begin{subfigure}[b]{0.48\textwidth}
        \centering
        \includegraphics[width=\linewidth]{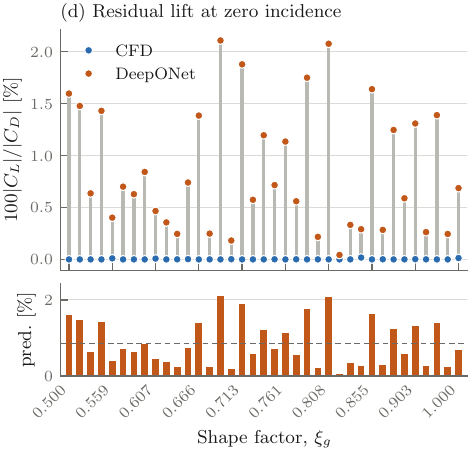}
        \caption{Residual lift at zero incidence.}
        \label{fig:qoi_lift}
    \end{subfigure}
 
    \caption[Global quantities of interest across the geometry family]%
    {Global quantities of interest over the $37$ geometries, ordered by shape
    factor $\xi_g$ and equally spaced. Upper axes: reference and surrogate,
    joined by a stem whose length is the disagreement. Lower axes: the signed
    discrepancy, with the dashed line at the mean magnitude. Panel
    (\subref{fig:qoi_lift}) reports the predicted lift itself, in percentage
    points of drag, the reference being a discretisation residual rather than
    a physical load.}
    \label{fig:global_qoi_four}
\end{figure}

The field errors of the preceding section establish that the surrogate
reproduces the flow; they do not establish that it can stand in for the solver
inside a design loop. What a design loop consumes is not the field but a small
number of scalars reduced from it, and the reduction can either amplify or
cancel the field error depending on how it is formed.
\Cref{fig:global_qoi_four} therefore evaluates the four quantities that enter
the present study, the drag coefficient, the mean and maximum wall heat
flux, and the residual lift at zero angle of attack on all $37$ geometries, each
obtained by passing the predicted and the reference wall distributions through
the same post-processing routine, so that the comparison isolates the
surrogate's error from any difference in reduction.
\Cref{tab:global_qoi_metrics} collects the resulting statistics.
 
 
The drag coefficient is reproduced to a mean relative error of $1.16\%$, with
a median of $1.06\%$ and a worst case of $3.01\%$, the slenderest
geometry of the integer family and an endpoint of the sampled range. The mean
wall heat flux is recovered to $1.62\%$ and the peak wall heat flux to
$5.29\%$, the latter reaching $8.05\%$. The ordering is consistent
with the wall distributions of \Cref{fig:wall_qoi} and has a simple
explanation: the drag and the mean flux are integrals over the wetted surface,
and the oscillatory part of the wall residual largely cancels under
integration, whereas the peak flux is the value at a single point --- the
stagnation point, where the profile turns over within a few per cent of the
arc length --- and inherits the local error in full. A thermal-protection
constraint written on the peak flux therefore carries roughly three times the
uncertainty of one written on the integrated load, and should be posed with
that margin.
 
 
The character of the error matters as much as its size, and here the two
thermal quantities behave differently from the drag. For both the mean and the
peak heat flux the signed mean equals the mean magnitude to four figures ---
$+1.62\%$ and $+5.29\%$ --- which is to say that the surrogate over-predicts
the wall heating on every one of the $37$ geometries without exception. This
is a systematic offset rather than scatter, and three consequences follow.
First, it is largely removable: a single multiplicative correction would
reduce the mean peak-flux error from $5.29\%$ to the residual scatter about
the offset, which is under $1.6$ percentage points at the $95$th percentile.
Second, it cancels in the comparison of two designs, so the ranking of
candidate geometries is sharper than the absolute error suggests --- a point
on which the design study of \Cref{sec:pareto_optimization} depends. Third,
over-prediction is the conservative direction for a thermal-protection margin,
so the bias does not invalidate a feasibility assessment made with the
surrogate. The drag error leans the same way but less completely, its signed
mean of $+0.84\%$ being $73\%$ of the mean magnitude.
 
 
Relative error measures how close the prediction comes; it does not measure
how much of the geometry-to-geometry variation the prediction actually
resolves, and for a design study the second question is the operative one.
Normalising the root-mean-square error by the range the reference data span
gives $5.31\%$, $7.79\%$ and $23.17\%$ for the three quantities, and the
coefficients of determination about the line of equality --- not about a
fitted line, which any monotone rescaling would drive to unity --- are
$0.9610$, $0.9085$ and $0.2597$. The last figure is the one to weigh. The
surrogate accounts for about a quarter of the variation in $q_{w,\max}$ across
the family: the peak flux varies little between these geometries, the error
does not, and what remains after the systematic offset is removed is
comparable to the spread being resolved. The drag and the mean flux are not
subject to this difficulty, and it is on those two that quantitative
conclusions are drawn in what follows; the peak flux is reported, and used as
a constraint, but its ordering across nearby designs should be confirmed
against the solver rather than taken from the surrogate alone.
 
 The residual lift stands apart from the other three. At zero incidence a
symmetric body produces no lift, and the reference values are the symmetry
residual of the discretisation rather than a physical load, so no useful
relative error can be formed from them; panel~(\subref{fig:qoi_lift}) reports
the predicted value itself, which is the asymmetry the surrogate introduces.
It averages $0.862$ percentage points of drag and reaches $2.11$. The
surrogate predicts the upper and lower surfaces independently and nothing in
its construction enforces the symmetry of the geometry, so a symmetry that the
governing equations and the discretisation both preserve to round-off is
recovered only to within the pointwise accuracy of the prediction. The
magnitude is small enough to be of no consequence for the loads themselves,
but it is the quantity on which the feasibility constraint of
\Cref{sec:pareto_optimization} is imposed, and it should be read there as a
trust region on the surrogate rather than as a physical requirement. Imposing
the symmetry by construction --- predicting on a half-body and mirroring ---
would remove the diagnostic and the constraint together.
 
 \begin{table}[htbp]
  \centering
  \caption{Accuracy of the surrogate in the four global quantities of
  interest over the $37$ geometries. The signed mean is reported alongside the
  mean magnitude because the two coincide for both heat fluxes, indicating a
  systematic offset rather than scatter. NRMSE normalises the root-mean-square
  error by the range the reference data span, and $R^2$ is taken about the
  line of equality rather than about a fitted line. The residual lift is a
  symmetry diagnostic whose reference value is a discretisation residual, so
  it is reported in percentage points of drag and admits neither a relative
  error nor a meaningful $R^2$.}
  \label{tab:global_qoi_metrics}
  \begin{tabular}{lrrrrrr}
    \toprule
    Quantity & Signed mean & Mean & Median & Max & NRMSE & $R^2_{1:1}$ \\
    \midrule
    $C_D$ [\%]            & $+0.84$ & 1.16 & 1.06 & 3.01 & 5.31  & 0.9610 \\
    $\overline{q}_w$ [\%] & $+1.62$ & 1.62 & 1.35 & 3.35 & 7.79  & 0.9085 \\
    $q_{w,\max}$ [\%]     & $+5.29$ & 5.29 & 5.35 & 8.05 & 23.17 & 0.2597 \\
    \midrule
    Residual lift [p.p.]  & $+0.86$ & 0.86 & --- & 2.11 & --- & --- \\
    \bottomrule
  \end{tabular}
\end{table}
 
 \subsection{Three-objective Pareto optimization and robustness assessment}
\label{sec:pareto_optimization}
The trained surrogate was used in place of repeated high-fidelity flow
evaluations during design exploration \cite{forrester2008}. Since the
tangent-ogive family is parameterized by the single scalar shape factor
$\xi_g$ of \Cref{eq:ogive_radius}, a dense deterministic sweep of $1001$
candidate geometries was preferred to a stochastic population-based
optimizer. In one dimension this strategy allows the complete design
trajectory to be evaluated directly, and it avoids introducing additional
variability from the optimization algorithm itself. The three objectives were
chosen to represent complementary aerodynamic and thermal requirements:
\begin{equation}
    \min_{\xi_g}
    \left[
        C_D(\xi_g),\;
        q_{w,\max}(\xi_g),\;
        \dot Q'(\xi_g)
    \right],
    \qquad
    \dot Q'(\xi_g)
    =
    \int_{\Gamma_w(\xi_g)} q_w(\xi_g,\ell)\,\mathrm{d}\ell ,
    \label{eq:three_objective_problem}
\end{equation}
where $C_D$ is the drag coefficient, $q_{w,\max}$ the maximum wall heat flux,
and $\dot Q'$ the wall-integrated heat load per unit span. The latter two are
intentionally retained as separate objectives, because a geometry that reduces
a local thermal peak does not necessarily minimize the total thermal load
integrated over the complete surface. The optimization is subject to the
zero-incidence symmetry criterion
\begin{equation}
    \frac{|C_L(\xi_g)|}{|C_D(\xi_g)|}\le 0.01 ,
    \label{eq:lift_constraint}
\end{equation}
which is used primarily as a consistency check for the nominally symmetric
zero-angle-of-attack configuration.
 
Standard Pareto dominance was used to identify non-dominated feasible designs
\cite{miettinen1999,deb2001}. A feasible design $\xi_{g,a}$ is dominated by
another feasible design $\xi_{g,b}$ if
\begin{equation}
    f_i(\xi_{g,b})\le f_i(\xi_{g,a}),\qquad i=1,2,3,
    \label{eq:dominance_3obj}
\end{equation}
with strict inequality in at least one objective. Of the $1001$ candidate
geometries, $591$ satisfied \Cref{eq:lift_constraint} and $200$ of those were
non-dominated.

The sweep spans $0.50\le\xi_g\le1.00$, whereas the surrogate was trained on
$0.50\le\xi_g\le0.95$ (\Cref{eq:ogive_range}); the uppermost $100$ candidates
are therefore extrapolations beyond the sampled range, of which
$N$ of the $200$ non-dominated designs lie outside it. The design
selected below is interior to the training range and is unaffected, but
results quoted for the extreme-$\xi_g$ end of the front should be read with
the extrapolation-fold errors of \Cref{sec:loo_cfd} in mind.
 
Because the Pareto set contains many admissible trade-offs, a single
representative design was selected using the normalized distance to the ideal
point. Each objective was scaled as
\begin{equation}
    \widehat f_i(\xi_g)
    =
    \frac{f_i(\xi_g)-f_i^{\min}}
         {f_i^{\max}-f_i^{\min}},
    \label{eq:objective_normalisation}
\end{equation}
and the compromise design was defined by
\begin{equation}
    \xi_g^\star
    =
    \arg\min_{\xi_g\in\mathcal P}
    \left(
        \sum_{i=1}^{3}\widehat f_i(\xi_g)^2
    \right)^{1/2}.
    \label{eq:ideal_compromise}
\end{equation}
This criterion gives
\begin{equation}
    \boxed{\xi_g^\star=0.825},
\end{equation}
with
\begin{equation}
    C_D=0.548764,\qquad
    q_{w,\max}=2.56305\times10^6~{\rm W\,m^{-2}},\qquad
    \dot Q'=2.706738\times10^6~{\rm W\,m^{-1}}.
    \label{eq:baseline_pareto_result}
\end{equation}
 
\begin{figure}[tbp]
    \centering
    \includegraphics[width=0.68\linewidth]{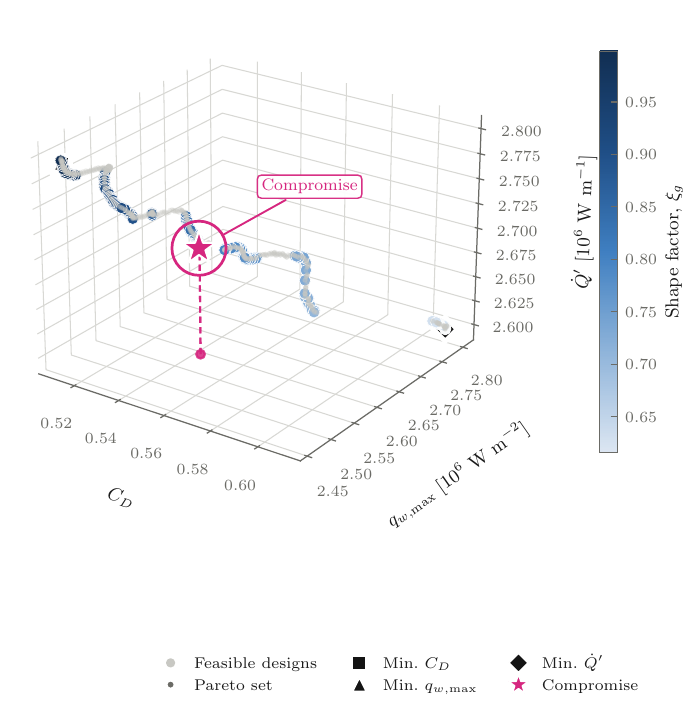}
    \caption{Three-objective Pareto structure in
    $(C_D,q_{w,\max},\dot Q')$ space. Grey points denote feasible but
    dominated designs; coloured points denote the non-dominated set and are
    coloured by the shape factor $\xi_g$.}
    \label{fig:pareto3d}
\end{figure}

The three-dimensional Pareto structure is shown in \Cref{fig:pareto3d}. The
non-dominated points form a one-dimensional trajectory in the
three-objective space, as expected for a design family controlled by a single
scalar parameter. Importantly, the trajectory is not degenerate: the
peak-heating and integrated-heating objectives vary differently along the
front, so the third objective adds information that is not contained in a
two-objective drag--peak-heating plot.

\section{Results II: Waverider Surface Loads and Angle-of-Attack Optimisation}
\label{sec:results_waverider_surface}
This section and the next report the two halves of the waverider study, and
they are presented in order of difficulty. The surface loads come first
because they are obtained with the two-step construction of
\Cref{subsec:two_step_operator} applied directly: a trunk on the surface
mesh, a branch on the angle of attack, and nothing else. No shock-aligned
coordinate and no residual correction are used here, and none is needed. The
volume field of \Cref{sec:results_waverider_volume} is the case in which that
same construction is not sufficient, and the additions it requires are
motivated by the contrast with what follows.
 
The surrogates evaluated here are those of \Cref{subsec:waverider_surface};
the validation protocol and the error measures are given in
\Cref{subsec:exp_waverider_surface_protocol,%
subsec:exp_waverider_surface_metrics}. Accuracy is reported first, then the
multi-objective selection of an operating incidence that the surrogates
support.
 
\subsection{Surface Surrogate Accuracy}
\label{sec:accuracy_head}
 
\label{sec:accuracy}
 
\begin{figure}
  \centering
  \includegraphics[width=\textwidth]{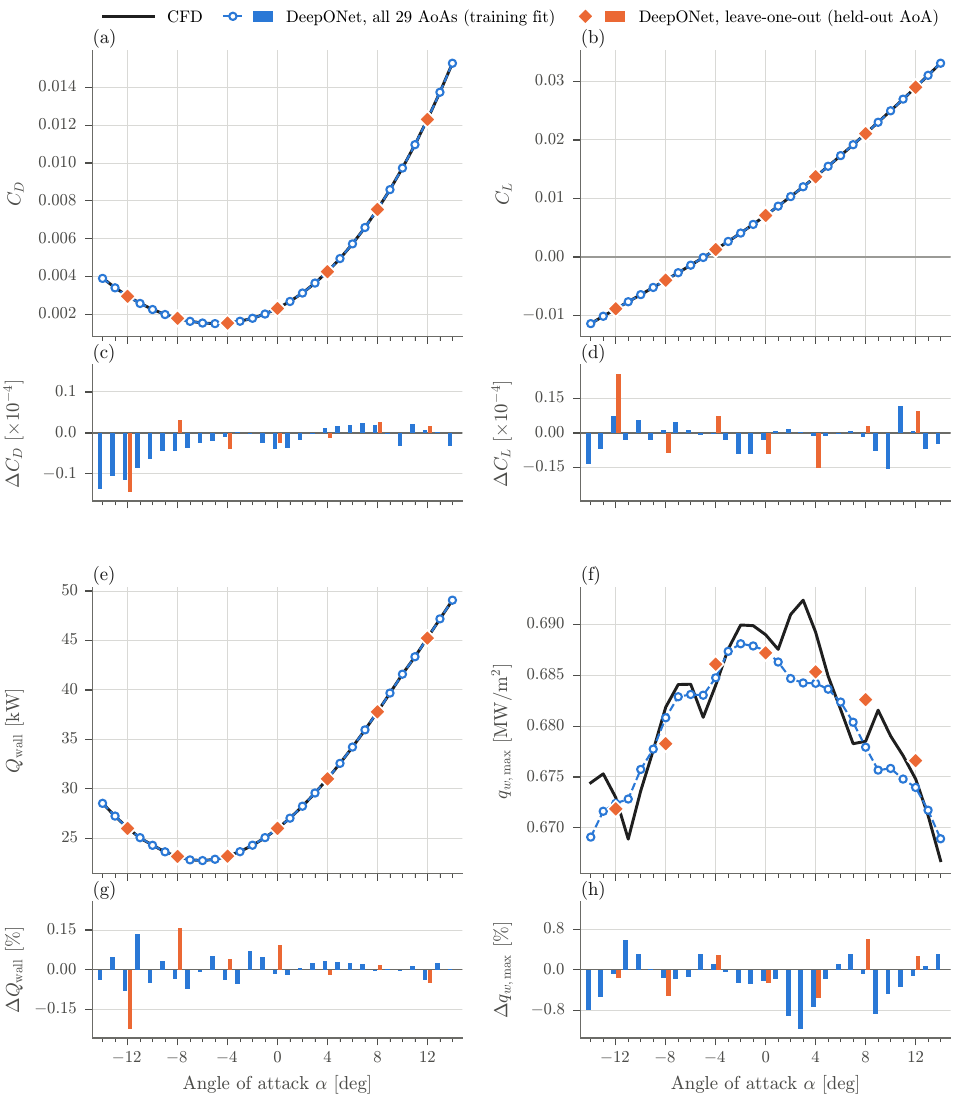}
  \caption{Quantities of interest for the 3D AEDC waverider versus angle of attack,
    with the corresponding surrogate errors. (a)~Drag coefficient $C_D$ and (c)~its
    error; (b)~lift coefficient $C_L$ and (d)~its error; (e)~integrated wall heat load
    $Q_\mathrm{wall}=\sum_e q_{w,e}A_e$ and (g)~its relative error; (f)~peak wall heat
    flux $q_{w,\max}$ and (h)~its relative error. Errors are signed,
    $\Delta(\cdot)=(\cdot)_\mathrm{DeepONet}-(\cdot)_\mathrm{CFD}$; $\Delta C_D$ and
    $\Delta C_L$ are absolute (in units of $10^{-4}$), the heat quantities relative to
    CFD. Black line: CFD. Blue: DeepONet trained on all 29 AoAs, evaluated at its
    training AoAs (training fit). Orange: leave-one-out prediction at each AoA excluded
    from training (generalization).}
  \label{fig:qoi_comparison}
\end{figure}

\Cref{fig:qoi_comparison} shows the four quantities of interest over the full
angle-of-attack range, together with the signed errors
\eqref{eq:qoi_error} at each angle. The drag polar has its minimum,
$C_D \approx 1.5\times10^{-3}$, at the zero-lift angle
$\alpha \approx -5^\circ$. The lift coefficient increases almost linearly
with $\alpha$, at about $1.6\times10^{-3}$ per degree, and the maximum
lift-to-drag ratio, $L/D \approx 3.3$, occurs near
$\alpha \approx 2$--$3^\circ$. The integrated heat load $Q_\mathrm{wall}$
follows the drag trend, with a minimum near $\alpha \approx -6^\circ$ and a
twofold increase towards $\alpha = +14^\circ$.
 
The all-angle model reproduces all four quantities closely
(\Cref{tab:summary}): the mean errors are
$|\Delta C_D| = 3.4\times10^{-6}$ and $|\Delta C_L| = 4.4\times10^{-6}$,
both below $0.05$ counts, and the error in $Q_\mathrm{wall}$ is at most
$0.13\%$. At the held-out angles the leave-one-out errors are larger, as
expected, but remain small beside the ranges the quantities span:
$|\Delta C_D| \le 1.45\times10^{-5}$,
$|\Delta C_L| \le 2.58\times10^{-5}$ and
$|\Delta Q_\mathrm{wall}| \le 0.22\%$ (\Cref{tab:qoi_errors}). The paired
bars in \Cref{fig:qoi_comparison}(c,d,g,h) show the generalisation gap at
each held-out angle directly: the orange leave-one-out bar may be compared
with the blue all-angle bar beside it, and with the all-angle mean evaluated
at the same angles in the middle column of \Cref{tab:summary}.
 
The peak heat flux is the least robust of the four. Over the $29$ CFD cases
$q_{w,\max}$ varies by only about $4\%$, and its location remains on the same
few nose and leading-edge cells throughout. Against that narrow variation the
surrogate error is not negligible: evaluated at the seven held-out angles,
the all-angle model gives a mean error of $0.22\%$ and the leave-one-out
model $0.38\%$, so generalisation costs a factor of $1.7$, and the worst
leave-one-out case reaches $0.61\%$. Over the full sweep the all-angle error
reaches $1.18\%$ at its worst angle, which is a quarter of the physical
variation in the quantity being predicted. As a pointwise maximum,
$q_{w,\max}$ is sensitive to local error at the hotspot, whereas
$Q_\mathrm{wall}$ averages error over the whole surface and is reproduced two
orders of magnitude more accurately. This is why $Q_\mathrm{wall}$, and not
$q_{w,\max}$, is the thermal objective in the multi-objective optimisation of
\Cref{sec:moo}; $q_{w,\max}$ is nevertheless evaluated and reported for every
design.
 
\Cref{tab:field_errors} reports the cell-wise field errors
\eqref{eq:field_error} at the held-out angles. For the all-angle model the
relative $\ell_2$ errors stay below $2.9\%$ for all four fields, with means
of $1.9\%$ for $q_w$ and $t_x$, $1.1\%$ for $t_y$ and $1.3\%$ for
$t_z$.
 
Under leave-one-out they are essentially unchanged. The means are $1.99\%$,
$1.76\%$, $1.09\%$ and $1.45\%$ against $1.92\%$, $1.91\%$, $1.08\%$ and
$1.34\%$ for the all-angle model, and the worst single case is $2.60\%$
against $2.87\%$; for $t_x$ the leave-one-out model is in fact the more
accurate of the two, both on average and at its worst angle. Removing one
angle from the training set therefore barely perturbs the learned basis: the
surface fields vary smoothly with incidence, and the remaining $28$ angles
determine the basis almost as well as all $29$. At field level the surrogate
interpolates in $\alpha$ at no measurable cost.
 
That does not carry over to every quantity derived from the fields. The
integrated quantities of interest are two orders of magnitude more accurate
than the fields they are formed from --- $Q_\mathrm{wall}$ to $0.09\%$ on
average, and $C_D$ and $C_L$ to a few parts in $10^{6}$ on coefficients of
order $10^{-3}$ --- because local field errors of either sign partly cancel
under the area-weighted sums
\eqref{eq:cd_cl}--\eqref{eq:thermal_qoi}. The peak heat flux is the
exception, since a pointwise maximum selects a single cell rather than
averaging over many: there the leave-one-out penalty is a factor of $1.7$
(\Cref{tab:summary}), where at field level there is none.

\begin{table}[t]
  \centering
  \caption{Relative $L_2$ error (\%) of the predicted surface fields on the 3D AEDC waverider at the held-out angles of attack: leave-one-out model (LOO), trained without that AoA, versus the model trained on all 29 AoAs (all).}
  \label{tab:field_errors}
  \small
  \begin{tabular}{rcccccccc}
  \toprule
   & \multicolumn{2}{c}{$q_w$} & \multicolumn{2}{c}{$t_x$} & \multicolumn{2}{c}{$t_y$} & \multicolumn{2}{c}{$t_z$} \\
  \cmidrule(lr){2-3} \cmidrule(lr){4-5} \cmidrule(lr){6-7} \cmidrule(lr){8-9}
  $\alpha$ [deg] & LOO & all & LOO & all & LOO & all & LOO & all \\
  \midrule
  $-12$ & 2.00 & 2.87 & 1.89 & 1.83 & 1.38 & 1.26 & 1.35 & 1.43 \\
  $-8$ & 2.44 & 2.62 & 1.48 & 1.46 & 1.35 & 1.32 & 1.36 & 1.26 \\
  $-4$ & 2.60 & 2.35 & 1.35 & 1.43 & 1.53 & 1.38 & 1.59 & 1.34 \\
  $0$ & 1.97 & 1.71 & 1.36 & 1.55 & 1.07 & 1.09 & 1.39 & 1.24 \\
  $+4$ & 1.76 & 1.37 & 1.55 & 1.92 & 0.85 & 0.89 & 1.35 & 1.22 \\
  $+8$ & 1.57 & 1.26 & 2.23 & 2.38 & 0.76 & 0.82 & 1.37 & 1.35 \\
  $+12$ & 1.59 & 1.25 & 2.48 & 2.80 & 0.66 & 0.80 & 1.75 & 1.55 \\
  \midrule
  Mean & 1.99 & 1.92 & 1.76 & 1.91 & 1.09 & 1.08 & 1.45 & 1.34 \\
  Max & 2.60 & 2.87 & 2.48 & 2.80 & 1.53 & 1.38 & 1.75 & 1.55 \\
  \bottomrule
  \end{tabular}
\end{table}
 
\begin{table}[t]
  \centering
  \caption{Leave-one-out errors in the quantities of interest at the held-out angles of attack. $|\Delta C_D|$ and $|\Delta C_L|$ are absolute errors with respect to CFD; $Q_\mathrm{wall}$ and $q_{w,\max}$ errors are relative (\%).}
  \label{tab:qoi_errors}
  \small
  \begin{tabular}{rcccccc}
  \toprule
  $\alpha$ [deg] & $C_D^{\mathrm{CFD}}$ & $|\Delta C_D|$ & $C_L^{\mathrm{CFD}}$ & $|\Delta C_L|$ & $Q_\mathrm{wall}$ err.\ [\%] & $q_{w,\max}$ err.\ [\%] \\
  \midrule
  $-12$ & $0.00296$ & $1.45\times10^{-5}$ & $-0.00888$ & $2.58\times10^{-5}$ & 0.22 & 0.17 \\
  $-8$ & $0.00178$ & $3.05\times10^{-6}$ & $-0.00396$ & $8.68\times10^{-6}$ & 0.16 & 0.53 \\
  $-4$ & $0.00154$ & $3.96\times10^{-6}$ & $+0.00127$ & $7.14\times10^{-6}$ & 0.04 & 0.30 \\
  $0$ & $0.00231$ & $2.43\times10^{-6}$ & $+0.00710$ & $9.16\times10^{-6}$ & 0.09 & 0.26 \\
  $+4$ & $0.00425$ & $1.16\times10^{-6}$ & $+0.01372$ & $1.55\times10^{-5}$ & 0.02 & 0.57 \\
  $+8$ & $0.00754$ & $2.59\times10^{-6}$ & $+0.02108$ & $3.16\times10^{-6}$ & 0.02 & 0.61 \\
  $+12$ & $0.01231$ & $1.62\times10^{-6}$ & $+0.02898$ & $9.57\times10^{-6}$ & 0.05 & 0.27 \\
  \midrule
  Mean &  & $4.19\times10^{-6}$ &  & $1.13\times10^{-5}$ & 0.09 & 0.38 \\
  Max &  & $1.45\times10^{-5}$ &  & $2.58\times10^{-5}$ & 0.22 & 0.61 \\
  \bottomrule
  \end{tabular}
\end{table}

\begin{table}[t]
  \centering
  \caption{Training fit of the model trained on all 29 AoAs versus leave-one-out generalization. The middle column evaluates the all-AoA model at the same AoAs used as LOO hold-outs, for a like-for-like comparison.}
  \label{tab:summary}
  \small
  \resizebox{\textwidth}{!}{%
  \begin{tabular}{lccccc}
  \toprule
   & \multicolumn{2}{c}{All-AoA model, 29 AoAs} & All-AoA model & \multicolumn{2}{c}{LOO, held-out AoAs} \\
  \cmidrule(lr){2-3} \cmidrule(lr){4-4} \cmidrule(lr){5-6}
  Quantity & Mean & Max & Mean at held-out $\alpha$ & Mean & Max \\
  \midrule
  $q_w$ rel.\ $L_2$ [\%] & 1.93 & 3.06 & 1.92 & 1.99 & 2.60 \\
  $t_x$ rel.\ $L_2$ [\%] & 1.94 & 3.02 & 1.91 & 1.76 & 2.48 \\
  $t_y$ rel.\ $L_2$ [\%] & 1.09 & 1.53 & 1.08 & 1.09 & 1.53 \\
  $t_z$ rel.\ $L_2$ [\%] & 1.37 & 1.72 & 1.34 & 1.45 & 1.75 \\
  \midrule
  $|\Delta C_D|$ & $3.44\times10^{-6}$ & $1.38\times10^{-5}$ & $3.57\times10^{-6}$ & $4.19\times10^{-6}$ & $1.45\times10^{-5}$ \\
  $|\Delta C_L|$ & $4.44\times10^{-6}$ & $1.59\times10^{-5}$ & $2.27\times10^{-6}$ & $1.13\times10^{-5}$ & $2.58\times10^{-5}$ \\
  $Q_\mathrm{wall}$ rel.\ err.\ [\%] & 0.04 & 0.13 & 0.03 & 0.09 & 0.22 \\
  $q_{w,\max}$ rel.\ err.\ [\%] & 0.34 & 1.18 & 0.22 & 0.38 & 0.61 \\
  \bottomrule
  \end{tabular}
  }
\end{table}

\begin{figure}[t]   
  \centering
  \includegraphics[width=\textwidth]{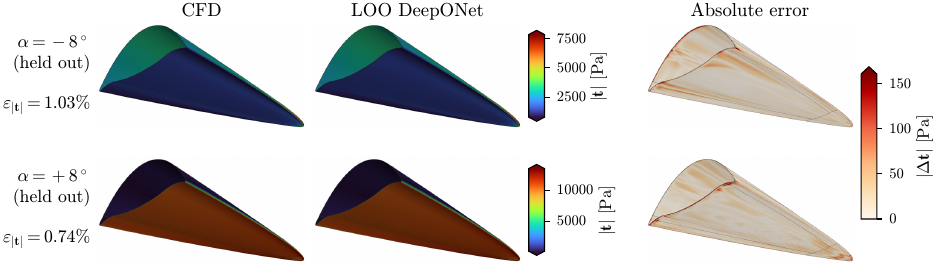}
  \caption{Surface traction magnitude $|\mathbf{t}|$ on the AEDC waverider at the
  two held-out angles of attack $\alpha=-8^\circ$ (top row) and $\alpha=+8^\circ$
  (bottom row). Left to right: CFD reference, leave-one-out (LOO) two-step
  DeepONet prediction (trained without the displayed angle), and the pointwise
  absolute error $\bigl||\mathbf{t}|_{\mathrm{LOO}}-|\mathbf{t}|_{\mathrm{CFD}}\bigr|$.
  The CFD and LOO panels in a row share one color scale (limits at the 1st and
  99th percentiles of that row); the error panels share one scale for both
  angles. The vehicle is viewed from below and behind, showing the windward
  surface. The relative error $\varepsilon_{|\mathbf{t}|}$ of
  Eq.~\eqref{eq:eps_t} over all $N_c$ skin cells is given at the left of each row.}
  \label{fig:traction_loo_3d}
\end{figure}
 
\begin{figure}[t]   
  \centering
  \includegraphics[width=\textwidth]{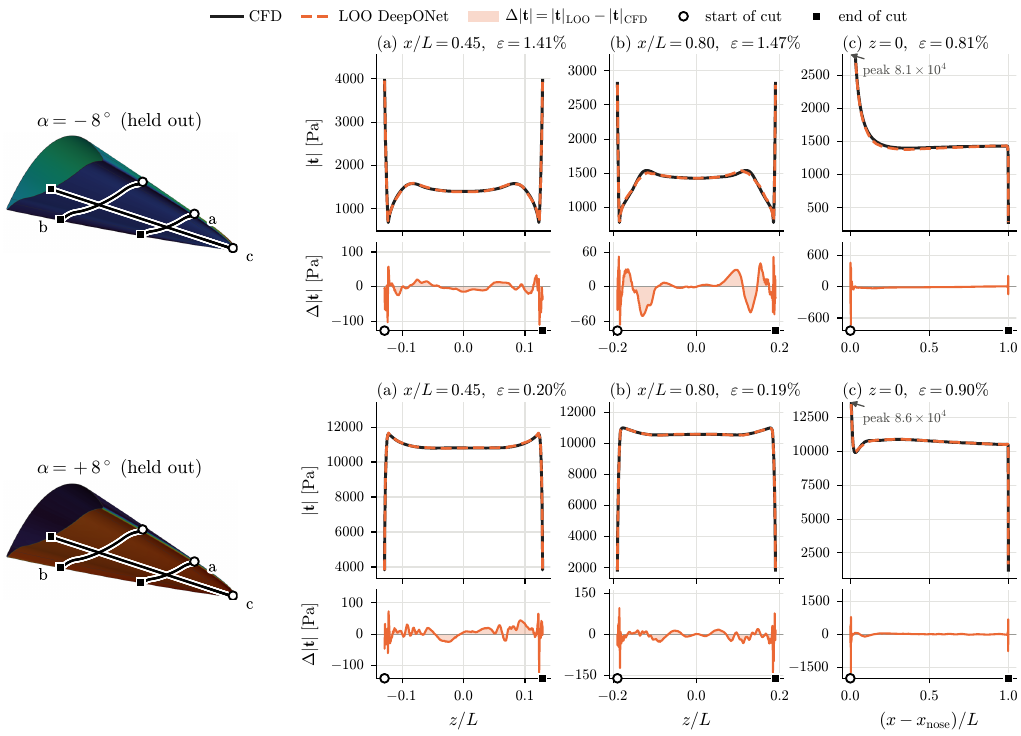}
  \caption{Traction magnitude along three cuts of the windward (lower) surface
  at the held-out angles $\alpha=-8^\circ$ (top block) and $\alpha=+8^\circ$
  (bottom block). Left: location of the cuts on the vehicle, with the start
  ($\circ$) and end ($\blacksquare$) of each profile marked. (a)~Spanwise cut
  at $x/L=0.45$ and (b)~at $x/L=0.80$, both running from the leading-edge tip at
  $-z$ to the tip at $+z$; (c)~centerline cut $z=0$, running from the nose to the
  base. Upper panels: $|\mathbf{t}|$ from CFD (solid) and the LOO DeepONet
  (dashed); lower panels: signed error
  $\Delta|\mathbf{t}|=|\mathbf{t}|_{\mathrm{LOO}}-|\mathbf{t}|_{\mathrm{CFD}}$.
  The relative error along each cut, $\varepsilon$ in Eq.~\eqref{eq:eps_t}
  restricted to the cells crossed by that cut, is given in the panel title. In
  (c) the nose stagnation peak is outside the plotted range; its value is
  annotated.}
  \label{fig:traction_loo_lines}
\end{figure}
 
\Cref{fig:traction_loo_3d} compares the surface traction magnitude
$|\mathbf{t}|=\bigl(t_x^2+t_y^2+t_z^2\bigr)^{1/2}$ predicted by the
leave-one-out two-step DeepONet with the CFD solution at $\alpha=-8^\circ$
and $\alpha=+8^\circ$. Each prediction comes from a model whose training set
excludes the displayed angle, so both rows are true out-of-sample tests. The
accuracy is summarised by the relative error
\begin{equation}
  \varepsilon_{|\mathbf{t}|}
  = \frac{\bigl\| |\mathbf{t}|_{\mathrm{LOO}} - |\mathbf{t}|_{\mathrm{CFD}} \bigr\|_2}
         {\bigl\| |\mathbf{t}|_{\mathrm{CFD}} \bigr\|_2},
  \qquad
  \|f\|_2^2 = \sum_{j=1}^{N_c} f_j^2,
  \label{eq:eps_t}
\end{equation}
evaluated over all $N_c$ skin cells. It equals
$\varepsilon_{|\mathbf{t}|}=1.03\%$ at $\alpha=-8^\circ$ and $0.74\%$ at
$\alpha=+8^\circ$. Both figures are smaller than the error in any individual
traction component at the same angle (\Cref{tab:field_errors}): at
$+8^\circ$, for instance, $t_x$ carries a leave-one-out error of $2.23\%$
while the magnitude carries $0.74\%$. The component errors are therefore
partly of opposite sign and cancel in the Euclidean norm --- the same
cancellation that makes the integrated loads more accurate than the fields,
appearing here one level lower.
 
The surrogate reproduces the large-scale distribution at both angles: the
change in windward loading between the two attitudes, and the high traction
along the leading edges and at the nose. The absolute error is small over
most of the surface and is concentrated in narrow bands along the leading
edges and near the nose, where $|\mathbf{t}|$ varies most steeply. The
error scale is capped at $150$~Pa, against surface values reaching
$7.5$~kPa at $\alpha=-8^\circ$ and above $10$~kPa at $\alpha=+8^\circ$, so
even in those bands the prediction departs from the reference by one to two
per cent of the peak loading.

\Cref{fig:traction_loo_lines} examines these regions quantitatively along
three cuts of the windward surface: two spanwise stations, $x/L=0.45$ and
$x/L=0.80$, and the centreline $z=0$. The cut locations and directions are
marked on the vehicle at left. Along the spanwise cuts the leave-one-out
prediction follows the CFD profile across the span, including the rise
towards both leading edges, and the signed error $\Delta|\mathbf{t}|$ stays
small in the interior, growing only near the leading-edge tips. The relative
errors along the cuts are $1.41\%$ and $1.47\%$ at $\alpha=-8^\circ$, and
$0.20\%$ and $0.19\%$ at $\alpha=+8^\circ$, for (a) and (b) respectively.
 
The sevenfold difference between the two attitudes is a matter of the
denominator rather than of the prediction. The error panels share a
$\pm100$~Pa ordinate at both angles, so the absolute discrepancy along the
cuts is essentially the same; what differs is the loading they are measured
against, which rises from roughly $1$--$4$~kPa at $\alpha=-8^\circ$ to
$4$--$12$~kPa at $\alpha=+8^\circ$. The ratio of the loadings and the ratio
of the relative errors agree to within a few per cent.
 
Along the centreline (c) the prediction captures the decay from the nose
stagnation peak to the nearly constant loading over the rest of the windward
surface, with $\varepsilon=0.81\%$ and $0.90\%$ at the two angles. That this
cut does \emph{not} improve between the attitudes, where the spanwise cuts
improve sevenfold, is consistent with the same explanation: panel~(c) is
dominated by the stagnation peak, annotated at $8.1\times10^{4}$~Pa and
$8.6\times10^{4}$~Pa, which differ by only a few per cent because stagnation
loading responds weakly to eight degrees of incidence. With the denominator
unchanged, the relative error is unchanged.
 
Taken together the two figures show that the leave-one-out error is
concentrated where the loading varies most sharply --- the
$\Delta|\mathbf{t}|$ traces are flat across the interior of every cut and
spike only at the leading-edge tips and the nose --- while over the bulk of
the windward surface, which carries most of the aerodynamic load, the
held-out traction field is reproduced to of order one per cent or better.
 
\subsection{Multi-objective optimization problem}
\label{sec:moo}
 
The trained surrogates are used to select the angle of attack that best
balances aerodynamic efficiency against aerothermal load. The design variable
is the angle of attack $\alpha$, restricted to the range covered by the CFD
training data, $\mathcal{A} = [-14^\circ, +14^\circ]$, so that the surrogate
is never evaluated in extrapolation. The optimization problem is
\begin{equation}
  \min_{\alpha \in \mathcal{A}}\;
  \mathbf{J}(\alpha) =
  \big(\, \hat C_D(\alpha),\; -\hat C_L(\alpha),\; \hat Q_\mathrm{wall}(\alpha) \,\big),
  \label{eq:moo}
\end{equation}
that is, minimize drag, maximize lift and minimize the integrated wall heat
load. Each objective is evaluated from the predicted surface fields: the
DeepONet surrogates for $q_w$, $t_x$, $t_y$ and $t_z$ are evaluated
at $\alpha$ on the $N_c$ surface cells, and the predictions are integrated
through \eqref{eq:force}, \eqref{eq:wind_axes} and \eqref{eq:thermal_qoi}.
The surrogate thus replaces a full CFD solution at each design point, while
the objectives remain consistent with the CFD definitions used in
Section~\ref{sec:accuracy}.
 
The thermal objective is the integrated heat load $Q_\mathrm{wall}$ rather
than the peak heat flux $q_{w,\max}$, for reasons that
\Cref{fig:qoi_comparison}(f,h) makes plain. Over the full incidence range
$q_{w,\max}$ varies by only about $4\%$, from $0.668$ to
$0.693$~MW\,m$^{-2}$, and its location remains on the same few nose and
leading-edge cells throughout.
 
More tellingly, that variation is not a smooth trend. The CFD curve in
panel~(f) is jagged: it carries two separate local maxima near
$\alpha=-1^\circ$ and $\alpha=+2^\circ$ with a dip between them, and a
further bump near $\alpha=+8^\circ$, with adjacent angles differing by
several tenths of a per cent in no systematic direction. This is the
signature of a pointwise maximum taken over discrete surface cells: as the
attitude changes, the hotspot migrates and the maximum jumps from one cell
to the next rather than sliding continuously. The surrogate, which is smooth
in $\alpha$ by construction, tracks the mean trend and cuts across this
structure, sitting below the reference throughout the interval from
$\alpha=0^\circ$ to about $+10^\circ$; panel~(h) shows the resulting errors
there to be predominantly one-signed and to reach about $1.2\%$.
 
A Pareto front ordered by $q_{w,\max}$ would therefore rank designs partly
on the component of the signal the surrogate does not resolve.
$Q_\mathrm{wall}$ is free of this difficulty on both counts: it varies by a
factor of two over the same range, from $22.7$ to $49.1$~kW, and because it
averages the wall flux over the whole surface it is smooth in $\alpha$ and
is predicted to within $0.22\%$ at the held-out angles, with a mean error of
$0.09\%$ (\Cref{tab:qoi_errors}). The ratio of physical variation to
prediction error is about $500$ for $Q_\mathrm{wall}$ against about $6$ for
$q_{w,\max}$ --- two orders of magnitude --- which is what makes one a usable
design objective and the other a reported diagnostic. $q_{w,\max}$ is
nevertheless evaluated and reported for every design.
 
A design $\alpha_1$ dominates $\alpha_2$, written $\alpha_1 \prec \alpha_2$,
if $J_k(\alpha_1) \le J_k(\alpha_2)$ for all $k$ and
$J_k(\alpha_1) < J_k(\alpha_2)$ for at least one $k$. The solution of
\eqref{eq:moo} is the Pareto set
$\mathcal{P} = \{\alpha \in \mathcal{A} : \nexists\, \alpha' \in \mathcal{A},\ \alpha' \prec \alpha\}$.
Because the design space is one-dimensional and the surrogate is
deterministic and inexpensive, $\mathcal{P}$ is computed exactly up to grid
resolution, by an exhaustive scan rather than a population-based heuristic
such as NSGA-II. The scan evaluates $\mathbf{J}$ at $561$ uniformly spaced
angles ($\Delta\alpha = 0.05^\circ$), each requiring four branch evaluations
and one matrix--vector product per field with the precomputed trunk basis.
Every scanned design is then tested for dominance against all others. The
complete scan takes $220$ seconds on a single CPU, against
$2.05\times10^{4}$ core-hours --- $20$ hours of wall-clock time on $1024$
cores --- for one converged CFD solution at a single incidence. The resulting front is checked for consistency: no front member is
dominated, and every excluded design is dominated by a front member. Figure~\ref{fig:moo_pareto} shows the outcome: panels~(a)--(d) give the three
objectives and the lift-to-drag ratio against $\alpha$, with the
Pareto-optimal designs distinguished from the dominated ones, and
panels~(e)--(f) show the same designs in objective space.
 
The structure of the CFD data indicates the expected form of the solution.
Below the minimum-heat-load angle, $\alpha_{Q,\min} = -6.20^\circ$,
increasing $\alpha$ simultaneously reduces drag, increases lift and reduces
heat load, so every design there is dominated by one further up. Above it the
three objectives no longer move together. Between $\alpha_{Q,\min}$ and the
minimum-drag angle $-5.00^\circ$, drag and lift both continue to improve
while the heat load rises; beyond $-5.00^\circ$ all three increase with
$\alpha$. In either case no objective can be improved without worsening
another, so every design above $\alpha_{Q,\min}$ is non-dominated. The Pareto
set is therefore expected to span $\alpha \in [\alpha_{Q,\min}, 14^\circ]$,
which the scan confirms:
$\mathcal{P} = [-6.20^\circ,\, +14.00^\circ]$, some $40\%$ of the
range covered by the CFD training data. The maximum lift-to-drag design,
$\alpha = +2.20^\circ$ with $L/D = 3.30$, lies in the interior of the front.
 
\subsection{Selection of a preferred design}
\label{sec:preference}
 
\begin{figure}[t]
  \centering
  \includegraphics[width=\textwidth]{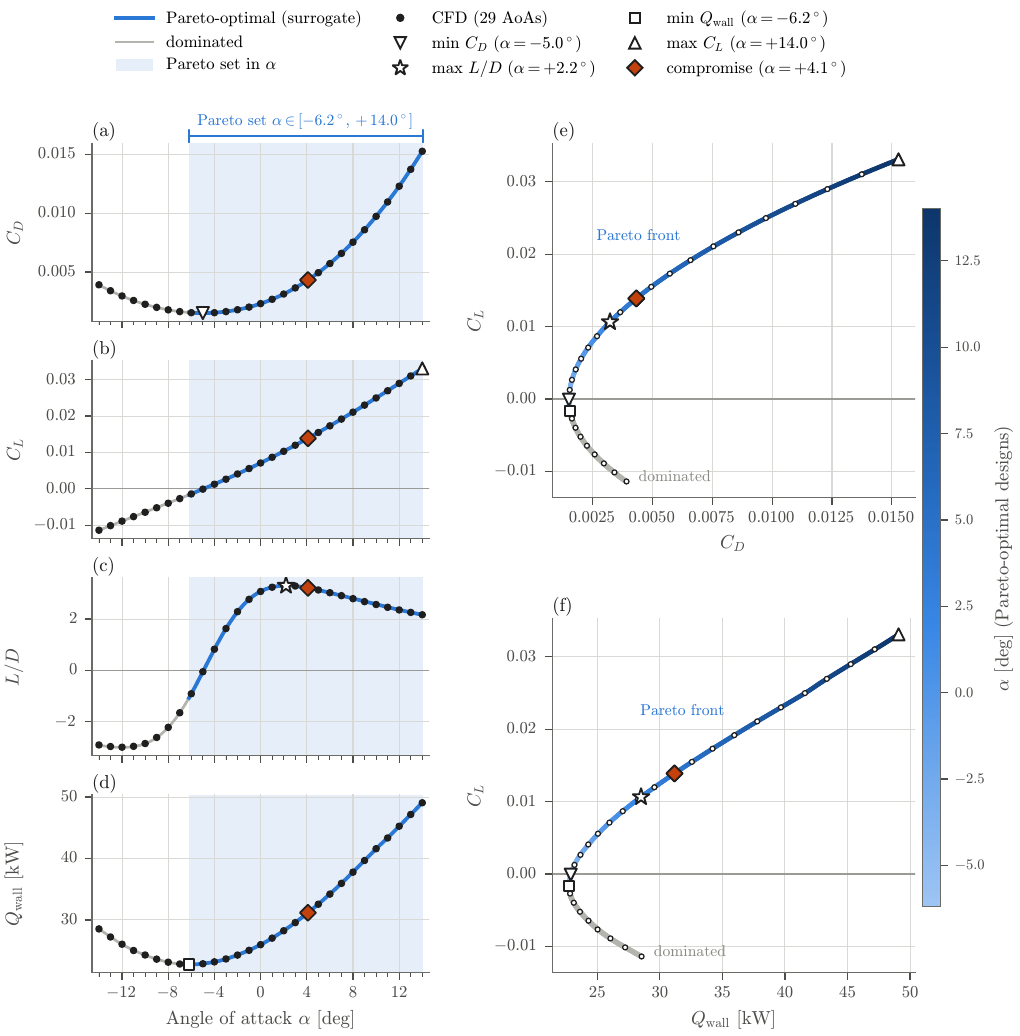}
  \caption{Angle-of-attack optimization of the 3D AEDC waverider: minimize $C_D$,
    maximize $C_L$, minimize the integrated wall heat load $Q_\mathrm{wall}$. (a)--(d)~Objectives and lift-to-drag ratio
    versus $\alpha$ from the surrogate scan ($\Delta\alpha = 0.05^\circ$);
    blue: Pareto-optimal designs, gray: dominated designs, shaded band: Pareto set;
    black dots: CFD at the 29 training AoAs. (e)~Drag polar and (f)~lift versus heat
    load in objective space; the Pareto branch is colored by $\alpha$, dominated
    designs are gray and open circles are CFD. Markers identify the key designs of
    Table~\ref{tab:moo_designs}.}
  \label{fig:moo_pareto}
\end{figure}

The Pareto set of Fig.~\ref{fig:moo_pareto} spans
$\alpha\in[-6.20^\circ,\,+14.00^\circ]$. Every angle in this interval
is optimal in the Pareto sense: none of $C_D$, $C_L$ and $Q_\mathrm{wall}$ can
be improved without degrading at least one of the others. The optimization by
itself therefore does not single out an operating point. Choosing one requires
an explicit statement of how the three objectives are to be traded, that is, a
preference. We express it with compromise programming, which selects the
Pareto-optimal design closest to an ideal but unattainable target.
 
\paragraph{Preference function.}
All objectives are written in minimization form,
$\mathbf{f}(\alpha)=\bigl(C_D,\,-C_L,\,Q_\mathrm{wall}\bigr)$, and normalized
over the Pareto set $\mathcal{P}$:
\begin{equation}
  \tilde f_i(\alpha)=\frac{f_i(\alpha)-u_i}{n_i-u_i}\in[0,1],\qquad
  u_i=\min_{\alpha\in\mathcal{P}}f_i(\alpha),\quad
  n_i=\max_{\alpha\in\mathcal{P}}f_i(\alpha).
  \label{eq:fnorm}
\end{equation}
The vector $\mathbf{u}$ is the utopia point. It collects the best value of each
objective, which no single angle attains simultaneously because the three
optima lie at different angles. The vector $\mathbf{n}$ is the nadir point,
the worst value of each objective on the front. After normalization the
objectives are dimensionless and span the same range. A weight therefore
expresses relative importance only, independent of the very different
magnitudes of $C_D$ ($\sim10^{-3}$), $C_L$ ($\sim10^{-2}$) and
$Q_\mathrm{wall}$ ($\sim10^{4}$\,W). The preferred angle minimizes the
weighted $L_p$ distance to the utopia point,
\begin{equation}
  \alpha^\ast=\arg\min_{\alpha\in\mathcal{P}} d_p(\alpha),\qquad
  d_p(\alpha)=\Bigl(\sum_{i\in\{D,L,T\}} w_i\,\tilde f_i(\alpha)^{p}\Bigr)^{1/p},
  \qquad \sum_i w_i=1,\; w_i\ge 0.
  \label{eq:pref}
\end{equation}
The weights $w_D$, $w_L$ and $w_T$ set the relative importance of drag, lift
and heat load. The exponent $p$ controls how imbalances between the objectives
are penalized:
\begin{itemize}
  \item $p=1$ is the weighted sum, which rewards a design that excels in one
        objective even if it is poor in another;
  \item $p=2$ is the Euclidean distance;
  \item $p\to\infty$ gives the weighted Tchebycheff function
        $\max_i w_i\tilde f_i$, which minimizes the worst weighted
        shortfall.
\end{itemize}
Because the minimization in Eq.~\eqref{eq:pref} is restricted to $\mathcal{P}$,
the selected design is Pareto-optimal for every choice of $\mathbf{w}$ and $p$.
We take equal weights, $w_D=w_L=w_T=1/3$, and $p=2$ as the baseline, so that no
objective is favored and imbalances are penalized moderately.
 
\begin{figure}[t]
  \centering
  \includegraphics[width=\textwidth]{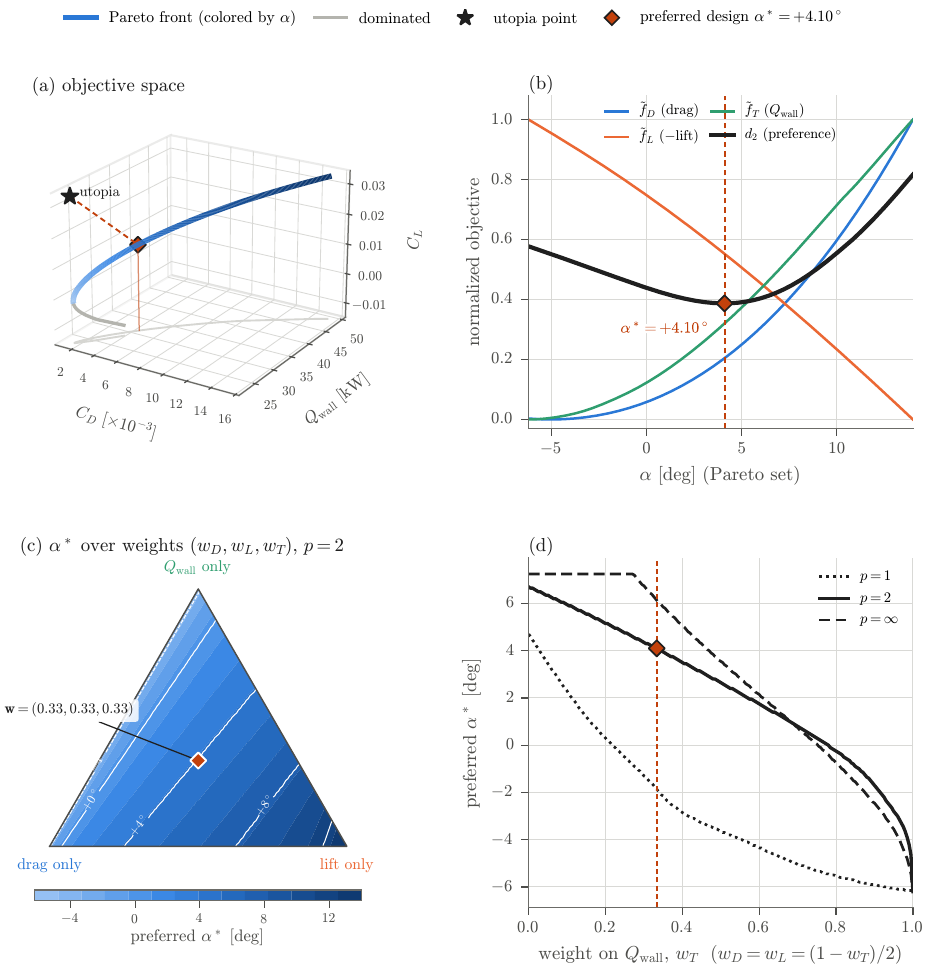}
  \caption[Selection of a single design from the Pareto front]%
  {Selection of a single design from the Pareto front by compromise
  programming. (a)~The $\alpha$ sweep in the objective space
  $(C_D,Q_\mathrm{wall},C_L)$: the Pareto-optimal branch is coloured by
  $\alpha$ and the dominated branch is grey, the star marks the utopia
  point and the diamond the preferred design, joined by the dashed segment
  whose length is the normalised weighted distance being minimised.
  (b)~The three normalised objectives and the preference function $d_2$
  along the Pareto set; $\tilde f_D$ and $\tilde f_T$ rise with $\alpha$
  while $\tilde f_L$ falls, and $d_2$ is minimised at
  $\alpha^\ast=+4.10^\circ$ (dashed line).
  (c)~The preferred angle over all weight vectors $(w_D,w_L,w_T)$ for
  $p=2$, each point of the simplex coloured by the resulting
  $\alpha^\ast$; the corners recover the single-objective optima and the
  diamond marks the equal-weight baseline.
  (d)~The preferred angle against the weight on the heat load, with
  $w_D=w_L=(1-w_T)/2$, for the three metrics $p=1$, $2$ and $\infty$.
  Panels (a), (b) and the marker in (d) use the baseline preference,
  $w_D=w_L=w_T=1/3$ and $p=2$.}
  \label{fig:moo_preference}
\end{figure}

\paragraph{Geometric interpretation (Fig.~\ref{fig:moo_preference}a).}
In the three-dimensional objective space $(C_D,Q_\mathrm{wall},C_L)$, the
$\alpha$ sweep traces a single curve, since the design has one degree of
freedom. Its Pareto-optimal part, colored by $\alpha$, is the Pareto front.
The dominated part (gray) corresponds to $\alpha<-6.20^\circ$, where
lowering the angle further increases drag and heat load and reduces lift. The
utopia point (star) lies off the curve, at the minimum drag, minimum heat load
and maximum lift of the front. The preferred design (diamond) is the point of
the front closest to it in the normalized, weighted metric; the dashed segment
marks this distance.
 
\paragraph{How $\alpha^\ast$ is selected (Fig.~\ref{fig:moo_preference}b).}
Along the Pareto set, $\tilde f_D$ and $\tilde f_T$ increase with $\alpha$:
drag and heat load are smallest near the lower end of the set, at the
minimum-drag and minimum-heat-load attitudes ($\alpha=-5.00^\circ$ and
$-6.20^\circ$), and largest at $\alpha=+14.00^\circ$. The normalized lift
objective $\tilde f_L$ decreases monotonically, because lift grows with
$\alpha$ across the whole set. The preference function $d_2$ combines these
opposing trends into a single convex-shaped curve. Its minimum lies at
$\alpha^\ast=+4.10^\circ$, where the normalized objectives take the values
$\tilde f_D=0.20$, $\tilde f_T=0.32$ and $\tilde f_L=0.56$: the design
concedes $20\%$ of the drag range and $32\%$ of the heat-load range in order
to capture $44\%$ of the available lift. Physically, $\alpha^\ast$ lies
slightly above the maximum-$L/D$ attitude, $\alpha=+2.20^\circ$
(Table~\ref{tab:moo_designs}). Compared with that design, it gives $29\%$
more lift for a $9\%$ higher heat load, while $L/D$ drops by only $2.4\%$
(from $3.30$ to $3.22$). The compromise thus trades a small loss of
aerodynamic efficiency for a substantial gain in lift, at a moderate thermal
penalty.
 
\begin{table}[!htbp]
  \centering
  \caption{Key designs of the angle-of-attack optimization of the 3D AEDC waverider, from the surrogate trained on all 29 AoAs. The Pareto set is $[-6.20^\circ, +14.00^\circ]$. The LOO column gives the range of the optimal angle over the 7 surrogates trained with one CFD snapshot removed.}
  \label{tab:moo_designs}
  \small
  \begin{tabular}{lcccccc}
  \toprule
  Design & $\alpha^\ast$ [deg] & LOO $\alpha^\ast$ [deg] & $C_D$ & $C_L$ & $L/D$ & $Q_\mathrm{wall}$ [kW] \\
  \midrule
  Minimum drag & $-5.00$ & $-5.05$ to $-5.00$ & $0.00150$ & $-0.00008$ & $-0.06$ & $22.88$ \\
  Maximum $L/D$ & $+2.20$ & $+2.20$ to $+2.30$ & $0.00322$ & $+0.01064$ & $+3.30$ & $28.50$ \\
  Minimum heat load & $-6.20$ & $-6.15$ to $-6.05$ & $0.00155$ & $-0.00166$ & $-1.07$ & $22.74$ \\
  Maximum lift & $+14.00$ & $+14.00$ to $+14.00$ & $0.01528$ & $+0.03310$ & $+2.17$ & $49.08$ \\
  \bottomrule
  \end{tabular}
\end{table}
 
\paragraph{Dependence on the weights (Fig.~\ref{fig:moo_preference}c).}
The preferred angle depends on the weights, and panel~(c) shows this
dependence over all admissible preferences. Each point of the triangle is a
weight vector $(w_D,w_L,w_T)$, and its color is the resulting $\alpha^\ast$.
The corners recover the single-objective optima: minimum drag at
$-5.00^\circ$, maximum lift at $+14.00^\circ$ and minimum heat load at
$-6.20^\circ$. The contours of $\alpha^\ast$ run nearly parallel to the
drag--heat edge, so $\alpha^\ast$ is governed mainly by the lift weight $w_L$
relative to $w_D+w_T$. Redistributing weight between drag and heat load moves
$\alpha^\ast$ much less, because $C_D$ and $Q_\mathrm{wall}$ increase with
$\alpha$ together over all but the lowest $1.2^\circ$ of the Pareto set and
therefore pull in the same direction. The
equal-weight baseline (diamond) lies in a region where $\alpha^\ast$ varies
smoothly, so moderate changes of the weights shift the recommended angle
gradually rather than abruptly.
 
\paragraph{Dependence on the metric (Fig.~\ref{fig:moo_preference}d).}
Panel~(d) follows $\alpha^\ast$ as the heat-load weight $w_T$ increases from 0
to 1, with the remaining weight split equally between drag and lift, for
$p=1$, $2$ and $\infty$. In all three cases $\alpha^\ast$ decreases as
$w_T$ increases: placing more emphasis on the heat load moves the vehicle
toward the minimum-$Q_\mathrm{wall}$ attitude, and at $w_T=1$ all three
metrics coincide at that attitude, $\alpha=-6.20^\circ$, as they must. The
metric matters in between, however.
\begin{itemize}
  \item For $p=1$ (weighted sum), $\alpha^\ast$ falls fastest and reaches
        low-lift attitudes already at moderate $w_T$. A weighted sum favors
        designs that are excellent in some objectives at the expense of
        others, and is known to sample the front
        unevenly~\cite{dasdennis1997,marlerarora2010}.
  \item For $p=\infty$ (Tchebycheff), $\alpha^\ast$ stays at a constant
        $+7.2^\circ$ up to $w_T\approx0.27$, where the lift
        shortfall remains the worst weighted term, and then decreases. The
        plateau is characteristic of the metric: only the largest weighted
        shortfall enters the objective, so changes in the others leave the
        minimiser untouched until the ordering
        switches~\cite{bowman1976,steuerchoo1983}.
  \item The Euclidean choice $p=2$ lies between the two and varies most
        smoothly.
\end{itemize}
At equal weights the three metrics give $\alpha^\ast=-1.8^\circ$
($p=1$), $+4.10^\circ$ ($p=2$) and $+6.2^\circ$ ($p=\infty$). This
spread of some $8^\circ$ shows that the recommended angle is a consequence of
the stated preference rather than an intrinsic property of the vehicle.
Reporting the full maps of panels~(c) and~(d) allows a designer with
different priorities to read off the corresponding operating point directly.
 
\begin{figure}
  \centering
  \includegraphics[width=\textwidth]{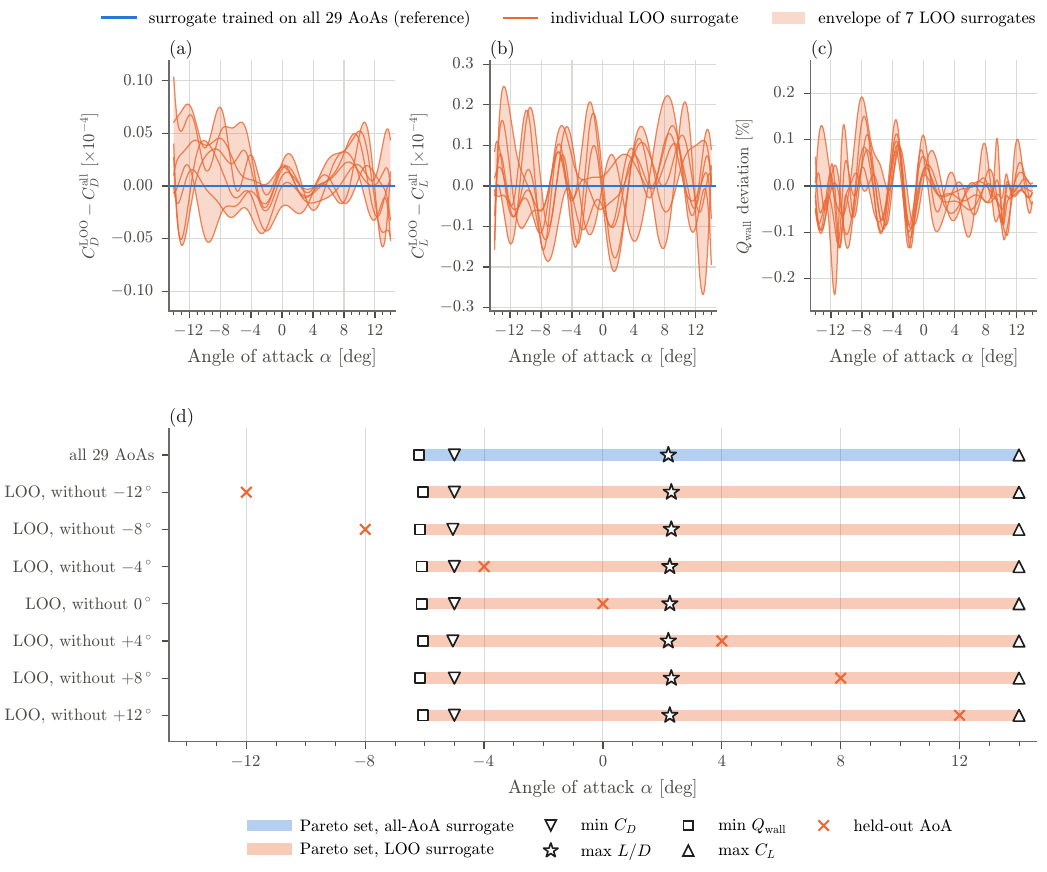}
  \caption{Robustness of the optimization to the training data. (a)--(c)~Objectives
    from the surrogate trained on all 29 AoAs (line) and the min--max envelope of the
    7 leave-one-out surrogates (band), each trained with one CFD snapshot
    removed; dots are CFD. (d)~Pareto set (bars) and key design angles (markers)
    found by each surrogate; $\times$ marks the AoA removed from training.}
  \label{fig:moo_robustness}
\end{figure}

\paragraph{Robustness.}
Finally, the preferred design is insensitive to the training data of the
surrogate. Figure~\ref{fig:moo_robustness}(a)--(c) compares the objectives
predicted by the all-angle surrogate with the min--max envelope of the seven
leave-one-out surrogates, each trained with one CFD snapshot withheld: the
envelope is narrow across the whole incidence range, so the objective curves
on which the optimization acts barely move when a snapshot is removed.
Panel~(d) carries this through to the optimization itself, giving the Pareto
set and the key design angles returned by each surrogate. Repeating the
selection with each of the seven gives
$\alpha^\ast\in[+4.10^\circ,+4.20^\circ]$ (Table~\ref{tab:moo_designs}), a
spread of $0.10^\circ$, or two scan increments. The recommendation therefore
does not hinge on any single CFD snapshot.
 
Two things are worth carrying forward from this section. The first is that
the surface surrogates required no special treatment: the construction of
\Cref{subsec:two_step_operator}, applied to each surface field in turn,
reproduces the held-out loads to within a few per cent and the integrated
quantities to a few parts in a thousand, which is accurate enough that the
Pareto front above is limited by the physics rather than by the surrogate.
The second is why that was possible. The surface fields vary smoothly with
incidence over a fixed mesh; the loading pattern shifts and strengthens, but
nothing in it moves discontinuously. A trunk expressed in the surface
coordinates can therefore learn one basis that serves every angle.
 
\section{Results III: Waverider Volume Fields and Shock Reconstruction}
\label{sec:results_waverider_volume}
The volume field is a harder problem than the surface loads, and it is the
one place in this work where the baseline construction had to be extended
rather than merely applied. The reason is the leading-edge shock. On the
surface the fields vary smoothly with incidence, so a single trunk basis in
the surface coordinates serves the whole operating range; in the volume the
dominant feature of the flow is a thin curved discontinuity that moves,
tilts and changes curvature as the angle of attack changes, and a trunk
expressed in Cartesian coordinates alone would have to represent the same
jump at a different place for every $\alpha$. Applied unchanged, the
two-step construction of \Cref{subsec:two_step_operator} therefore smears
the shock: the basis averages neighbouring discontinuity positions, and the
error that remains is not spread through the domain but concentrated in the
shock layer, where a loss formed from uniformly sampled points gives it
little weight.
 
Two additions were made in response, and both are reported here because
their necessity is specific to this case.
\begin{enumerate}
    \item \emph{A learned shock-aligned coordinate.} A frozen auxiliary
          network supplies a normalised distance to the shock surface, from
          which a bounded coordinate is formed that places the discontinuity
          at approximately the same value for every angle of attack. The
          trunk is then asked to represent a feature that does not move
          (\Cref{sec:shock_deeponet_method}).
    \item \emph{A hotspot-aware residual correction.} Shock alignment
          reduces the error but leaves it localised rather than uniform. A
          gated mixture of experts, driven by a learned hotspot probability
          that uses only quantities available at inference, is fitted to
          what remains (\Cref{sec:hotspot_moe_method}).
\end{enumerate}
Both are data-driven devices rather than physical constraints: no
Rankine--Hugoniot condition and no governing-equation residual is imposed.
Neither is used for the surface fields of
\Cref{sec:results_waverider_surface}, where they would add cost without
purpose, and the contrast between the two sections is the clearest evidence
in this work of where the extra machinery earns its place.
 
The subsections below recall only as much of the method of
\Cref{subsec:waverider_method} as the results require.
 
\subsection{Why the three-dimensional shock is difficult}
\label{sec:shock_difficulty}
 
A waverider is designed so that its leading-edge shock remains attached
\cite{Nonweiler1959,Bowcutt1987}. Consequently, the dominant nonsmooth
feature of the flow is a thin, curved shock surface
$\Gamma_s(\alpha)$ whose position, inclination, and curvature vary with
the angle of attack $\alpha$. This presents a particularly difficult
problem for operator learning. Neural networks with smooth activation
functions are biased toward low-frequency components
\cite{Rahaman2019}, while a conventional DeepONet \cite{Lu2021}
expressed only in fixed Cartesian coordinates must represent a family
of sharp features that translate and deform throughout the
three-dimensional domain. In this representation, a moving
discontinuity generally requires a large number of trunk basis
functions. This behavior is closely related to the slow decay of the
Kolmogorov $n$-width for transport-dominated solution manifolds, which
also limits classical linear reduced-order models
\cite{Ohlberger2016,Peherstorfer2022}.
 
The difficulty is amplified in three dimensions. In one dimension, a
shock can be viewed as a moving jump location; for the present
waverider problem, however, the discontinuity is a two-dimensional
surface embedded in a three-dimensional flow field. Small changes in
$\alpha$ modify not only the location of this surface but also its
orientation and curvature. If several such shock locations are
represented by a smooth neural basis in the physical coordinates,
regression tends to average neighboring discontinuity locations,
resulting in an artificially broadened or \emph{smeared} shock.
Moreover, the shock occupies only a small fraction of the computational
domain. A loss formed from uniformly sampled points is therefore
dominated by the much larger smooth-flow region and can provide
insufficient weight to errors in the shock layer. Finally, the finite
set of 29 available CFD conditions samples the continuous motion of
the shock only sparsely.
 
For these reasons, we use density as the primary field for assessing
the proposed shock-aware operator architecture. Across the bow shock,
density undergoes a pronounced and highly localized change, making
shock displacement and shock smearing readily visible in both the
density field and its gradient. Density therefore provides a
particularly stringent test of whether an operator-learning model can
represent a moving three-dimensional discontinuity. The objective of
the present study is not to construct a density-specific architecture;
rather, density is used as a demanding proof-of-concept for the
shock-aligned representation. Once the shock geometry has been learned,
the same formulation can be applied to other thermodynamic and flow
variables, including pressure, temperature, and velocity components.
 
\begin{figure}[t]
  \centering
  \includegraphics[width=\linewidth]{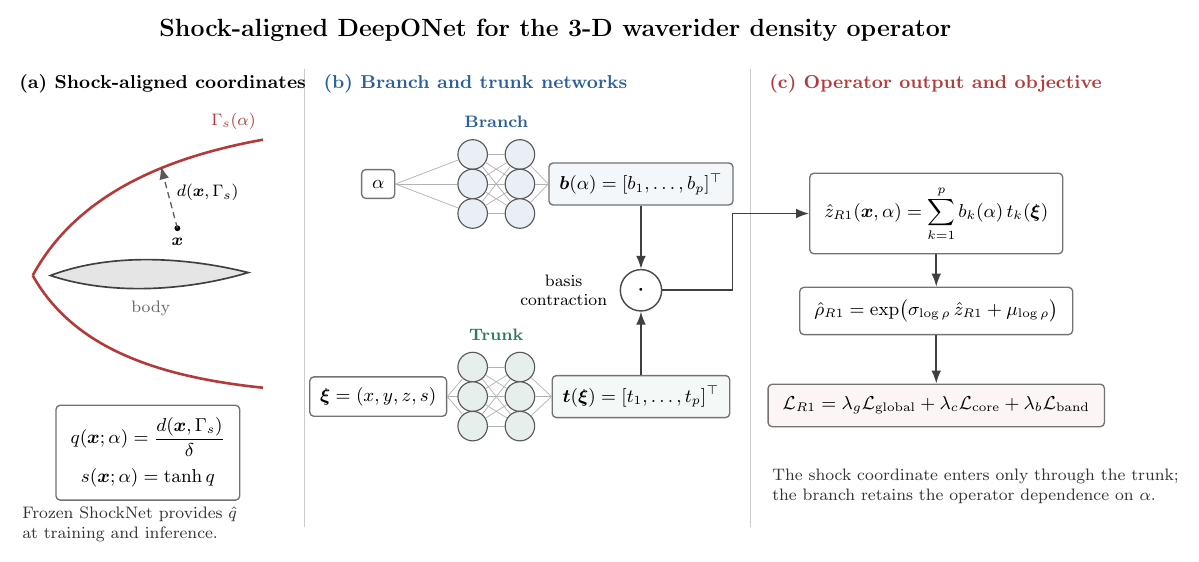}
  \caption{Shock-aligned DeepONet. (a)~A frozen ShockNet predicts the
  normalized shock distance $\widehat q$, which defines the bounded
  shock coordinate $s=\tanh\widehat q$. (b)~The branch network encodes
  the global operating condition $\alpha$, whereas the trunk network
  encodes the shock-aligned local coordinate
  $\bm{\xi}=(x,y,z,s)$. (c)~Their DeepONet contraction produces the
  standardized log-density $\widehat z_{R1}$. Because the physical
  shock corresponds approximately to $s=0$ for every $\alpha$, the
  trunk does not need to represent the shock motion solely through
  Cartesian coordinates.}
  \label{fig:deeponet}
\end{figure}

\begin{figure}[t]
  \centering
  \includegraphics[width=\linewidth]{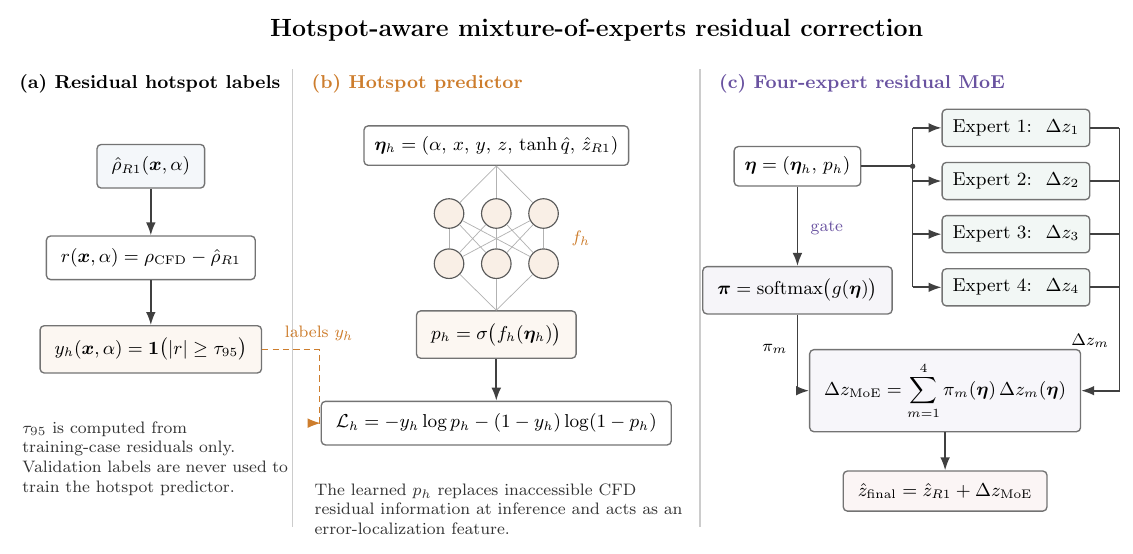}
  \caption{Hotspot-aware residual correction. (a)~Residuals above the
  training 95th percentile $\tau_{95}$ are labelled as hotspots.
  (b)~A classifier predicts the hotspot probability $p_h$ using
  inference-available quantities only. (c)~A gated mixture of four
  experts \cite{Jacobs1991} predicts the log-density correction
  $\Delta z_{\rm MoE}$, which is added to the frozen Stage-R1
  prediction.}
  \label{fig:hotspot_moe}
\end{figure}

We address the above difficulties in two stages. First, we learn the
density operator in a coordinate system attached to the shock, so that
the discontinuity occurs at approximately the same coordinate for all
values of $\alpha$ (Fig.~\ref{fig:deeponet}). Second, after the global
operator has been learned, we identify the spatial regions in which
the remaining prediction error concentrates and apply a learned
hotspot-aware residual correction (Fig.~\ref{fig:hotspot_moe}). The
complete training and inference procedure is summarized in
Fig.~\ref{fig:pipeline}.

\subsection{Density data and shock-aligned DeepONet}
\label{sec:shock_deeponet_method}
 
The dataset consists of 29 three-dimensional CFD solutions spanning
angles of attack from $-14^\circ$ to $14^\circ$. All cases are mapped
to a common unstructured FEBrick mesh containing approximately
$5.08\times10^{7}$ nodal locations. The solver-native density fields
are initially available at cell centers and are interpolated to the
common mesh nodes before surrogate training, yielding a consistent
nodal representation
\begin{equation}
    \rho=\rho(\bm{x};\alpha),
    \qquad
    \bm{x}=(x,y,z).
\end{equation}
Because the physical density is positive and exhibits a substantial
dynamic range across the compressed and uncompressed regions, the
learning target is the standardized logarithmic density
\begin{equation}
    z_{\rho}
    =
    \frac{\log\rho-\mu_{\log\rho}}
         {\sigma_{\log\rho}},
    \label{eq:waverider_logrho}
\end{equation}
where $\mu_{\log\rho}$ and $\sigma_{\log\rho}$ are evaluated using the
training cases only.
 
The central idea is to augment the physical coordinates with a learned
coordinate measuring position relative to the moving shock. Let
$\Gamma_s(\alpha)$ denote the shock surface for angle of attack
$\alpha$. From the CFD training data we construct a smooth,
normalized signed-distance coordinate
\begin{equation}
    q(\bm{x};\alpha)
    =
    \frac{d(\bm{x},\Gamma_s(\alpha))}{\delta},
    \label{eq:waverider_q}
\end{equation}
where $\delta$ is a characteristic shock-region length scale and the
sign distinguishes the two sides of the shock. A separate geometry
network, denoted here as ShockNet, learns
\begin{equation}
    (\alpha,\bm{x})
    \longmapsto
    \widehat q(\bm{x};\alpha).
\end{equation}
ShockNet is subsequently frozen, and its prediction is transformed into
the bounded coordinate
\begin{equation}
    s(\bm{x};\alpha)
    =
    \tanh\!\left[\widehat q(\bm{x};\alpha)\right].
    \label{eq:waverider_s}
\end{equation}
The important consequence is that the physical shock, although moving
in $(x,y,z)$ as $\alpha$ varies, remains associated with
$s\simeq0$. Thus, instead of requiring the trunk basis to represent a
different discontinuity location for every flow condition, the
operator is supplied with an approximately shock-attached coordinate.
 
The resulting surrogate retains the standard DeepONet
branch--trunk structure. The branch network encodes the global
operating parameter,
\begin{equation}
    \alpha
    \longmapsto
    \bm{b}(\alpha)
    =
    [b_1(\alpha),\ldots,b_p(\alpha)]^{T},
\end{equation}
whereas the trunk network receives the local shock-aligned coordinates
\begin{equation}
    \bm{\xi}
    =
    \left(x,y,z,s\right)
    \longmapsto
    \bm{t}(\bm{\xi})
    =
    [t_1(\bm{\xi}),\ldots,t_p(\bm{\xi})]^{T}.
\end{equation}
The standardized log-density prediction is given by the DeepONet
contraction
\begin{equation}
    \widehat z_{R1}(\bm{x};\alpha)
    =
    \sum_{k=1}^{p}
    b_k(\alpha)t_k(\bm{\xi}),
    \label{eq:waverider_deeponet}
\end{equation}
and the physical density follows from
\begin{equation}
    \widehat\rho_{R1}
    =
    \exp\!\left(
        \sigma_{\log\rho}\widehat z_{R1}
        +\mu_{\log\rho}
    \right).
    \label{eq:waverider_inverse_transform}
\end{equation}
 
Training is performed at the level of complete angle-of-attack cases,
rather than by randomly mixing points from all conditions into the
training and validation sets. In the development configuration used
here, 27 AoA cases are used for training and the
$\alpha=-6^\circ$ case is reserved for validation. Each optimization
batch combines samples from the global flow field with additional
samples from the shock core and its surrounding band. The density
objective is therefore written schematically as
\begin{equation}
    \mathcal{L}_{R1}
    =
    \lambda_{\mathrm{g}}\mathcal{L}_{\mathrm{global}}
    +
    \lambda_{\mathrm{c}}\mathcal{L}_{\mathrm{core}}
    +
    \lambda_{\mathrm{b}}\mathcal{L}_{\mathrm{band}},
    \label{eq:waverider_r1_loss}
\end{equation}
where the additional regional terms prevent the relatively small
shock-containing portion of the mesh from being overwhelmed by the
smooth background. This weighting does not impose an additional
physical model; it simply reallocates optimization effort toward the
most difficult part of the three-dimensional solution.
 
\subsection{Hotspot-aware residual correction}
\label{sec:hotspot_moe_method}
 
Although shock alignment substantially improves the representation of
the moving discontinuity, the remaining Stage-R1 error is highly
localized rather than uniformly distributed throughout the domain. We
therefore freeze the Stage-R1 surrogate and define the density residual
on the training cases as
\begin{equation}
    r(\bm{x};\alpha)
    =
    \rho_{\mathrm{CFD}}(\bm{x};\alpha)
    -
    \widehat\rho_{R1}(\bm{x};\alpha).
\end{equation}
Nodes in the upper tail of the residual magnitude distribution are
labelled as hotspots; in the present implementation, the training
hotspot target corresponds to residuals above the 95th percentile.
A classifier is then trained to predict the hotspot probability using
only quantities available during inference,
\begin{equation}
    p_h
    =
    \mathcal{H}
    \left(
       \alpha,\bm{x},
       \tanh\widehat q,
       \widehat z_{R1}
    \right).
    \label{eq:hotspot_probability}
\end{equation}
Consequently, the hotspot definition may use the CFD residual during
offline training, whereas no CFD target value is required when the
surrogate is deployed.
 
The predicted hotspot probability is supplied to a gated
mixture-of-experts residual model \cite{Jacobs1991}. For four experts,
the correction can be written as
\begin{equation}
    \Delta z_{\mathrm{MoE}}
    =
    \sum_{m=1}^{4}
    \pi_m(\bm{\eta})\,
    \Delta z_m(\bm{\eta}),
    \qquad
    \sum_{m=1}^{4}\pi_m=1,
    \label{eq:waverider_moe}
\end{equation}
where $\bm{\eta}$ contains the physical coordinates, angle of attack,
frozen shock coordinate, Stage-R1 prediction, and learned hotspot
probability. The final standardized log-density prediction becomes
\begin{equation}
    \widehat z_{\mathrm{final}}
    =
    \widehat z_{R1}
    +
    \Delta z_{\mathrm{MoE}}.
    \label{eq:waverider_final_prediction}
\end{equation}
This construction preserves the global operator learned by Stage-R1
while allocating additional model capacity specifically to regions in
which its residual remains difficult.
 
\subsection{Held-out volume reconstruction at $\alpha=-6^\circ$}
\label{sec:volume_validation_m6}
 
\begin{figure}[tbp]
  \centering
  \begin{minipage}[c]{0.27\linewidth}
    \begin{subfigure}[b]{\linewidth}
      \centering
      \includegraphics[width=\linewidth]{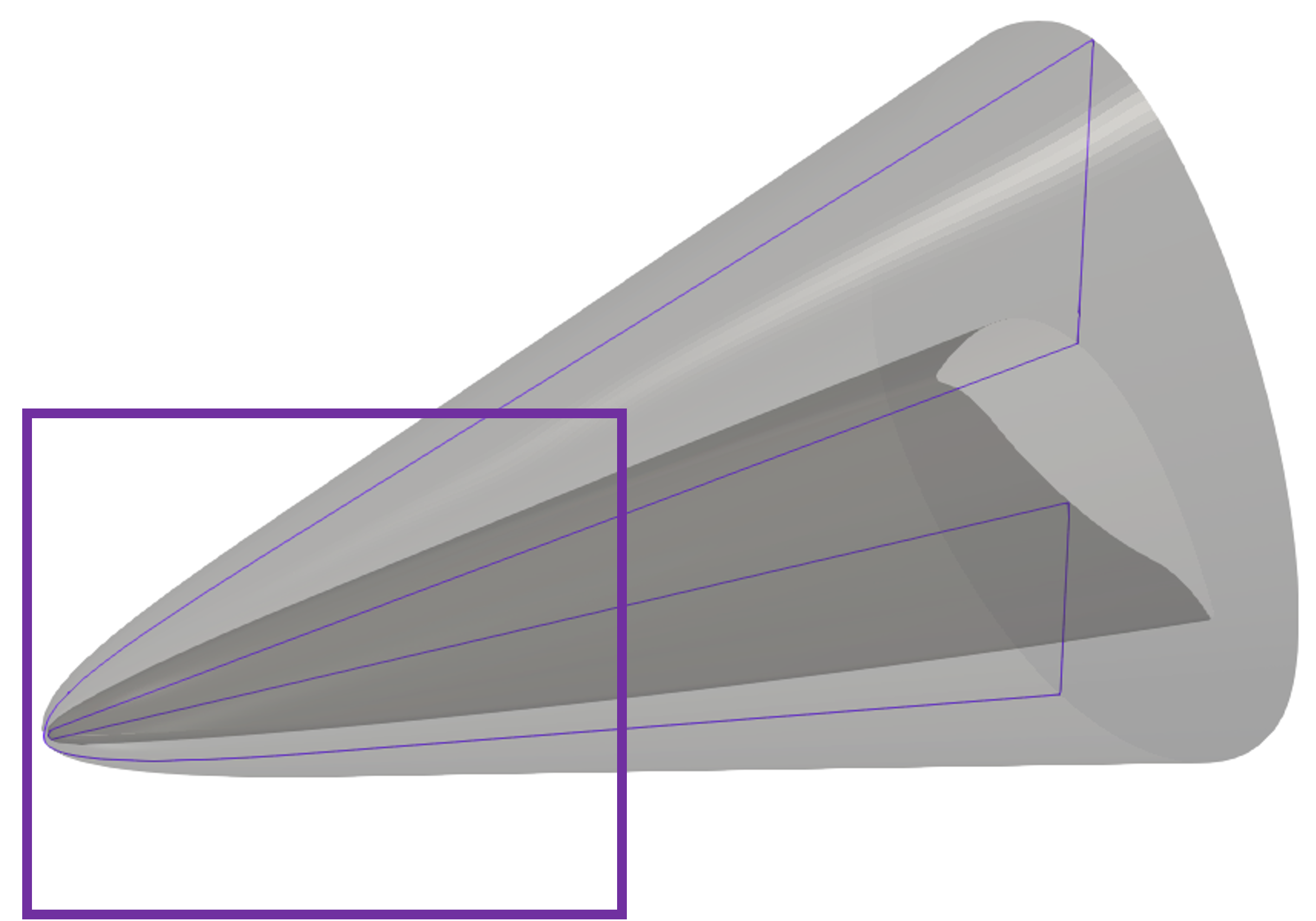}
      \caption{AEDC waverider with the pitch plane outlined in blue; the
      purple box marks the region shown in (b) and (c).}
      \label{fig:volume_locator}
    \end{subfigure}\\[0.6em]
    \begin{subfigure}[b]{\linewidth}
      \centering
      \includegraphics[width=0.60\linewidth]{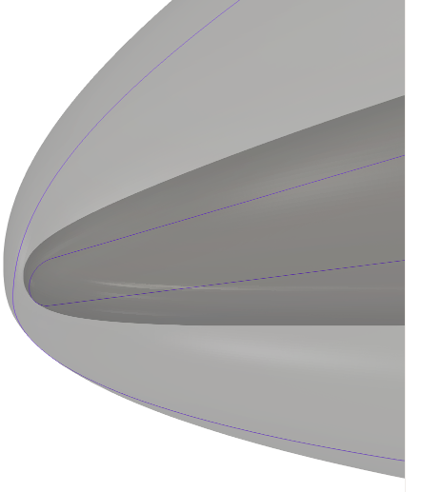}
      \caption{Nose detail of the boxed region.}
      \label{fig:volume_nose_zoom}
    \end{subfigure}
  \end{minipage}\hfill
  \begin{subfigure}[c]{0.70\linewidth}
    \centering
    \includegraphics[width=\linewidth]{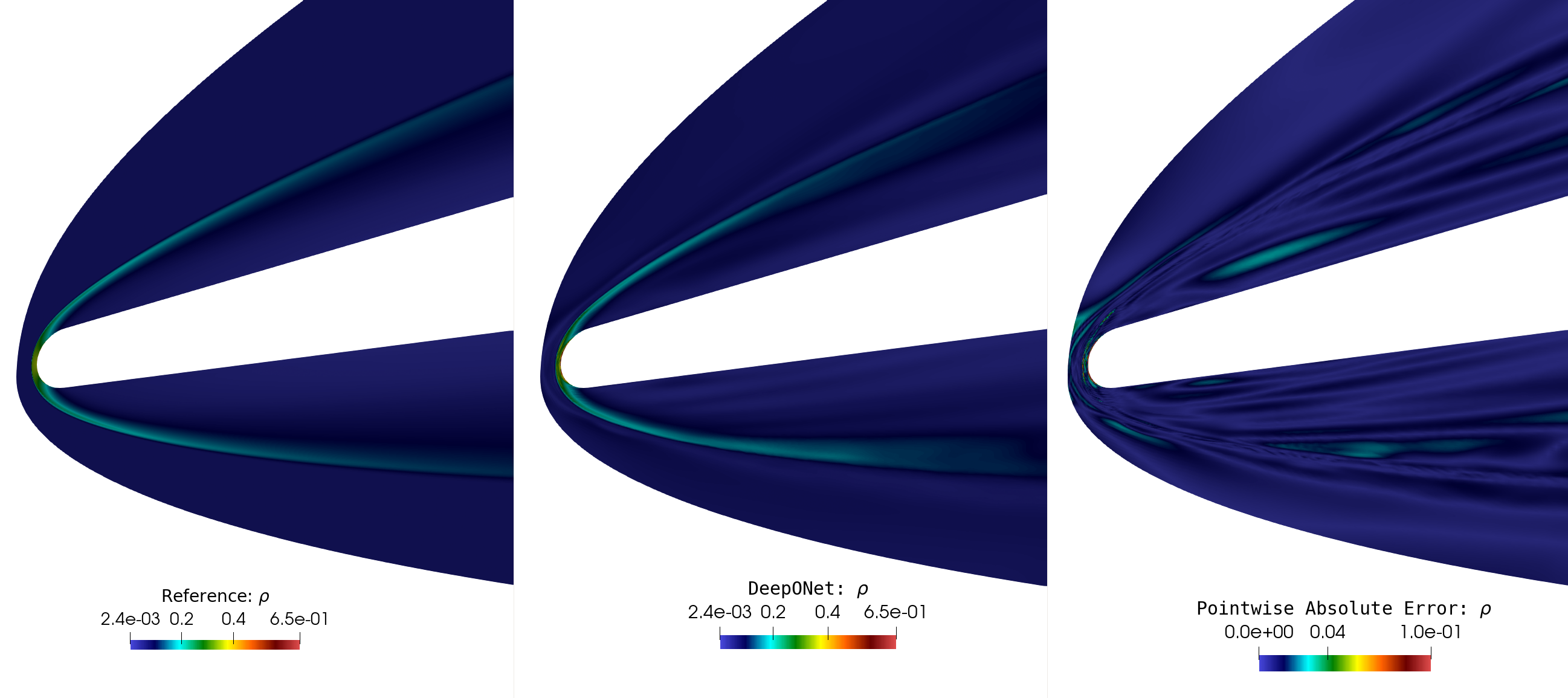}
    \caption{Density on the pitch plane over the boxed region. Left: CFD
    reference. Centre: DeepONet prediction, on the common scale
    $\rho\in[2.4\times10^{-3},\,6.5\times10^{-1}]$. Right:
    pointwise absolute error. The body is masked
    in white.}
    \label{fig:volume_density_zoom}
  \end{subfigure}
    \caption[Volume density field, shock-aligned DeepONet]%
  {Density prediction for the AEDC waverider at the held-out angle of attack
  $\alpha=-6^\circ$, excluded from training of both stages. The shock-aligned
  DeepONet (\Cref{sec:shock_deeponet_method}) with the hotspot-aware residual
  correction (\Cref{sec:hotspot_moe_method}) reproduces the thin, curved
  density layer at the leading edge at the correct position and inclination;
  the residual error is concentrated at the nose and in coherent striations
  trailing along both surfaces, and the relative $L_2$ error over the full
  three-dimensional field is $12.27\%$.}
  \label{fig:volume_density}
\end{figure}

The volume surrogate is assessed at the held-out angle of attack
$\alpha=-6^\circ$, which is excluded from training and used only for
validation. \Cref{fig:volume_locator} identifies the pitch plane used for the
comparison, while \Cref{fig:volume_nose_zoom} highlights the nose region of
interest. The three panels in \Cref{fig:volume_density_zoom} show the density
field over the boxed region of the pitch plane. The left panel presents the
CFD reference solution, the center panel shows the DeepONet prediction, and
the right panel displays the corresponding pointwise absolute error. The predicted field reproduces the thin, strongly curved
density layer that forms at the leading edge and then trails along the upper
and lower surfaces, resolving it as a layer of finite thickness rather than
smearing it across the intervening cells---the characteristic failure mode of
a smooth global basis when a moving discontinuity must be represented between
training conditions. Three variants are compared on the full three-dimensional field at
$\alpha=-6^\circ$, which together isolate what each stage of the construction
contributes. The frozen stage-one surrogate alone gives
$E_{L_2}^{R1}=14.45\%$; adding the unconstrained hotspot-aware
mixture-of-experts correction gives $E_{L_2}^{\rm MoE}=11.53\%$; and the
shock-protected hybrid, in which the correction is suppressed within the
shock region so that the stage-one representation is retained there, gives
$E_{L_2}^{\rm hybrid}=12.27\%$. The residual correction is therefore
responsible for the greater part of the improvement over stage one.
 
The spatial distribution of the error is as informative as its magnitude, and
it is clearly nonuniform. The largest residuals occur in a compact region near
the nose, where the density gradient is strongest and the stagnation streamline
meets the body, and along the leading-edge shock itself. Away from these
regions, the error does not decay to a structureless background but instead
organizes into weaker coherent striations that follow the upper and lower
surfaces downstream. Although their amplitude is considerably smaller than the
global peak error of approximately $1.0\times10^{-1}$ shown by the contour
scale, their spatial coherence is important. An error field that remains
correlated with the body surface does not cancel under spatial integration as
efficiently as an uncorrelated residual and may therefore contribute
systematically to derived or surface-integrated quantities. This distinction is
consistent with the broader separation between random and systematic near-wall
error discussed in \Cref{sec:intro}. 
 
Consequently, the solution close to the shock retains the Stage-R1 representation, whereas the learned hotspot-aware correction remains active in the downstream and far-from-shock regions in which structured residual error persists. The remaining striations are therefore unlikely to be attributable solely to shock-position error. Their persistence downstream suggests a
structured post-shock or near-surface contribution rather than random background noise. A more detailed distinction between expansion-wave characteristics, mesh-induced structure, and mixture-of-experts partitioning would require a dedicated directional analysis and is therefore not inferred from the density contour alone.
 
\subsection{Extension beyond density}
\label{sec:waverider_multifield_extension}
 
Although the results reported here are for density, the construction is not
density-specific. Its essential component is the learned shock-attached
coordinate, which describes the geometry of the moving discontinuity
independently of the output variable. Once that coordinate is available an
analogous operator follows for pressure, temperature or any velocity
component, either through independent branch--trunk pairs or through a shared
latent representation with variable-specific output heads. Density is used
here to establish that the representation resolves the most prominent moving
discontinuity in the flow before the framework is extended to the remaining
thermodynamic and velocity fields; because the shock coordinate is computed
once and can be shared, the marginal cost of each additional field is that of
a branch--trunk pair rather than of the full pipeline.

\section{Summary and Conclusions}
\label{sec:conclusions}
 
This work has developed DeepONet-based operator surrogates for two hypersonic
configurations and used them to carry out multi-objective design studies at a
sampling density that direct simulation could not reach. The two
configurations were chosen because they fail in different ways under a naive
construction, and the thread connecting them is a single device: the trunk is
supplied with a coordinate system in which the feature that varies between
cases is held fixed, leaving the branch to carry only the parametric
dependence that remains.
 
\paragraph{Three surrogates.}
This work constructs three operator surrogates. The first spans the
two-dimensional spherically blunted tangent-ogive family, in which the
geometry and the mesh change with the shape parameter and the canonical
mapping is what makes a single operator applicable across them. The second
and third both address the three-dimensional AEDC waverider, for which the
geometry and the mesh are common to all $29$ incidences and the difficulty is
instead that the leading-edge shock moves: one predicts the surface
quantities on the wall, the other the volume field. The ogive surrogate and
the waverider surface surrogate are the two that drive the design studies
reported below; the volume surrogate is a field-prediction result and does
not enter them.
 
\paragraph{Fixed geometry, moving shock: the waverider surface surrogate.}
The surface loads required no special treatment. The same two-step
construction applied directly to the wall heat flux and the traction
components recovers the held-out fields to $1.1$--$2.0\%$ on average, and the
integrated loads to $|\Delta C_D| \le 1.45\times10^{-5}$,
$|\Delta C_L| \le 2.58\times10^{-5}$ and $0.22\%$ in the wall heat load.
Notably, withholding an incidence costs essentially nothing at field level
--- the leave-one-out and all-angle errors agree to within a tenth of a
percentage point --- because the surface fields vary smoothly with incidence
and the remaining $28$ angles determine the basis almost as well as all $29$.
It is this accuracy, on the quantities that define the objectives, that
licenses the use of the surrogate in place of the solver for the waverider
design study.
 
\paragraph{The waverider volume surrogate.}
The volume field is where the baseline construction proved insufficient. A
trunk expressed in Cartesian coordinates alone must represent the same
discontinuity at a different place for every incidence, and in practice
averages neighbouring shock positions into a smeared front. Two additions
were made in response: a frozen ShockNet supplying a shock-aligned coordinate
that places the discontinuity at approximately the same value for every
angle, and a hotspot-aware mixture of experts fitted to the residual that
remains, which is localised rather than uniform. Together these recover the
density field to a relative $L_2$ error of $12.27\%$ at a held-out incidence,
on a reacting three-dimensional field of $5.08\times10^{7}$ nodes, against
$14.45\%$ for the stage-one surrogate without the residual correction; the
error analysis attributes what remains to coherent surface-aligned striations
downstream of the shock rather than to the representation of the shock
itself. That the surface surrogate needs none of this machinery and the
volume surrogate does not converge without it is the clearest evidence in
this work of where the extra construction earns its place.
 
\paragraph{Design studies.}
The ogive surrogate and the waverider surface surrogate were used in place of
repeated flow solutions. For the tangent ogive, a sweep of $1001$ candidate
geometries in the three objectives $C_D$, $q_{w,\max}$ and $\dot Q'$ returned
$200$ non-dominated designs from $591$ feasible ones, with a compromise
design at $\xi_g^\star = 0.825$. The complete sweep took approximately three
minutes on a single CPU, about $0.18$~s per candidate, against $20.7$
hours of wall-clock time on $400$ cores, or $8.3\times10^{3}$ core-hours,
for one converged reference solution. For the waverider, a $561$-angle scan at
$\Delta\alpha = 0.05^\circ$ gave a Pareto set spanning
$-6.20^\circ \le \alpha \le +14.00^\circ$, within which compromise
programming with equal weights selects $\alpha^\star = +4.10^\circ$ --- an
attitude giving $29\%$ more lift than the maximum-$L/D$ design for a $9\%$
higher thermal load, at a cost of only $2.4\%$ in $L/D$. Repeating the
selection with each of the seven leave-one-out surrogates moves
$\alpha^\star$ by at most two scan increments, so the recommendation does not
rest on any single CFD snapshot. Both design studies rest on surface and
integrated quantities throughout; the volume surrogate is not used to select
a design.
 
\paragraph{Cost.}
The three surrogates replace calculations of very different size, and it is
worth setting the online cost of each against the high-fidelity solution it
stands in for (\Cref{tab:cost_comparison}). A converged tangent-ogive
solution takes $20.7$ hours of wall-clock time on $400$ MPI tasks,
$8.3\times10^{3}$ core-hours; a converged waverider solution, on which both
the surface and the volume surrogates are trained, takes $20$ hours on
$1024$ cores, $2.05\times10^{4}$ core-hours. Against these, a surrogate
evaluation of the surface and integrated quantities at a new design point
costs a fraction of a second, and a reconstruction of the complete
$5.08\times10^{7}$-node volume field some four minutes. The
consequence is not that individual solutions are obtained more cheaply ---
the database had to be computed once, at no more than $3.1\times10^{5}$
core-hours for the ogive family and $5.9\times10^{5}$ for the waverider ---
but that the design space can be swept rather than sampled. Evaluating the
$1001$ ogive candidates by CFD would have required $8.3\times10^{6}$
core-hours, and the $561$-angle waverider scan $1.2\times10^{7}$; run end to
end on the same machines, each would take longer than two years and longer
than one year respectively. Both completed in minutes. That is the
difference between choosing a design from among the few geometries one can
afford to simulate and choosing it from a resolved Pareto front, and it is
the practical case for surrogate-driven hypersonic design.

\begin{table}[htbp]
  \centering
  \caption{Cost of a high-fidelity solution against the cost of a surrogate
  evaluation, for each of the three surrogates. Core-hours are wall-clock
  time multiplied by the number of cores. The ogive solutions were run with
  Eilmer on $400$ MPI tasks; the waverider solutions with US3D on the AFRL
  Warhawk system, $8$ standard nodes and $1024$ cores. Surrogate evaluation
  is on a single CPU for the surface and integrated quantities and on one
  GPU for the volume field.}
  \label{tab:cost_comparison}
  \small
  \begin{tabular}{lccc}
    \toprule
    & Ogive & Waverider surface & Waverider volume \\
    \midrule
    \multicolumn{4}{l}{\emph{High-fidelity solution, one case}}\\
    Wall-clock time  & $20.7$~h & $20$~h & $20$~h \\
    Cores            & $400$ & $1024$ & $1024$ \\
    Cost             & $8.3\times10^{3}$~core-h & $2.05\times10^{4}$~core-h
                     & $2.05\times10^{4}$~core-h \\
    \addlinespace
    \multicolumn{4}{l}{\emph{Surrogate evaluation, one design point}}\\
    Points evaluated & $1.8\times10^{5}$ & $2.7\times10^{5}$ & $5.08\times10^{7}$ \\
    Fields           & $14$ & $4$ & $1$ \\
    Wall-clock time  & $0.18$~s & $0.39$~s & $240$~s \\
    Speed-up         & $4.1\times10^{5}$ & $1.8\times10^{5}$ & $3.0\times10^{2}$ \\
    \addlinespace
    \multicolumn{4}{l}{\emph{Design study}}\\
    Design points    & $1001$ & $561$ & --- \\
    Surrogate scan   & $180$~s & $220$~s & --- \\
    Same scan by CFD & $8.3\times10^{6}$~core-h & $1.2\times10^{7}$~core-h & --- \\
    \bottomrule
  \end{tabular}
\end{table}

\paragraph{Limitations.}
Four qualifications bear on how these results should be used. First, the peak
wall heat flux is the least reliable quantity in both design studies and
should not be treated as a design objective on this evidence: for the ogive
family its coefficient of determination about the line of equality is
$0.2597$, meaning the surrogate resolves about a quarter of the
geometry-to-geometry variation, and for the waverider the surrogate error
reaches a quarter of the $4\%$ physical variation while failing to follow the
cell-to-cell structure that makes up much of it. The integrated heat load is
free of both difficulties and is the thermal objective used throughout.
Second, the zero-incidence lift constraint imposed in the ogive study is a
trust region on the surrogate rather than a physical requirement: the
reference lift is a discretisation residual, so the constraint excludes
designs on which the surrogate is internally inconsistent, not designs that
are aerodynamically unacceptable. Third, the ogive sweep spans
$0.50 \le \xi_g \le 1.00$ while training covered
$0.50 \le \xi_g \le 0.95$, so the uppermost designs are extrapolations; the
selected compromise is interior to the trained interval and unaffected, but
results quoted at the extreme end should be read with that in mind. Fourth,
the waverider volume surrogate covers the density field alone; the
construction is not density-specific, but its extension to pressure,
temperature and velocity remains to be demonstrated.
 
\paragraph{Outlook.}
Three directions follow naturally. Imposing the geometric symmetry on the
ogive surrogate by construction --- predicting on a half-body and mirroring
--- would remove the residual-lift diagnostic and the trust-region constraint
together, and would recover the $41\%$ of candidates that constraint
presently excludes. Extending the shock-aligned representation of the volume
surrogate to the remaining thermodynamic and velocity fields, through either
independent branch--trunk pairs or a shared latent representation with
variable-specific heads, would complete it, and reducing the surface-aligned
striations identified above is the accuracy increment that matters most
there. And the designs on both fronts remain to be confirmed with the solver:
for the waverider that is a few tens of calculations in place of the $561$
the scan evaluated, which is the practical form in which a surrogate of this
accuracy is best used --- to decide where the expensive calculations should
be spent, rather than to replace them.

\section*{Acknowledgments} This work was supported by DEVCOM Army Research Laboratory cooperative agreement W911NF-24-2-0150. K.~Shukla and G.~E.~Karniadakis acknowledge partial support from the U.S. Department of Energy, Office of Science, Advanced Scientific Computing Research program, through the Scientific Discovery through Advanced Computing (SciDAC) Institute ``LEADS: LEarning-Accelerated Domain Science'' (Subcontract 831126 under DE-AC05-76RL01830).G.~E.~Karniadakis further acknowledges support from the ONR Vannevar Bush Faculty Fellowship (N00014-22-1-2795). The authors gratefully acknowledge the High-Performance Computing Modernization Program (HPCMP) resources and support provided by the Department of War Supercomputing Resource Center (DSRC) as part of the 2022 Frontier Project, \emph{Large-Scale Integrated Simulations of Transient Aerothermodynamics in Gas Turbine Engines}. 

\section*{Code and Data Availability}
The surrogate code, trained models and post-processing routines will be available
upon acceptance. The high-fidelity CFD databases are available from
the corresponding author on reasonable request.
 
\newpage
\bibliographystyle{unsrtnat}
\bibliography{ref}

@techreport{gnoffo1989nonequilibrium,
  author      = {Gnoffo, Peter A. and Gupta, Roop N. and Shinn, Judy L.},
  title       = {Conservation Equations and Physical Models for
                 Hypersonic Air Flows in Thermal and Chemical Nonequilibrium},
  institution = {NASA Langley Research Center},
  number      = {NASA-TP-2867},
  address     = {Hampton, Virginia},
  year        = {1989},
  month       = feb
}

@article{passiatore2022thermochemical,
  author  = {Passiatore, D. and Sciacovelli, L. and
             Cinnella, P. and Pascazio, G.},
  title   = {Thermochemical Non-Equilibrium Effects in
             Turbulent Hypersonic Boundary Layers},
  journal = {Journal of Fluid Mechanics},
  volume  = {941},
  pages   = {A21},
  year    = {2022},
  doi     = {10.1017/jfm.2022.283}
}

@techreport{sutton1971,
  author      = {Sutton, Kenneth and Graves, Randolph A., Jr.},
  title       = {A General Stagnation-Point Convective-Heating Equation
                 for Arbitrary Gas Mixtures},
  institution = {NASA Langley Research Center},
  number      = {NASA-TR-R-376},
  address     = {Hampton, Virginia},
  year        = {1971},
  month       = nov
}

@article{seager2017shape,
  author  = {Seager, Christopher and Agarwal, Ramesh K.},
  title   = {Hypersonic Blunt-Body Shape Optimization for
             Reducing Drag and Heat Transfer},
  journal = {Journal of Thermophysics and Heat Transfer},
  volume  = {31},
  number  = {1},
  pages   = {48--55},
  year    = {2017},
  doi     = {10.2514/1.T4650}
}

@article{wu2019efficient,
  author  = {Wu, Hao and Ma, Peter C. and Ihme, Matthias},
  title   = {Efficient Time-Stepping Techniques for Simulating
             Turbulent Reactive Flows with Stiff Chemistry},
  journal = {Computer Physics Communications},
  volume  = {243},
  pages   = {81--96},
  year    = {2019},
  doi     = {10.1016/j.cpc.2019.04.016}
}

@article{scoggins2020mutation++,
  title={Mutation++: Multicomponent thermodynamic and transport properties for ionized gases in {C++}},
  author={Scoggins, James B and Leroy, Vincent and Bellas-Chatzigeorgis, Georgios and Dias, Bruno and Magin, Thierry E},
  journal={SoftwareX},
  volume={12},
  pages={100575},
  year={2020},
  publisher={Elsevier}
}

@book{anderson2006hypersonic,
  author    = {Anderson, John D.},
  title     = {Hypersonic and High-Temperature Gas Dynamics},
  edition   = {2nd},
  publisher = {American Institute of Aeronautics and Astronautics},
  address   = {Reston, VA},
  year      = {2006},
  doi       = {10.2514/4.861956}
}

@article{forrester2009recent,
  author  = {Forrester, Alexander I. J. and Keane, Andy J.},
  title   = {Recent Advances in Surrogate-Based Optimization},
  journal = {Progress in Aerospace Sciences},
  volume  = {45},
  number  = {1--3},
  pages   = {50--79},
  year    = {2009},
  doi     = {10.1016/j.paerosci.2008.11.001}
}

@article{kovachki2023neural,
  author  = {Kovachki, Nikola and Li, Zongyi and Liu, Burigede and Azizzadenesheli, Kamyar and Bhattacharya, Kaushik and Stuart, Andrew and Anandkumar, Anima},
  title   = {Neural Operator: Learning Maps Between Function Spaces with Applications to {PDEs}},
  journal = {Journal of Machine Learning Research},
  volume  = {24},
  number  = {89},
  pages   = {1--97},
  year    = {2023}
}

@article{mao2021deepmmnet,
  author  = {Mao, Zhiping and Lu, Lu and Marxen, Olaf and Zaki, Tamer A. and Karniadakis, George Em},
  title   = {{DeepM\&Mnet} for Hypersonics: Predicting the Coupled Flow and Finite-Rate Chemistry Behind a Normal Shock Using Neural-Network Approximation of Operators},
  journal = {Journal of Computational Physics},
  volume  = {447},
  pages   = {110698},
  year    = {2021},
  doi     = {10.1016/j.jcp.2021.110698}
}

@inproceedings{li2021fno,
  author    = {Li, Zongyi and Kovachki, Nikola and Azizzadenesheli, Kamyar and Liu, Burigede and Bhattacharya, Kaushik and Stuart, Andrew and Anandkumar, Anima},
  title     = {Fourier Neural Operator for Parametric Partial Differential Equations},
  booktitle = {International Conference on Learning Representations (ICLR)},
  year      = {2021}
}

@inproceedings{rahaman2019spectral,
  author    = {Rahaman, Nasim and Baratin, Aristide and Arpit, Devansh and Draxler, Felix and Lin, Min and Hamprecht, Fred and Bengio, Yoshua and Courville, Aaron},
  title     = {On the Spectral Bias of Neural Networks},
  booktitle = {Proceedings of the 36th International Conference on Machine Learning},
  series    = {Proceedings of Machine Learning Research},
  volume    = {97},
  pages     = {5301--5310},
  year      = {2019}
}

@article{nonweiler1959aerodynamic,
  author  = {Nonweiler, T. R. F.},
  title   = {Aerodynamic Problems of Manned Space Vehicles},
  journal = {Journal of the Royal Aeronautical Society},
  volume  = {63},
  number  = {585},
  pages   = {521--528},
  year    = {1959},
  doi     = {10.1017/S0368393100071662}
}

@article{jacobs1991adaptive,
  author  = {Jacobs, Robert A. and Jordan, Michael I. and Nowlan, Steven J. and Hinton, Geoffrey E.},
  title   = {Adaptive Mixtures of Local Experts},
  journal = {Neural Computation},
  volume  = {3},
  number  = {1},
  pages   = {79--87},
  year    = {1991},
  doi     = {10.1162/neco.1991.3.1.79}
}

@article{deb2002nsga2,
  author  = {Deb, Kalyanmoy and Pratap, Amrit and Agarwal, Sameer and Meyarivan, T.},
  title   = {A Fast and Elitist Multiobjective Genetic Algorithm: {NSGA-II}},
  journal = {IEEE Transactions on Evolutionary Computation},
  volume  = {6},
  number  = {2},
  pages   = {182--197},
  year    = {2002},
  doi     = {10.1109/4235.996017}
}

@book{karniadakis2013spectral,
  title={Spectral/hp element methods for computational fluid dynamics},
  author={Karniadakis, George and Sherwin, Spencer},
  year={2013},
  publisher={American Chemical Society}
}

@article{lu2021deeponet,
  author  = {Lu, Lu and Jin, Pengzhan and Pang, Guofei and
             Zhang, Zhongqiang and Karniadakis, George Em},
  title   = {Learning Nonlinear Operators via {DeepONet} Based on the
             Universal Approximation Theorem of Operators},
  journal = {Nature Machine Intelligence},
  volume  = {3},
  pages   = {218--229},
  year    = {2021},
  doi     = {10.1038/s42256-021-00302-5}
}

@article{lee2024training,
  author  = {Lee, Sanghyun and Shin, Yeonjong},
  title   = {On the Training and Generalization of Deep Operator Networks},
  journal = {SIAM Journal on Scientific Computing},
  volume  = {46},
  number  = {4},
  pages   = {C273--C296},
  year    = {2024},
  doi     = {10.1137/23M1598751}
}

@article{shukla2024deepII,
  title={Deep neural operators as accurate surrogates for shape optimization},
  author={Shukla, Khemraj and Oommen, Vivek and Peyvan, Ahmad and Penwarden, Michael and Plewacki, Nicholas and Bravo, Luis and Ghoshal, Anindya and Kirby, Robert M and Karniadakis, George Em},
  journal={Engineering Applications of Artificial Intelligence},
  volume={129},
  pages={107615},
  year={2024},
  publisher={Elsevier}
}

@article{dasdennis1997,
  author  = {Das, Indraneel and Dennis, J. E.},
  title   = {A closer look at drawbacks of minimizing weighted sums of
             objectives for {P}areto set generation in multicriteria
             optimization problems},
  journal = {Structural Optimization},
  volume  = {14},
  number  = {1},
  pages   = {63--69},
  year    = {1997},
  doi     = {10.1007/BF01197559},
}

@article{marlerarora2010,
  author  = {Marler, R. Timothy and Arora, Jasbir S.},
  title   = {The weighted sum method for multi-objective optimization:
             new insights},
  journal = {Structural and Multidisciplinary Optimization},
  volume  = {41},
  number  = {6},
  pages   = {853--862},
  year    = {2010},
  doi     = {10.1007/s00158-009-0460-7},
}

@incollection{bowman1976,
  author    = {Bowman, V. Joseph},
  title     = {On the relationship of the {T}chebycheff norm and the
               efficient frontier of multiple-criteria objectives},
  booktitle = {Multiple Criteria Decision Making},
  editor    = {Thiriez, Herv{\'e} and Zionts, Stanley},
  series    = {Lecture Notes in Economics and Mathematical Systems},
  volume    = {130},
  publisher = {Springer},
  address   = {Berlin},
  pages     = {76--86},
  year      = {1976},
}

@article{steuerchoo1983,
  author  = {Steuer, Ralph E. and Choo, Eng-Ung},
  title   = {An interactive weighted {T}chebycheff procedure for multiple
             objective programming},
  journal = {Mathematical Programming},
  volume  = {26},
  number  = {3},
  pages   = {326--344},
  year    = {1983},
  doi     = {10.1007/BF02591870},
}

@book{miettinen1999,
  author    = {Miettinen, Kaisa},
  title     = {Nonlinear Multiobjective Optimization},
  series    = {International Series in Operations Research \&
               Management Science},
  volume    = {12},
  publisher = {Kluwer Academic Publishers},
  address   = {Boston, MA},
  year      = {1999},
}

@book{deb2001,
  author    = {Deb, Kalyanmoy},
  title     = {Multi-Objective Optimization Using Evolutionary Algorithms},
  publisher = {John Wiley \& Sons},
  address   = {Chichester},
  year      = {2001},
}

@book{chankong1983,
  author    = {Chankong, Vira and Haimes, Yacov Y.},
  title     = {Multiobjective Decision Making: Theory and Methodology},
  publisher = {North-Holland},
  address   = {New York},
  year      = {1983},
}

@article{Jacobs1991,
  author  = {Jacobs, Robert A. and Jordan, Michael I. and Nowlan, Steven J.
             and Hinton, Geoffrey E.},
  title   = {Adaptive mixtures of local experts},
  journal = {Neural Computation},
  volume  = {3},
  number  = {1},
  pages   = {79--87},
  year    = {1991},
  doi     = {10.1162/neco.1991.3.1.79},
}

@article{li2023geofno,
  author  = {Li, Zongyi and Huang, Daniel Zhengyu and
             Liu, Burigede and Anandkumar, Anima},
  title   = {Fourier Neural Operator with Learned Deformations
             for {PDE}s on General Geometries},
  journal = {Journal of Machine Learning Research},
  volume  = {24},
  number  = {388},
  pages   = {1--26},
  year    = {2023}
}

@inproceedings{liu_schiaffini2024localized,
  author    = {Liu-Schiaffini, Miguel and Berner, Julius and
               Bonev, Boris and Kurth, Thorsten and
               Azizzadenesheli, Kamyar and Anandkumar, Anima},
  title     = {Neural Operators with Localized Integral and
               Differential Kernels},
  booktitle = {Proceedings of the 41st International Conference
               on Machine Learning},
  series    = {Proceedings of Machine Learning Research},
  volume    = {235},
  pages     = {32576--32594},
  publisher = {PMLR},
  year      = {2024}
}

@inproceedings{lanthaler2023discontinuities,
  author    = {Lanthaler, Samuel and Molinaro, Roberto and
               Hadorn, Patrik and Mishra, Siddhartha},
  title     = {Nonlinear Reconstruction for Operator Learning
               of {PDE}s with Discontinuities},
  booktitle = {The Eleventh International Conference on
               Learning Representations},
  year      = {2023}
}

@article{peyvan2026fusion,
  author  = {Peyvan, Ahmad and Kumar, Varun and
             Karniadakis, George Em},
  title   = {{Fusion-DeepONet}: A Data-Efficient Neural Operator for
             Geometry-Dependent Hypersonic and Supersonic Flows},
  journal = {Journal of Computational Physics},
  volume  = {544},
  pages   = {114432},
  year    = {2026},
  doi     = {10.1016/j.jcp.2025.114432}
}

@inproceedings{wu2024transolver,
  author    = {Wu, Haixu and Luo, Huakun and Wang, Haowen and
               Wang, Jianmin and Long, Mingsheng},
  title     = {Transolver: A Fast Transformer Solver for
               {PDE}s on General Geometries},
  booktitle = {Proceedings of the 41st International Conference
               on Machine Learning},
  series    = {Proceedings of Machine Learning Research},
  volume    = {235},
  pages     = {53681--53705},
  publisher = {PMLR},
  year      = {2024}
}

@misc{shukla2024waverider,
  author        = {Shukla, Khemraj and Ratchford, Jasmine and
                   Bravo, Luis and Oommen, Vivek and
                   Plewacki, Nicholas and Ghoshal, Anindya and
                   Karniadakis, George Em},
  title         = {Deep Operator Learning-Based Surrogate Models for
                   Aerothermodynamic Analysis of {AEDC} Hypersonic Waverider},
  year          = {2024},
  eprint        = {2405.13234},
  archivePrefix = {arXiv},
  primaryClass  = {physics.flu-dyn},
  doi           = {10.48550/arXiv.2405.13234},
  note          = {71st JANNAF Propulsion Meeting,
                   Oklahoma City, Oklahoma, 6--10 May 2024}
}

@techreport{keener1977side,
  title={Side forces on a tangent ogive forebody with a fineness ratio of 3.5 at high angles of attack and mach numbers from 0.1 to 0.7},
  author={Keener, Earl R and Chapman, Gary T and Cohen, Lee and Taleghani, Jamshid},
  year={1977}
}

@techreport{wing1971tangent,
  title={Tangent ogive nose aerodynamic heating program: NQLD019},
  author={Wing, LD},
  year={1971}
}

@article{gibbons2023eilmer,
  title={Eilmer: an open-source multi-physics hypersonic flow solver},
  author={Gibbons, Nicholas N and Damm, Kyle A and Jacobs, Peter A and Gollan, Rowan J},
  journal={Computer Physics Communications},
  volume={282},
  pages={108551},
  year={2023},
  publisher={Elsevier}
}

@article{redding2023thermochemical,
  title={Thermochemical non-equilibrium hypersonic flow over a rectangular cavity embedded on a compression ramp},
  author={Redding, Jeremy and Gamertsfelder, Jacob and Bravo, Luis and Khare, Prashant},
  journal={Physics of Fluids},
  volume={35},
  number={12},
  year={2023},
  publisher={AIP Publishing},
  DOI = {https://doi.org/10.1063/5.0172435},
  url = {https://pubs.aip.org/aip/pof/article/35/12/126101/2925575/Thermochemical-non-equilibrium-hypersonic-flow},
}

@article{redding2025vortex,
  title={Vortex behavior in a dynamically moving hypersonic cavity},
  author={Redding, Jeremy and Khare, Prashant and Bravo, Luis},
  journal={Physics of Fluids},
  volume={37},
  number={5},
  year={2025},
  publisher={AIP Publishing}
}

@article{redding2025investigation,
  title={Investigation of scramjet inlet unstart/restart behavior induced by high angle of attack and dynamic isolator camber motion},
  author={Redding, Jeremy and Cavanaugh, Emma and Bravo, Luis and Murugan, Muthuvel and Narayanaswamy, Venkat},
  journal={Physics of Fluids},
  volume={37},
  number={6},
  year={2025},
  publisher={AIP Publishing}
}

@inproceedings{US3D,
booktitle = {53rd AIAA Aerospace Sciences Meeting},
author = {Candler, Graham and Johnson, Heath and Nompelis, Ioannis and Gidzak, Vladimyr and Subbareddy, Pramod and Barnhardt, Michael},
year = {2015},
month = {01},
title = {Development of the US3D Code for Advanced Compressible and Reacting Flow Simulations},
journal = {53rd AIAA Aerospace Sciences Meeting},
	url = {https://arc.aiaa.org/doi/abs/10.2514/6.2015-1893},
doi = {10.2514/6.2015-1893}
}

@techreport{gupta1990review,
  title={A review of reaction rates and thermodynamic and transport properties for an 11-species air model for chemical and thermal nonequilibrium calculations to 30000 K},
  author={Gupta, Roop N and Yos, Jerrold M and Thompson, Richard A and Lee, Kam-Pui},
  instituion={NASA},
  type={NASA Technical Memorandum},
  number={101528},
  year={1990}
}

@article{fritsch1980monotone,
  author  = {Fritsch, F. N. and Carlson, R. E.},
  title   = {Monotone Piecewise Cubic Interpolation},
  journal = {SIAM Journal on Numerical Analysis},
  volume  = {17},
  number  = {2},
  pages   = {238--246},
  year    = {1980},
  doi     = {10.1137/0717021}
}

@book{forrester2008,
  author    = {Alexander I. J. Forrester and Andr{\'a}s S{\'o}bester and Andy J. Keane},
  title     = {Engineering Design via Surrogate Modelling: A Practical Guide},
  publisher = {John Wiley \& Sons},
  year      = {2008},
  doi       = {10.1002/9780470770801},
  isbn      = {9780470060681}
}

@article{stone1974,
  author  = {M. Stone},
  title   = {Cross-Validatory Choice and Assessment of Statistical Predictions},
  journal = {Journal of the Royal Statistical Society: Series B (Methodological)},
  volume  = {36},
  number  = {2},
  pages   = {111--133},
  year    = {1974},
  doi     = {10.1111/j.2517-6161.1974.tb00994.x}
}

@article{Nonweiler1959,
  author  = {Nonweiler, T. R. F.},
  title   = {Aerodynamic Problems of Manned Space Vehicles},
  journal = {Journal of the Royal Aeronautical Society},
  volume  = {63},
  number  = {585},
  pages   = {521--528},
  year    = {1959},
  doi     = {10.1017/S0368393100071662}
}

@inproceedings{Bowcutt1987,
  author    = {Bowcutt, Kevin G. and Anderson, John D. and Capriotti, Diego},
  title     = {Viscous Optimized Hypersonic Waveriders},
  booktitle = {25th AIAA Aerospace Sciences Meeting},
  year      = {1987},
  note      = {AIAA Paper 87-0272},
  doi       = {10.2514/6.1987-272}
}

@inproceedings{Rahaman2019,
  author    = {Rahaman, Nasim and Baratin, Aristide and Arpit, Devansh and Draxler, Felix and Lin, Min and Hamprecht, Fred and Bengio, Yoshua and Courville, Aaron},
  title     = {On the Spectral Bias of Neural Networks},
  booktitle = {Proceedings of the 36th International Conference on Machine Learning},
  series    = {Proceedings of Machine Learning Research},
  volume    = {97},
  pages     = {5301--5310},
  year      = {2019}
}

@article{Lu2021,
  author  = {Lu, Lu and Jin, Pengzhan and Pang, Guofei and Zhang, Zhongqiang and Karniadakis, George Em},
  title   = {Learning Nonlinear Operators via {DeepONet} Based on the Universal Approximation Theorem of Operators},
  journal = {Nature Machine Intelligence},
  volume  = {3},
  number  = {3},
  pages   = {218--229},
  year    = {2021},
  doi     = {10.1038/s42256-021-00302-5}
}

@inproceedings{Ohlberger2016,
  author    = {Ohlberger, Mario and Rave, Stephan},
  title     = {Reduced Basis Methods: Success, Limitations and Future Challenges},
  booktitle = {Proceedings of the Conference Algoritmy},
  pages     = {1--12},
  year      = {2016},
  eprint    = {1511.02021},
  archivePrefix = {arXiv}
}

@article{Peherstorfer2022,
  author  = {Peherstorfer, Benjamin},
  title   = {Breaking the {Kolmogorov} Barrier with Nonlinear Model Reduction},
  journal = {Notices of the American Mathematical Society},
  volume  = {69},
  number  = {5},
  pages   = {725--733},
  year    = {2022},
  doi     = {10.1090/noti2475}
}

@article{roohi2025shockaware,
  author  = {Roohi, Ehsan and Mahdavi, Amirmehran},
  title   = {Shock-Aware Physics-Guided {Fusion-DeepONet} Operator for
             Rarefied Micro-Nozzle Flows},
  journal = {arXiv preprint arXiv:2510.17887},
  year    = {2025}
}

@article{roohi2026shockcentered,
  author  = {Roohi, Ehsan and Mahdavi, Amirmehran},
  title   = {Shock-Centered Low-Rank Structure and Neural-Operator
             Representation of Rarefied Micro-Nozzle Flows},
  journal = {arXiv preprint arXiv:2605.12723},
  year    = {2026}
}

@article{reiss2018spod,
  author  = {Reiss, Julius and Schulze, Philipp and Sesterhenn, J{\"o}rn and
             Mehrmann, Volker},
  title   = {The Shifted Proper Orthogonal Decomposition: A Mode
             Decomposition for Multiple Transport Phenomena},
  journal = {SIAM Journal on Scientific Computing},
  volume  = {40},
  number  = {3},
  pages   = {A1322--A1344},
  year    = {2018}
}

@article{krah2024robustspod,
  author  = {Krah, Philipp and Marmin, Arthur and Zorawski, Beata and
             Reiss, Julius and Schneider, Kai},
  title   = {A Robust Shifted Proper Orthogonal Decomposition: Proximal
             Methods for Decomposing Flows with Multiple Transports},
  journal = {SIAM Journal on Scientific Computing},
  year    = {2025},
  doi     = {10.1137/24M164392X}
}

@article{gowrachari2025registration,
  author  = {Gowrachari, Harshith and Stabile, Giovanni and Rozza, Gianluigi},
  title   = {Model Reduction for Transport-Dominated Problems via
             Cross-Correlation Based Snapshot Registration},
  journal = {arXiv preprint arXiv:2501.01299},
  year    = {2025}
}

@article{taddei2020registration,
  author  = {Taddei, Tommaso},
  title   = {A Registration Method for Model Order Reduction: Data
             Compression and Geometry Reduction},
  journal = {SIAM Journal on Scientific Computing},
  volume  = {42},
  number  = {2},
  pages   = {A997--A1027},
  year    = {2020}
}

@article{zahr2018shocktracking,
  author  = {Zahr, Matthew J. and Persson, Per-Olof},
  title   = {An Optimization-Based Approach to High-Order Accurate
             Discretization of Conservation Laws with Discontinuous
             Solutions},
  journal = {Journal of Computational Physics},
  volume  = {365},
  pages   = {105--134},
  year    = {2018}
}

@techreport{kammeyer1994waverider,
  author      = {Kammeyer, Mark E. and Gillum, Mark J.},
  title       = {Design Validation Tests on a Realistic Hypersonic Waverider
                 at {Mach} 10, 14, and 16.5 in the {Naval Surface Warfare
                 Center} Hypervelocity Wind Tunnel No.~9},
  institution = {Naval Surface Warfare Center, Dahlgren Division},
  number      = {NSWCDD/TR-93/198},
  year        = {1994}
}

@inproceedings{norris2006mach,
  title={Mach 8 High Reynolds Number Static Stability Capability Extension Using a Hypersonic Waverider at AEDC Tunnel 9},
  author={Norris, Joseph},
  booktitle={14th AIAA/AHI Space Planes and Hypersonic Systems and Technologies Conference},
  pages={2815},
  year={2006},
  doi={10.2514/6.2006-2815},
  series={AIAA Paper 2006-2815}
}

@article{shukla2026fixed,
  title={Fixed and Adaptive Topological DeepONets: Functional Measurements on Hausdorff Locally Convex Spaces},
  author={Shukla, Khemraj and Karniadakis, George Em},
  journal={arXiv preprint arXiv:2608.06428},
  year={2026}
}

@article{de2025deep,
  title={Deep operator neural network model predictive control},
  author={de Jong, Thomas O and Shukla, Khemraj and Lazar, Mircea},
  journal={IEEE Open Journal of Control Systems},
  volume={4},
  pages={501--517},
  year={2025},
  publisher={IEEE}
}

@article{jiang2026complex,
  title={Complex valued deep operator network ({DeepONet}) for three dimensional {Maxwell}'s equations: {$G \in \mathbb{C}^{m \times n}$}},
  author={Jiang, Qile and Salvadori, Marc and Ota, Dale and Shankar, Vijaya and Shukla, Khemraj},
  journal={Journal of Computational Physics},
  pages={114993},
  year={2026},
  publisher={Elsevier}
}

@article{bora2026inception,
  title={Inception UNet (IUNet) and multi-scale multi-branch (M\&M) UNet plug-in neural operators for online bias correction in E3SM atmosphere model},
  author={Bora, Aniruddha and Zhang, Shixuan and Shukla, Khemraj and Harrop, Bryce E and Karniadakis, George and Leung, L Ruby},
  journal={Journal of Computational Physics},
  pages={115294},
  year={2026},
  publisher={Elsevier}
}

@inproceedings{cobelli2023physics,
  title={Physics informed neural networks for wind field modeling in wind farms},
  author={Cobelli, P and Shukla, K and Nesmachnow, S and Draper, M},
  booktitle={Journal of Physics: Conference Series},
  volume={2505},
  number={1},
  pages={012051},
  year={2023},
  organization={IOP Publishing}
}

@article{shukla2025neurosem,
  title={NeuroSEM: A hybrid framework for simulating multiphysics problems by coupling PINNs and spectral elements},
  author={Shukla, Khemraj and Zou, Zongren and Chan, Chi Hin and Pandey, Additi and Wang, Zhicheng and Karniadakis, George Em},
  journal={Computer Methods in Applied Mechanics and Engineering},
  volume={433},
  pages={117498},
  year={2025},
  publisher={Elsevier}
}

@article{shukla2024comprehensive,
  title={A comprehensive and FAIR comparison between MLP and KAN representations for differential equations and operator networks},
  author={Shukla, Khemraj and Toscano, Juan Diego and Wang, Zhicheng and Zou, Zongren and Karniadakis, George Em},
  journal={Computer Methods in Applied Mechanics and Engineering},
  volume={431},
  pages={117290},
  year={2024},
  publisher={Elsevier}
}

@article{zhang2026turbulence,
  title={Turbulence closure in Reynolds-averaged Navier--Stokes and flow inference around a cylinder using physics-informed neural networks and sparse experimental data},
  author={Zhang, Zhen and Shukla, Khemraj and Wang, Zhicheng and Morales, Anthony and K{\"a}ufer, Theo and Salauddin, Sheikh and Walters, Nathan and Barrett, David and Ahmed, Kareem and Triantafyllou, Michael S and others},
  journal={Journal of Fluid Mechanics},
  volume={1034},
  pages={A16},
  year={2026},
  publisher={Cambridge University Press}
}

@article{peyvan2023high,
  title={High-order methods for hypersonic flows with strong shocks and real chemistry},
  author={Peyvan, Ahmad and Shukla, Khemraj and Chan, Jesse and Karniadakis, George},
  journal={Journal of Computational Physics},
  volume={490},
  pages={112310},
  year={2023},
  publisher={Elsevier}
}

@article{peyvan2026h3pc,
  title={H3PC: Hypersonic, high-order, high-performance code with adaptive mesh refinement and real chemistry},
  author={Peyvan, Ahmad and Shukla, Khemraj and Karniadakis, George Em},
  journal={Journal of Computational Physics},
  pages={115394},
  year={2026},
  publisher={Elsevier}
}

@article{shukla2026uncertainty,
  title={Uncertainty Quantification in PINNs for Turbulent Flows: Bayesian Inference and Repulsive Ensembles},
  author={Shukla, Khemraj and Zou, Zongren and Kaeufer, Theo and Triantafyllou, Michael and Karniadakis, George Em},
  journal={arXiv preprint arXiv:2604.17156},
  year={2026}
}

@article{chan2018discretely,
  title={On discretely entropy conservative and entropy stable discontinuous Galerkin methods},
  author={Chan, Jesse},
  journal={Journal of Computational Physics},
  volume={362},
  pages={346--374},
  year={2018},
  publisher={Elsevier}
}

@article{ranocha2021adaptive,
  title={Adaptive numerical simulations with Trixi. jl: A case study of Julia for scientific computing},
  author={Ranocha, Hendrik and Schlottke-Lakemper, Michael and Winters, Andrew R and Faulhaber, Erik and Chan, Jesse and Gassner, Gregor J},
  journal={arXiv preprint arXiv:2108.06476},
  year={2021}
}
\appendix

\section{Definitions of the Quantities of Interest}
\label{app:qoi}
 
This appendix gives the post-processing used to reduce a predicted or
reference flow field to the aerodynamic and thermal quantities reported in
\Cref{sec:results_ogive,sec:results_waverider_surface}. The same routine is
applied to the CFD fields and to the surrogate predictions.
 
The aerodynamic and thermal quantities of interest (QoIs) are evaluated directly from the
CFD reference fields and the corresponding surrogate predictions on the common structured
grid used for all geometries.  For each of the 37 ogive geometries, the post-processing
procedure uses the fields
\[
p,\qquad U,\qquad V,\qquad T,\qquad T_m,\qquad \mu,\qquad k,\qquad k_m,
\]
where $p$ is pressure, $(U,V)$ are the Cartesian velocity components, $T$ is the
translational temperature, $T_m$ is the modal temperature, $\mu$ is the dynamic viscosity,
$k$ is the translational thermal conductivity, and $k_m$ is the modal thermal conductivity.
Both the CFD and surrogate QoIs are computed with the same wall geometry, surface normals,
reference quantities, quadrature, and integration procedure so that the reported differences
reflect only differences in the predicted flow fields.

\subsection{Quantities Evaluated from the Predicted Fields}
\label{subsec:exp_qoi_metrics}
 
 
Field-level agreement does not necessarily imply equivalent accuracy in the
integrated or extremal quantities used for engineering assessment. The
predicted tangent-ogive fields are therefore mapped to the corresponding
physical geometry and used to reconstruct the drag coefficient $C_D$, lift
coefficient $C_L$, wall heat-flux distribution $q_w$, and peak wall heat flux
$q_{w,\max}$.
 
For the CFD cases considered here, the turbulent thermal conductivity is zero.
The conductive wall heat flux is reconstructed as
\begin{equation}
q_{w,\mathrm{cond}}
=
k_w
\left.
\frac{\partial T}{\partial n}
\right|_w
+
k_{m,w}
\left.
\frac{\partial T_m}{\partial n}
\right|_w,
\label{eq:exp_wall_heat_flux}
\end{equation}
where $n$ denotes the physical surface-normal direction and the subscript $w$
denotes wall evaluation. The predicted viscosity field is retained for
reconstruction of viscous stresses and their contribution to the integrated
surface forces.
 
\begin{figure}
    \centering
    \includegraphics[width=\linewidth]{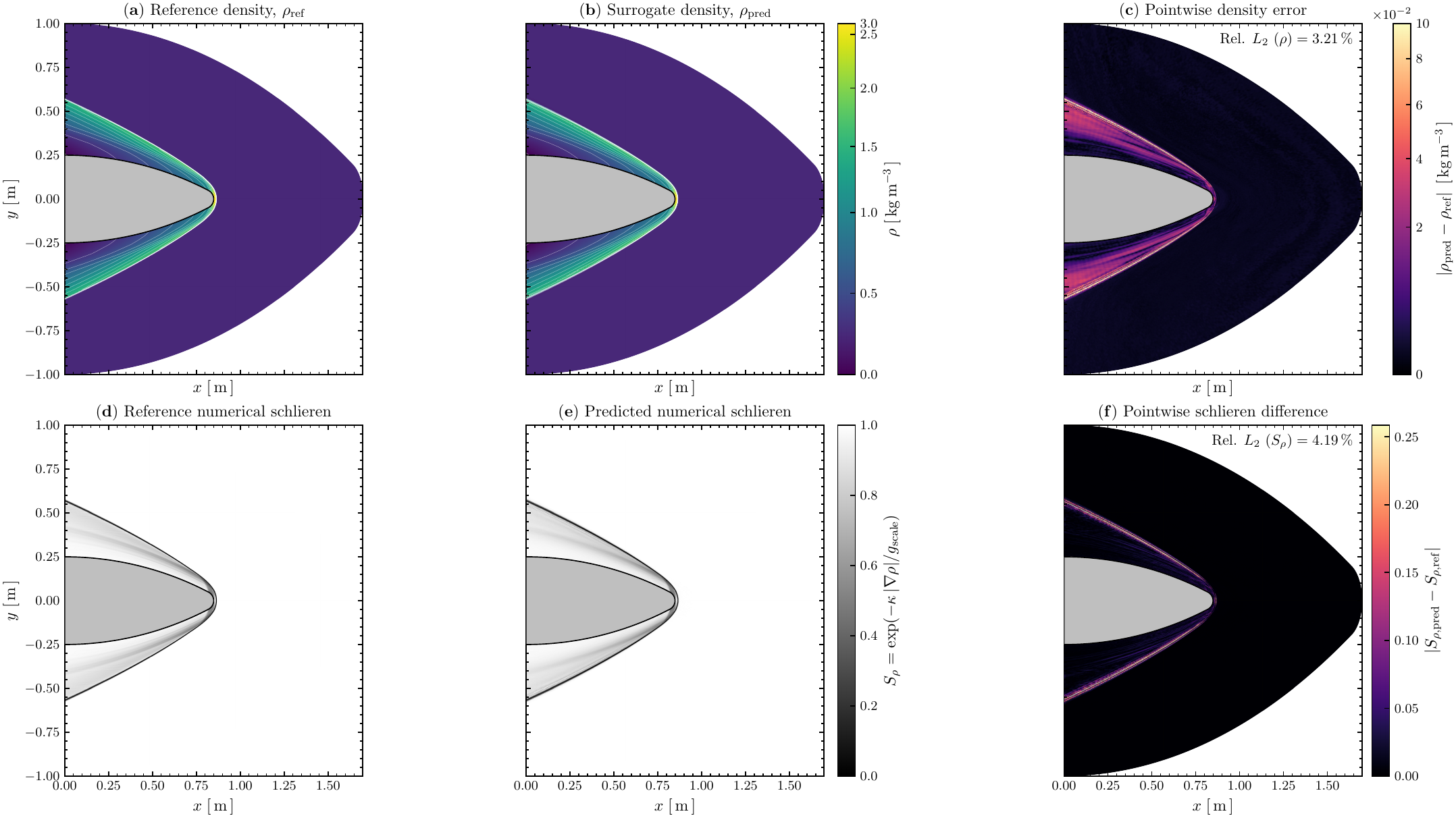}
    \caption{Representative interior LOO density-field prediction for
    $\xi_g=0.83156$.}
    \label{fig:run7_density_interpolation}
\end{figure}
 
For nonzero scalar quantities, relative errors are evaluated as
\begin{equation}
E_{C_D}
=
100\,
\frac{
\left|
\widehat{C}_D-C_D
\right|
}{
\left|C_D\right|
},
\qquad
E_{q_{\max}}
=
100\,
\frac{
\left|
\widehat{q}_{w,\max}-q_{w,\max}
\right|
}{
q_{w,\max}
}.
\label{eq:exp_drag_heat_errors}
\end{equation}
Because the lift coefficient may approach zero, its absolute error is used,
\begin{equation}
E_{C_L}^{\mathrm{abs}}
=
\left|
\widehat{C}_L-C_L
\right|.
\label{eq:exp_lift_absolute_error}
\end{equation}
The wall heat-flux distribution is assessed using
\begin{equation}
E_{q_w}
=
100\,
\frac{
\left\|
\widehat{q}_w-q_w
\right\|_{2,\Gamma_w}
}{
\left\|
q_w
\right\|_{2,\Gamma_w}
},
\label{eq:exp_wall_heat_flux_error}
\end{equation}
and the error in the peak location is
\begin{equation}
E_{\mathrm{peak,loc}}
=
\left\|
\widehat{\mathbf{x}}_{q_{w,\max}}
-
\mathbf{x}_{q_{w,\max}}
\right\|_2.
\label{eq:exp_peak_location_error}
\end{equation}
 
\begin{figure}
    \centering
    \includegraphics[width=\linewidth]{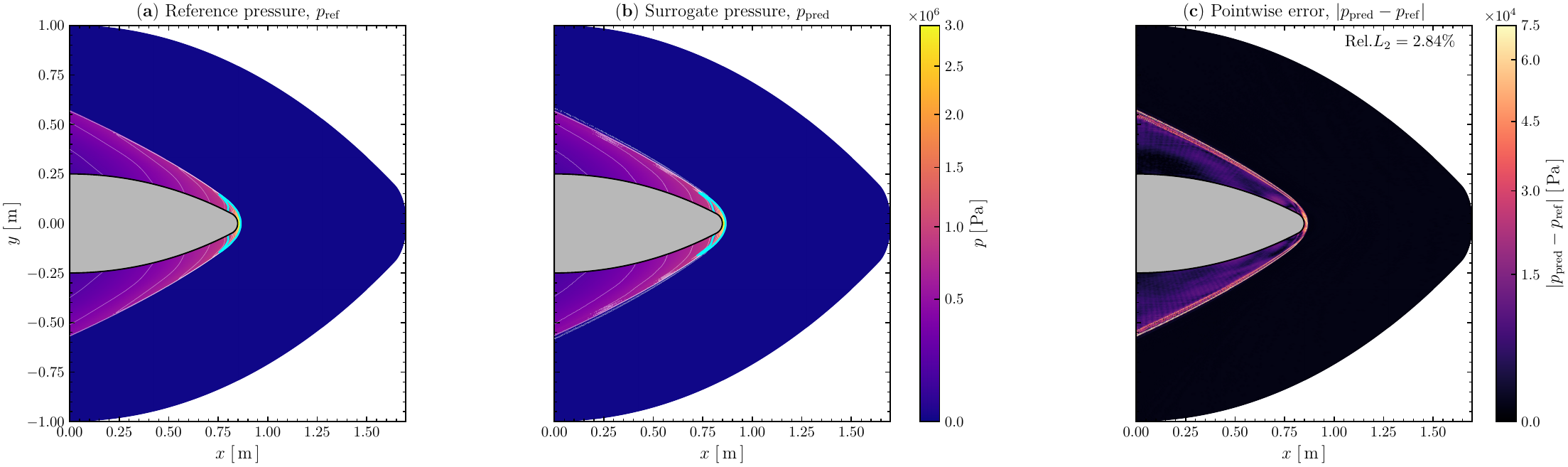}
    \caption{Representative interior LOO pressure-field prediction for
    $\xi_g=0.83156$.}
    \label{fig:run7_pressure_interpolation}
\end{figure}

\subsection{Wall geometry, tangent, and normal vectors}
\label{app:wall_geometry}
 
The canonical structured grid has dimensions $600\times 300$, with the first coordinate
corresponding to the surface direction and the second coordinate corresponding to the
wall-normal direction.  The wall is therefore represented by the $\eta=0$ grid line and
contains $N_w=600$ points.  Let the wall coordinates be
\[
\mathbf{x}_w(s)=\bigl(x_w(s),y_w(s)\bigr),
\]
where $s$ denotes the computational surface coordinate.  The unit tangent vector is
constructed from the physical wall coordinates as
\[
\mathbf{t}
=
\frac{\partial_s\mathbf{x}_w}
{\left\|\partial_s\mathbf{x}_w\right\|}
=
(t_x,t_y),
\]
and the corresponding unit normal is
\[
\mathbf{n}=(n_x,n_y).
\]
The sign of $\mathbf{n}$ is chosen such that it points from the solid surface into the fluid.
The physical wall arc-length coordinate $\ell$ is obtained from
\[
d\ell
=
\sqrt{(dx_w)^2+(dy_w)^2},
\]
and all local aerodynamic and thermal quantities are integrated using this physical arc length.
 
\subsection{Pressure and viscous surface traction}
\label{app:traction}
 
The pressure contribution to the surface traction is evaluated using gauge pressure,
\[
\mathbf{f}_p
=
-(p-p_\infty)\mathbf{n},
\]
where $p_\infty$ is the freestream pressure.  The viscous contribution is
\[
\mathbf{f}_\mu
=
\boldsymbol{\tau}\mathbf{n},
\]
where the compressible Newtonian stress tensor is
\[
\boldsymbol{\tau}
=
\begin{bmatrix}
\tau_{xx} & \tau_{xy}\\
\tau_{xy} & \tau_{yy}
\end{bmatrix},
\]
with
\begin{align}
\tau_{xx}
&=
\mu
\left(
2\frac{\partial U}{\partial x}
-\frac{2}{3}\nabla\cdot\mathbf{u}
\right),\\
\tau_{yy}
&=
\mu
\left(
2\frac{\partial V}{\partial y}
-\frac{2}{3}\nabla\cdot\mathbf{u}
\right),\\
\tau_{xy}
&=
\mu
\left(
\frac{\partial U}{\partial y}
+
\frac{\partial V}{\partial x}
\right),
\end{align}
and
\[
\nabla\cdot\mathbf{u}
=
\frac{\partial U}{\partial x}
+
\frac{\partial V}{\partial y}.
\]
The wall-normal derivatives and the velocity-gradient information required for the viscous
stress are reconstructed in physical space from the first several grid layers adjacent to the
wall.  A quadratic least-squares fit is used locally, with the no-slip wall condition
$U=V=0$ imposed at the surface.
 
The total surface traction is therefore
\[
\mathbf{f}
=
\mathbf{f}_p+\mathbf{f}_\mu.
\]
Let $\mathbf{e}_D$ and $\mathbf{e}_L$ denote the unit drag and lift directions, respectively.
The local drag and lift densities are
\[
D'(\ell)=\mathbf{f}\cdot\mathbf{e}_D,
\qquad
L'(\ell)=\mathbf{f}\cdot\mathbf{e}_L.
\]
The corresponding global forces are obtained by integration over the complete wall,
\[
D
=
\int_{\Gamma_w}
D'(\ell)\,d\ell,
\qquad
L
=
\int_{\Gamma_w}
L'(\ell)\,d\ell.
\]
Pressure and viscous contributions are also retained separately:
\[
D=D_p+D_\mu,
\qquad
L=L_p+L_\mu.
\]
The total drag coefficient and lift coefficient are defined as
\[
C_D
=
\frac{D}{q_\infty S_{\mathrm{ref}}},
\qquad
C_L
=
\frac{L}{q_\infty S_{\mathrm{ref}}},
\]
where $q_\infty$ is the freestream dynamic pressure and $S_{\mathrm{ref}}$ is the reference
area used consistently for both CFD and surrogate post-processing.  In the present
implementation,
\[
q_\infty = 1.6304862928\times 10^6~\mathrm{Pa}.
\]
 
\subsection{Wall heat flux}
\label{app:heat_flux}
 
The conductive wall heat flux is evaluated using the two-temperature form
\[
\boxed{
q_w
=
k_w
\left.
\frac{\partial T}{\partial n}
\right|_w
+
k_{m,w}
\left.
\frac{\partial T_m}{\partial n}
\right|_w
}
\]
rather than combining the two conductivities into a single coefficient.

Although the energy equation \eqref{eq:energyconservation} carries the
diffusive enthalpy flux $\sum_i h_i\mathbf{J}_i$, that term does not
contribute at the wall under the boundary conditions used here. The surface
is impermeable, so there is no mass injection and no convective enthalpy
flux through it, and it is treated as non-catalytic, so it neither promotes
nor inhibits recombination and the individual species diffusion fluxes
vanish there,
\[
\left.\mathbf{J}_i\cdot\mathbf{n}\right|_w = 0
\quad\Longrightarrow\quad
\left.\sum_i h_i\,\mathbf{J}_i\cdot\mathbf{n}\right|_w = 0 .
\]
With radiative transport neglected, the entire energy flux to the surface is
therefore conductive, and the two-temperature expression above is the
complete wall heat flux rather than a truncation of it. The assumption is not
innocuous and is stated as a modelling choice: at a catalytic wall the
dissociation energy released by surface recombination enters through exactly
the diffusive term set to zero here, and can be comparable to the conductive
contribution, so the present heat loads are those of the non-catalytic limit
and are lower than a fully catalytic surface would experience.

The translational and modal contributions are therefore
\[
q_{w,T}
=
k_w
\left.
\frac{\partial T}{\partial n}
\right|_w,
\qquad
q_{w,m}
=
k_{m,w}
\left.
\frac{\partial T_m}{\partial n}
\right|_w,
\]
so that
\[
q_w=q_{w,T}+q_{w,m}.
\]
 
The wall-normal derivatives are reconstructed in physical distance from the first five
wall-normal grid points using a quadratic least-squares approximation,
\[
\phi(d)
=
a_0+a_1d+a_2d^2,
\]
where $d$ is the physical distance measured along the local normal.  The wall derivative is
then
\[
\left.
\frac{\partial \phi}{\partial n}
\right|_w
=
a_1.
\]
For the translational temperature, the known isothermal wall condition
\[
T_w=216.65~\mathrm{K}
\]
is imposed directly in the fit.  No hard wall value is imposed on $T_m$ unless such a modal
wall condition is explicitly available from the flow solver.
 
The mean wall heat flux is defined by
\[
\bar q_w
=
\frac{1}{L_w}
\int_{\Gamma_w}
q_w(\ell)\,d\ell,
\]
where
\[
L_w
=
\int_{\Gamma_w}d\ell
\]
is the total wall arc length.  The maximum wall heat flux is
\[
q_{w,\max}
=
\max_{\Gamma_w}q_w(\ell).
\]
 
\subsection{Nusselt number}
\label{app:nusselt}
 
A common normalization is used for the CFD and surrogate predictions of each geometry.
The local Nusselt number is defined as
\[
Nu(\ell)
=
\frac{q_w(\ell)L_{\mathrm{ref}}}
{k_{\mathrm{ref}}\Delta T_{\mathrm{ref}}},
\]
where $L_{\mathrm{ref}}$ is the axial length of the corresponding ogive,
\[
L_{\mathrm{ref}}
=
x_{w,\max}-x_{w,\min},
\]
and the reference conductivity is extracted from the CFD farfield as
\[
k_{\mathrm{ref}}
=
k_\infty+k_{m,\infty}.
\]
The global Nusselt number is constructed from the mean wall heat flux,
\[
\boxed{
Nu_{\mathrm{global}}
=
\frac{\bar q_w L_{\mathrm{ref}}}
{k_{\mathrm{ref}}\Delta T_{\mathrm{ref}}}
},
\]
while the maximum local Nusselt number is
\[
Nu_{\max}
=
\max_{\Gamma_w}Nu(\ell).
\]
 
Because the freestream temperature and wall temperature are equal in the present data set,
$T_\infty-T_w=0$ and cannot be used as the normalization temperature difference.  Instead,
a fixed recovery-temperature scale is used only for non-dimensionalization.  Using a
calorically perfect-gas reference with
\[
\gamma=1.4,\qquad R=287.05~\mathrm{J/(kg\,K)},\qquad Pr=0.72,
\]
the freestream Mach number is
\[
M_\infty
=
\frac{U_\infty}
{\sqrt{\gamma R T_\infty}},
\]
the laminar recovery factor is
\[
r=\sqrt{Pr},
\]
and the recovery temperature is
\[
T_r
=
T_\infty
\left[
1+
r\frac{\gamma-1}{2}M_\infty^2
\right].
\]
The reference temperature difference is then
\[
\Delta T_{\mathrm{ref}}
=
T_r-T_w.
\]
For the present freestream conditions,
\[
M_\infty \approx 10.9991,
\qquad
T_r \approx 4664.661~\mathrm{K},
\qquad
\Delta T_{\mathrm{ref}}
\approx 4448.011~\mathrm{K}.
\]
This recovery temperature is used strictly as a fixed non-dimensionalization scale; it is not
interpreted as a fully coupled reacting, nonequilibrium adiabatic-wall temperature.
 
\subsection{Zero-angle-of-attack lift and symmetry diagnostics}
\label{app:symmetry}
 
All geometries considered here are symmetric and are evaluated at zero angle of attack.
Accordingly, the physically expected global lift is approximately zero,
\[
L\approx 0,
\qquad
C_L\approx 0,
\]
because the upper- and lower-surface transverse force contributions cancel.  A conventional
relative error in lift,
\[
\frac{|L_{\mathrm{sur}}-L_{\mathrm{CFD}}|}
{|L_{\mathrm{CFD}}|},
\]
is therefore ill-conditioned when $L_{\mathrm{CFD}}\rightarrow 0$.  To avoid a misleadingly
large percentage caused by division by an almost-zero reference, the residual lift is instead
normalized by the drag scale:
\[
\boxed{
E_{L/D}
=
100\frac{|L|}{|D|}
}
\]
and equivalently
\[
\boxed{
E_{C_L/C_D}
=
100\frac{|C_L|}{|C_D|}.
}
\]
Because $C_L$ and $C_D$ share the same normalization,
\[
\frac{|C_L|}{|C_D|}
=
\frac{|L|}{|D|}.
\]
 
The surrogate lift error is also reported relative to the CFD drag magnitude,
\[
\boxed{
E_{L,\mathrm{drag}}
=
100
\frac{|L_{\mathrm{sur}}-L_{\mathrm{CFD}}|}
{|D_{\mathrm{CFD}}|}
}
\]
and, equivalently,
\[
\boxed{
E_{C_L,C_D}
=
100
\frac{|C_{L,\mathrm{sur}}-C_{L,\mathrm{CFD}}|}
{|C_{D,\mathrm{CFD}}|}
}.
\]
These metrics provide a stable measure of spurious transverse force at zero angle of attack.
 
To identify the source of any residual lift, the pressure and viscous contributions are
reported separately through
\[
100\frac{|L_p|}{|D|}
\qquad\text{and}\qquad
100\frac{|L_\mu|}{|D|}.
\]
This separation is useful because pressure asymmetry can generate a finite global lift even
when the overall field prediction error is small.
 
\subsection{Upper--lower pressure symmetry}
\label{app:pressure_symmetry}
 
For each upper-surface wall point $(x_i,y_i)$, a corresponding lower-surface point is selected
by geometric reflection across the symmetry axis,
\[
(x_i,y_i)
\longrightarrow
(x_i,-y_i),
\]
followed by nearest-point matching on the discrete wall grid.  Symmetry is assessed using
gauge pressure,
\[
p'=p-p_\infty,
\]
so that the freestream offset does not artificially reduce the apparent mismatch.
 
Let $\mathbf{p}'_u$ and $\mathbf{p}'_l$ denote the paired upper- and lower-surface gauge
pressure vectors.  The relative $L_2$ pressure-symmetry error is defined as
\[
\boxed{
E_{p,\mathrm{sym}}
=
\frac{
\|\mathbf{p}'_u-\mathbf{p}'_l\|_2
}{
\sqrt{
\frac{1}{2}
\left(
\|\mathbf{p}'_u\|_2^2+
\|\mathbf{p}'_l\|_2^2
\right)
}
}.
}
\]
The reported percentage is $100E_{p,\mathrm{sym}}$.  A dimensional symmetry measure is also
reported as
\[
\boxed{
E_{p,\mathrm{RMS}}
=
\sqrt{
\frac{1}{N_p}
\sum_{i=1}^{N_p}
\left(
p'_{u,i}-p'_{l,i}
\right)^2
},
}
\]
where $N_p$ is the number of paired wall points.  These diagnostics are computed separately
for the CFD and surrogate fields.
 
\subsection{CFD--surrogate comparison metrics}
\label{app:comparison_metrics}
 
For global quantities whose CFD reference value is not close to zero, the relative error is
reported as
\[
E_Q
=
100
\frac{
|Q_{\mathrm{sur}}-Q_{\mathrm{CFD}}|
}{
|Q_{\mathrm{CFD}}|
}.
\]
This definition is used for quantities such as total drag, drag coefficient, mean heat flux,
maximum heat flux, global Nusselt number, and maximum Nusselt number.  Since the CFD and
surrogate Nusselt numbers use the same $L_{\mathrm{ref}}$, $k_{\mathrm{ref}}$, and
$\Delta T_{\mathrm{ref}}$, the relative error in $Nu_{\mathrm{global}}$ is identical to the
relative error in $\bar q_w$, while the relative error in $Nu_{\max}$ is identical to that
in $q_{w,\max}$.
 
For lift-related quantities at zero angle of attack, the drag-scaled metrics described above
are used instead of a conventional relative error because the physical reference value is
approximately zero.

\subsection{Waverider Aerodynamic Axes and Force Convention}
\label{app:axes}
 
\label{sec:axes}
 
The aerodynamic coefficients of the AEDC waverider are obtained by integrating
the wall traction over the vehicle skin (nose, leading edge, upper and lower
panels). The integrated force is
\begin{equation}
  \mathbf{F}(\alpha) = \sum_{e=1}^{N_c} \mathbf{t}_e(\alpha)\, A_e ,
  \label{eq:force}
\end{equation}
where $\mathbf{t}_e = (t_x, t_y, t_z)_e$ is the total
(pressure plus viscous) traction reported by the flow solver in cell $e$.
The mesh is fixed in the body frame, and the angle of attack enters only
through the freestream direction. Drag and lift are therefore the projections
of $\mathbf{F}$ onto the wind axes,
\begin{equation}
  \hat{\mathbf{e}}_D = (\cos\alpha,\ \sin\alpha,\ 0), \qquad
  \hat{\mathbf{e}}_L = (-\sin\alpha,\ \cos\alpha,\ 0),
  \label{eq:wind_axes}
\end{equation}
which define the $x$--$y$ plane as the pitch plane, with lift positive along
$+y$. This choice was not assumed. It was established from the CFD data
before any surrogate was trained, using three independent checks.
 
\paragraph{(i) Lateral symmetry identifies $z$ as the spanwise axis.}
The waverider is symmetric about its $x$--$y$ plane. Over all 29 CFD cases,
the integrated side force satisfies $|F_z| \le 1.3\times10^{-5}$\,N, whereas
$|F_y|$ reaches $4.4\times10^{3}$\,N, a ratio below $10^{-8}$. A vanishing
force component can only be spanwise. Had the pitch plane been $x$--$z$, $F_z$
would carry the lift and vary strongly with $\alpha$.
 
\paragraph{(ii) Only $F_y$ responds to pitch.}
$F_y$ increases monotonically from $-1.48\times10^{3}$\,N at
$\alpha = -14^\circ$ to $4.44\times10^{3}$\,N at $\alpha = +14^\circ$, about
$2.1\times10^{2}$\,N per degree, and changes sign between $-5^\circ$ and
$-4^\circ$. $F_z$ shows no dependence on $\alpha$. The surface fields agree.
The RMS of $t_y$ varies by a factor of $4.8$ across the AoA range. Its
minimum is shifted to $\alpha \approx -6^\circ$ by the up--down asymmetry of
the waverider, whose compression surface lies underneath. The RMS wall heat
flux reaches its minimum at nearly the same angle, $\alpha \approx -7^\circ$,
consistent with a windward/leeward exchange in the $x$--$y$ plane. In
contrast, $t_z$ is nearly symmetric about $\alpha = 0$ and varies by only
a factor of $1.5$, as expected for the spanwise traction on the left and
right panels.
 
\paragraph{(iii) Sign and physical consistency.}
With the solver's traction taken as the force on the body ($F = +\sum \mathbf{t}\,A$),
\eqref{eq:wind_axes} gives $C_D > 0$ at every AoA and a monotonically
increasing $C_L(\alpha)$. Zero lift occurs at $\alpha_{0} \approx -5^\circ$,
where $C_L = -8\times10^{-5}$. Minimum drag, $C_D \approx 1.5\times10^{-3}$,
occurs at the same angle, as required by a drag polar
$C_D \approx C_{D,0} + k\,(C_L - C_{L,\min D})^2$. At $\alpha = 0$ the wind
and body axes coincide, and the computed drag and lift equal $F_x$ and $F_y$
exactly, which checks the implementation of \eqref{eq:wind_axes}. Reversing
the force sign would give negative drag; choosing the $x$--$z$ plane would
give a lift coefficient that is identically zero. Both alternatives are
therefore ruled out by the data.
 
\section{Robustness of the Three-Objective Selection}
\label{app:robustness}
 
The four checks referred to in \Cref{sec:pareto_optimization}.
 
To assess whether the selected design depends strongly on a particular
multiobjective formulation or surrogate realization, four complementary
robustness tests were performed in addition to the direct Pareto analysis.
 
\paragraph{$\epsilon$-constraint reconstruction.}
The first check reformulated the problem using the classical
$\epsilon$-constraint approach \cite{chankong1983,miettinen1999}.  Drag was
retained as the scalar objective, while the two thermal quantities were
converted into explicit upper-bound constraints:
\begin{equation}
    \begin{aligned}
        \min_{\xi}\quad & C_D(\xi),\\
        \text{subject to}\quad
        &q_{w,\max}(\xi)\le\epsilon_q,\\
        &\dot Q'(\xi)\le\epsilon_Q,\\
        &100|C_L|/|C_D|\le1\%.
    \end{aligned}
    \label{eq:epsilon_constraint}
\end{equation}
The thermal bounds were varied systematically over normalized levels between
0.20 and 0.95.  Twenty-seven combinations admitted at least one feasible
design.  As the bounds were relaxed, the selected solution moved smoothly from
$\xi=0.708$ to $\xi=0.988$.  Representative solutions progressed through
$\xi=0.708$, $0.7265$, $0.7875$, $0.845$, $0.915$, and $0.988$ as the
admissible thermal region expanded.  This progression is important: the
$\epsilon$-constraint method does not return a single fixed optimum because the
constraints deliberately select different points along the trade-off.  The
smooth migration of the solution therefore provides an independent
reconstruction of the same Pareto structure obtained by direct non-dominated
sorting.
 
\paragraph{Weighted Tchebycheff scalarization.}
A second independent scalarization was constructed using the normalized
weighted Tchebycheff metric \cite{miettinen1999},
\begin{equation}
    \min_{\xi}
    \max_{i=1,2,3}
    \left[
        w_i\widehat f_i(\xi)
    \right],
    \qquad
    w_i>0,\qquad
    \sum_{i=1}^{3}w_i=1.
    \label{eq:tchebycheff}
\end{equation}
Thirty-six different weight combinations were examined.  As expected, the
selected design shifted when the relative importance of drag, peak heating,
and integrated heating was changed, giving
$\xi\in[0.6275,0.9180]$.  The purpose of this test is therefore not to force
all weight combinations to recover the same design, but to determine whether
the central compromise remains in the same region when a mathematically
different scalarization is used.  The mean selected parameter over the full
weight sweep was
\begin{equation}
    \overline{\xi}_{\rm Tch}=0.8188,
\end{equation}
which is close to the direct Pareto compromise $\xi^\star=0.825$.  Thus, the
location of the central compromise is not tied to the normalized-ideal
selection criterion alone.
 
\paragraph{Local design robustness.}
The third check examined whether the selected compromise corresponds to a
sharp numerical point or to a broader stable region of the design space.
Neighboring designs were evaluated around $\xi^\star=0.825$.  For a
perturbation of approximately $-0.01$, the changes in $C_D$, $q_{w,\max}$, and
$\dot Q'$ were $+0.31\%$, $+0.58\%$, and $+0.03\%$, respectively.  For a
perturbation of approximately $+0.01$, the corresponding changes were
$-0.03\%$, $-0.11\%$, and $+0.31\%$.  Even over a wider neighborhood of
approximately $\xi^\star\pm0.02$, the largest observed variation among the
three objectives was only about $1.2\%$.  This behavior indicates that the
selected design lies in a smooth compromise region rather than at an isolated
numerical spike.  It also shows directly that the three objectives do not all
improve in the same direction near the optimum, which is consistent with the
multiobjective structure observed in Fig.~\ref{fig:pareto3d}.
 
\paragraph{Leave-one-out surrogate robustness.}
The fourth check quantified sensitivity to the particular surrogate
realization.  The complete optimization was repeated with five independently
trained leave-one-out models \cite{stone1974}.  Three folds satisfied the
nominal $1\%$ symmetry criterion and yielded strict-feasible compromise
parameters
\begin{equation}
    \xi^\star_{\rm LOO}
    =
    \{0.8250,\;0.8675,\;0.7805\},
\end{equation}
with
\begin{equation}
    \overline{\xi^\star_{\rm LOO}}=0.8243,
    \qquad
    \sigma_{\xi^\star_{\rm LOO}}=0.0435  .
    \label{eq:loo_summary}
\end{equation}
The mean strict-feasible LOO compromise is therefore essentially identical to
the baseline value $\xi^\star=0.825$.  For the remaining two folds, the minimum
residual-lift levels were only slightly above the nominal tolerance,
approximately $1.24\%$ and $1.07\%$.  Since residual lift is used here as a
symmetry-quality diagnostic rather than as one of the optimized aerodynamic--
thermal objectives, these two folds were retained in the broader diagnostic
surrogate-robustness assessment without redefining the original constraint.
The fold-to-fold variability therefore provides a practical estimate of the
sensitivity of the selected design to the trained surrogate while preserving
the production optimization criterion.
 
\begin{table}[t]
\centering
\caption{Summary of the three-objective optimization and the four robustness
checks.}
\label{tab:optimization_validation_summary}
\begin{tabular}{@{}lcc@{}}
\toprule
Analysis & Selected/representative $\xi$ & Main observation \\
\midrule
Direct Pareto
& 0.825
& Baseline three-objective compromise \\
 
$\epsilon$-constraint
& 0.708--0.988
& Smoothly reconstructs the trade-off as bounds vary \\
 
Weighted Tchebycheff
& mean 0.8188
& Central compromise remains near the baseline \\
 
Local perturbation
& $0.825\pm0.01$
& Objective changes remain below $0.6\%$ \\
 
LOO surrogates
& $0.8243\pm0.0435$
& Mean strict-feasible compromise matches baseline \\
\bottomrule
\end{tabular}
\end{table}
 
Taken together, the five analyses provide complementary evidence that the
reported compromise is not an artifact of a single ranking rule,
scalarization, local sample, or surrogate realization.  The direct Pareto
analysis establishes the three-way aerodynamic--thermal trade-off; the
$\epsilon$-constraint and Tchebycheff formulations reproduce the same central
design region from different mathematical perspectives; the local
perturbation study demonstrates smooth objective variation around the selected
design; and the LOO study shows that independently trained surrogates recover
a mean strict-feasible compromise essentially identical to the baseline
result.  These tests establish internal robustness of the surrogate-based
optimization while remaining distinct from an independent high-fidelity
validation of the optimized geometry.

\section{Inference Procedure for the Volume Surrogate}
\label{app:volume_inference}
 
The volume surrogate of \Cref{subsec:waverider_method} is assembled from four
trained components, three of which are frozen before the fourth is fitted.
This appendix states the order in which they are evaluated at a new angle of
attack, and what each requires, since no CFD quantity is available at that
point.
 
\begin{figure}[t]
  \centering
  \includegraphics[width=\linewidth]{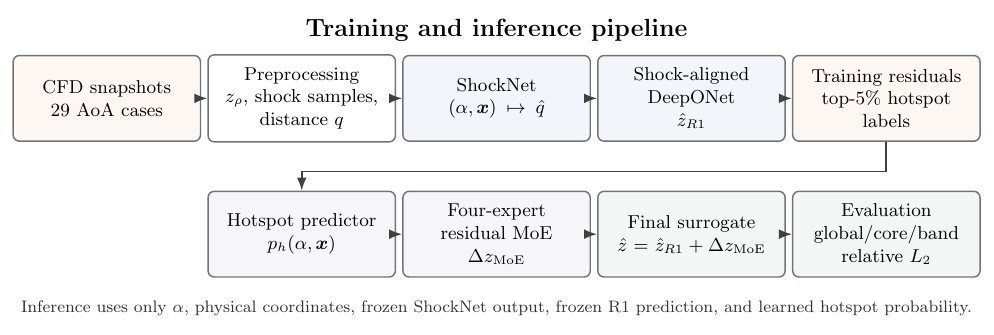}
  \caption{Training and inference pipeline. CFD-derived shock and
  residual information are used only during offline training. At
  inference, the surrogate requires only the angle of attack
  $\alpha$ and spatial coordinates $\bm{x}$; shock location and
  hotspot probability are supplied by the frozen learned components.}
  \label{fig:pipeline}
\end{figure}
 
Given an unseen angle $\alpha^\star$ and the nodal coordinates
$\mathbf{x}$ of the fixed mesh, the prediction proceeds as follows.
 
\begin{enumerate}
  \item \emph{Shock coordinate.} The frozen ShockNet returns the normalised
        distance $\widehat d(\mathbf{x};\alpha^\star)$ to the shock surface,
        from which the bounded coordinate $\chi = \tanh \widehat d$ is
        formed. ShockNet takes only $\alpha^\star$ and $\mathbf{x}$, so no
        reference solution is needed to locate the shock; this is what
        permits the shock-aligned representation to be used at an angle for
        which no CFD exists.
 
  \item \emph{Stage-one field.} The trunk is evaluated on the augmented
        coordinates $\boldsymbol{\zeta} = (x,y,z,\chi)$ and the branch on
        $\alpha^\star$, and the standardised log-density follows from the
        contraction
        $\widehat z_{R1} = \mathbf{Q}\,\mathcal{B}_\phi(\alpha^\star)$,
        with $\mathbf{Q}$ the orthonormalised basis stored after training.
 
  \item \emph{Hotspot probability.} The classifier returns
        $p_h$ from $(\alpha^\star, \mathbf{x}, \chi, \widehat z_{R1})$. Every
        argument is available at this point; in particular $p_h$ stands in
        for the stage-one residual, which cannot be formed at an unseen angle
        because it requires the reference field.
 
  \item \emph{Residual correction.} The gate and the four experts are
        evaluated on $(\alpha^\star, \mathbf{x}, \chi, \widehat z_{R1}, p_h)$
        and combined into $\Delta z_{\mathrm{MoE}}$.
 
  \item \emph{Inverse transform.} The physical density follows from
        \begin{equation}
          \widehat\rho(\mathbf{x};\alpha^\star)
          = \exp\!\Big[\varsigma_{\rho}\big(\widehat z_{R1}
            + \Delta z_{\mathrm{MoE}}\big) + m_{\rho}\Big],
          \label{eq:app_volume_inverse}
        \end{equation}
        with $m_\rho$ and $\varsigma_\rho$ the training-set statistics of
        $\log\rho$.
\end{enumerate}
 
Three points follow from this ordering. First, every CFD-derived quantity ---
the shock surface, the residual and the hotspot labels --- enters only during
offline training; at inference the surrogate consumes an angle and a set of
coordinates and nothing else. Second, the components must be evaluated in the
order given, since each takes the output of the last as an input, so the
correction cannot be applied without first forming the stage-one field.
Third, inference is inexpensive relative to the calculation it replaces.
Evaluating the full chain --- ShockNet, trunk and branch, hotspot classifier
and mixture of experts --- over the $5.08\times10^{7}$ mesh nodes at one
angle of attack takes approximately $240$~s. Each
reference solution, by contrast, was run on the AFRL Warhawk system on $8$
standard nodes, $1024$ cores in total, for $20$ hours of wall-clock time,
or $2.05\times10^{4}$ core-hours per converged case. Reconstructing the
full volume field at a new incidence is therefore some $300$ times faster in
wall-clock time than computing it, and cheaper by a further factor of $1024$
in resources, since the surrogate occupies one device rather than $1024$
cores. The margin is far smaller than for the surface quantities --- the
$561$-angle scan of \Cref{sec:results_waverider_surface} takes $220$~s in
total, where the equivalent set of converged solutions would require
approximately $1.2\times10^{7}$ core-hours --- because the volume surrogate
reconstructs $5.08\times10^{7}$ nodes through a seven-component chain rather
than $2.7\times10^{5}$ surface cells through one. It is accordingly a tool
for field reconstruction at a chosen incidence rather than for sweeping the
design space, which is consistent with its role in this work.

\end{document}